\documentclass{aa}  
\usepackage{graphicx}
\usepackage{txfonts}
\usepackage{rotating}
\usepackage{multirow} 
\usepackage{appendix}
\usepackage{latexsym}
\usepackage{lscape}
\usepackage{afterpage}

\newcommand{\kms}{$\mbox{km\,s}^{-1}$\,}
\newcommand{\nskms}{$\mbox{km\,s}^{-1}$}

\begin{document} 

\title{Unexplored outflows in nearby low luminosity AGNs:\\ the case of NGC\,1052}

\author{S.~Cazzoli\inst{1} \and L.~Hermosa~Mu{\~n}oz\inst{1} \and I.~M{\'a}rquez\inst{1}  \and  J.~Masegosa\inst{1}  \and  {\'A}.~Castillo-Morales\inst{2,3} \and A.~Gil de Paz\inst{2,3} \and L.~Hern{\'a}ndez-Garc{\'i}a\inst{4,5} \and F.~La Franca\inst{6} \and C.~Ramos 
Almeida\inst{7,8}  }

\institute{IAA-CSIC --  IAA - Instituto de Astrof{\'i}sica de Andaluc{\'i}a (CSIC), Apdo. 3004, 18008, Granada, Spain
\email{sara@iaa.es}
\and
Departamento de F{\'i}sica de la Tierra y Astrof{\'i}sica, Universidad Complutense de Madrid, E-28040 Madrid, Spain
\and
Instituto de F{\'i}sica de Part{\'i}culas y del Cosmos IPARCOS, Facultad de Ciencias F{\'i}sicas, Universidad Complutense de Madrid, E-28040 Madrid, Spain
\and
Millennium Institute of Astrophysics, Nuncio Monse\~{n}or S{\'o}tero Sanz 100, Providencia, Santiago, Chile
\and
Instituto de F{\'i}sica y Astronom{\'i}a,  Universidad de Valpara{\'i}so, Av. Gran Breta\~{n}a 1111, Playa Ancha, Chile
\and
Dipartimento di Matematica e Fisica, Universita Roma Tre, via della Vasca Navale 84, I-00146 Roma, Italy
\and
Instituto de Astrof{\'i}sica de Canarias, C/V{\'i}a L{\'a}ctea s/n, E-38205 La Laguna, Tenerife, Spain  
\and
Universidad de La Laguna, Dep.to de Astrof{\'i}sica, Av.da Astrof{\'i}sico F. S{\'a}nchez s/n, E-38206 La Laguna, Tenerife, Spain.
}

   \date{Received day month year; accepted day month year}


 \abstract
 {Multi-phase outflows play a central role in galaxy evolution shaping the properties of galaxies. Understanding outflows and their effects in low luminosity AGNs, such as LINERs, is essential. Indeed these latter bridge the gap between normal and active galaxies, being the most numerous AGN population in the local Universe.}
{Our goal is to analyse the kinematics and ionisation mechanisms of the multi-phase gas of NGC\,1052, the prototypical LINER, in order to detect and map ionised and neutral phases of the putative outflow.} 
{We obtained Very Large Telescope MUSE and Gran Telescopio Canarias MEGARA optical integral field spectroscopy data for NGC\,1052. Besides stellar kinematics maps, by modelling spectral lines with multiple Gaussian components, we obtained flux, kinematic, and excitation maps of both ionised and neutral gas. } 
{The stars are distributed in a dynamically hot disc (V/$\sigma$\,$\sim$\,1.2), with a centrally peaked velocity dispersion map ($\sigma_{c\rm}$\,=\,201\,$\pm$\,10\,\nskms) and large observed velocity amplitudes ($\Delta$V\,=\,167\,$\pm$\,19\,\nskms). The ionised gas, probed by the primary component is detected up to $\sim$\,30$\arcsec$ ($\sim$\,3.3\,kpc) mostly in the polar direction with blue and red velocities ($\mid$V$\mid$\,$<$\,250\,\kms). The velocity dispersion map shows a notable enhancement ($\sigma$\,$>$\,90\,\nskms) crossing the galaxy along the major axis of rotation in the central 10\arcsec. The secondary component has a bipolar morphology, velocity dispersion larger than 150\,\kms and velocities up to 660\,\nskms. A third component is detected with MUSE (barely with MEGARA)  but not spatially resolved. The BLR-component (used to model the broad H$\alpha$ emission only) has a full width half maximum of 2427\,$\pm$\,332 and 2350\,$\pm$\,470\,\kms for MUSE and MEGARA data, respectively.  The maps of the NaD absorption indicate optically thick neutral gas with complex kinematics. The velocity field is consistent with a slow rotating disc ($\Delta$V\,=\,77\,$\pm$\,12\,\nskms) but the velocity dispersion map is off-centred without any counterpart in the (centrally peaked) flux map.}
{We found evidence of an ionised gas outflow (secondary component) with mass of 1.6\,$\pm$\,0.6 $\times$\,10$^{5}$ M$_{\sun}$, and mass rate of 0.4\,$\pm$\,0.2 M$_{\sun}$yr$^{-1}$. The outflow is propagating in a cocoon of gas with enhanced turbulence and might be triggering the onset of kpc-scale buoyant bubbles (polar emission), both probed by the primary component. Taking into account the energy and kinetic power of the outflow (1.3\,$\pm$\,0.9 $\times$ 10$^{53}$ erg and 8.8\,$\pm$\,3.5 $\times$ 10$^{40}$ erg\,s$^{-1}$, respectively)  as well as its alignment with both the jet and the cocoon, and that the gas is collisionally ionised (due to gas compression), we consider that the most likely power source of the outflow is the jet, although some contribution from the AGN is possible. The hints of the presence of a neutral gas outflow are weak.}
    

\titlerunning{Outflows in LINERs, the case of NGC\,1052}
\authorrunning{S. Cazzoli et al.}

\maketitle

\section{Introduction}
\label{S_introduction}

\noindent Outflows, produced by Active Galactic Nuclei (AGNs) and intense episodes of star-formation, have been proposed to play a crucial role in regulating the built up of stellar mass and black-hole mass growth through negative and positive feedback (see e.g. \citealt{Kormendy2013} and reference therein).  Recently, it has been shown that outflows can be driven also by radio-jets (e.g. \citealt{Morganti2005,Harrison2014,Morganti2018, Jarvis2019,Molyneux2019, Venturi2021}). The outflows might result as an important source of feedback as they evolve and heat the interstellar medium (ISM) 
preventing the cooling of the gas on possibly large scales.\\
The different gas phases of outflows have been widely studied in different galaxies populations (\citealt{Veilleux2005, Veilleux2020}, for  reviews) mostly via long-slit spectroscopy (e.g. \citealt{Heckman2000, Rupke2002, Arribas2014, VillarMartin2018,Rose2018, LHG2019, Saturni2021}) and Integral Field Spectroscopy  observations  (IFS, e.g. \citealt{Cazzoli2014, Cresci2015,   CRA2017, Maiolino2017, Bosch2019, Perna2020, Perna2021, Comeron2021}). \\
\noindent To date, the vast majority of studies of multi-phase outflows  and feedback have focussed on local luminous and ultra luminous infrared galaxies (U/LIRGs, e.g. \citealt{Rupke2013,Cazzoli2014,Cazzoli2016, PereiraSantaella2016, PereiraSantaella2020, Fluetsch2021}) and luminous AGNs  (e.g. quasars or Seyferts galaxies; \citealt{Feruglio2010, MullerSanchez2011, Fiore2017, Brusa2018, Venturi2018, Cazzoli2020}). These works demonstrated the power of the 3D IFS in studies of this kind. For example, the wealth of optical and infrared (IR) IFS-data enable the exploration of possible scaling relations between AGN properties, host galaxy properties, and outflows (e.g. \citealt{Kang2018}, \citealt{Fluetsch2019, Kakkad2020, RuschelDutra2021, Avery2021, Luo2021}, \citealt{Singha2021}, and references therein). \\
For low-luminosity AGNs, as low-ionisation nuclear emission line regions (LINERs), no systematic search for outflows has been done yet. Except for individual discoveries (e.g. \citealt{Dopita2015, Raimundo2021}) the only systematic studies are by \citet{Cazzoli2018} and \citet{LHM2020} about ionised gas outflows in type-1 and type-2 LINERs, respectively. These two latter works, where 30 LINERS were studied on the basis of optical long-slit spectroscopy, showed that multi-phase outflows are common in LINERS (detection rate: 60$\%$, \citealt{Cazzoli2018}), showing an intriguing ionisation structure  in which low ionisation lines (e.g. [O\,I]$\lambda\lambda$6300,6364) behave differently than high ionisation lines (e.g. [O\,III]$\lambda\lambda$4959,5007). Most of these  spectroscopically-identified outflows show in their \textit{HST}-H$\alpha$ $\lambda$6563 image \citep{Pogge2000, Masegosa2011, LHM2022} a large-scale biconical or bubble-like shape along with evident spatially resolved sub-structures, such as gas $\sim$\,20-70 pc wide clumps.\\
A 3D description of multi-phase outflows and the quantification of their feedback (mass, energy and their rates) in low-luminosity AGNs, as LINERs, is lacking. The exploration of outflows and feedback for this AGN-family is crucial to improve our understanding of galaxy evolution as these sources are thought to bridge the gap between normal and luminous AGNs, and belongs to the most numerous AGN population in the local Universe (\citealt{Ho2008}, for a review). \\
\noindent NGC\,1052 (MCG-01-07-034, PKS\,0238-084) is considered as the prototypical LINER in the local Universe (z\,$\sim$\,0.005). Table\,\ref{T_properties} summarises the basic properties for this object.\\ 
\noindent There are four previous IFS studies focussing on NGC\,1052: \citet{Sugai2005}, \citet{Dopita2015} and \citet{Dahmer2019a,Dahmer2019b}. \citet{Sugai2005} probed the bulk of the outflow with channel maps of the [O\,III] emission line thanks to Kyoto3DII/Subaru data over the innermost 3$\arcsec$\,$\times$\,3$\arcsec$. \citet{Dopita2015} (D15, hereafter) analysed the stellar and gas kinematics within the inner 25$\arcsec$\,$\times$\,38$\arcsec$ using WiFeS/ANU data. The authors mapped the emission line properties on scales of hundred of parsecs (spatial sampling $\sim$\,1$\farcs$3) mainly studying shocks, with no detailed information on the properties of the different kinematic components. \citet{Dahmer2019a,Dahmer2019b} (DH19a,b, hereafter) mapped optical and NIR lines in the inner 3$\farcs$5\,$\times$\,5\arcsec (similar to the work by \citealt{Sugai2005}) exploiting GMOS/GEMINI data. The richness of tracers provided by the combination of multi-wavelength data offers a more detailed view than the previous works of the complex kinematics in NGC\,1052. Nevertheless the large scale (i.e. kpc-scale) emission is not covered by the GEMINI data set. Summarising, all these IFS-based works support the presence in NGC\,1052 of an emission line outflow possibly extended on kpc-scales.\\
\noindent In this paper, we use spectral and spatial capabilities of MUSE/VLT  and MEGARA/GTC optical IFS observations to build up, for the first time, a comprehensive picture of both stellar and ISM components in NGC\,1052 of the outflow, with a tens-of-pc resolution.\\
\noindent  This paper is organised as follows. In Section\,\ref{S_datared}, the data and observations are presented, as well as, the data reduction. In Section\,\ref{S_data_analysis}, we present the spectroscopic analysis: stellar subtraction, line modelling and maps generation. Section\,4 highlights the main observational results. In Section\,5, we discuss the stellar kinematics and dynamics, and the ionised and neutral gas properties with special emphasis on outflow properties and  its possible connection with the radio jet. In this Section we also estimate the black hole mass, and compare the full width half maximum (FWHM) of the unresolved BLR component with previous estimates. The main conclusions are presented in Section\,\ref{S_conclusions}. In Appendix\,\ref{Appendix_A}, we summarise the procedure to account for background sources. In Appendix\,\ref{Appendix_B} we present the kinematic, flux-intensity maps and fluxes-ratios from our IFS data set. Appendix\,\ref{Appendix_C} is devoted to present the 1D position-velocity and position-dispersion diagrams aimed at comparing gas and stellar motions
along the three major axes, i.e. major and minor axes of the host galaxy, and the radio jet.\\

\noindent All images and spectral maps are oriented following the standard criterion, so the north is up and east to the left. \\
Throughout the whole paper, angular dimensions will be converted into physical distances using the scale distance from the Local Group, i.e. 110 pc/$\arcsec$ (see Table\,\ref{T_properties}).

\begin{table}
\caption[Subsample]{General properties of  NGC\,1052.}
\begin{center}
\tiny{\begin{tabular}{l  c  c}
\hline \hline
Properties              & value             &  References\\
\hline
R.A. (J\,2000)                     &  02$^{h}$41$^{m}$04$^{s}$.799          & NED \\
Decl. (J\,2000)               &   -08$^{d}$15$^{m}$20$^{s}$.751         & NED \\
z                       &  0.00504           & NED \\	
V$_{\rm sys}$  [\nskms]   &  	1532 $\pm$ 6  & \citet{Karachentsev1996} \\
D [Mpc]     &    22.6 $\pm$ 1.6          & NED \\
Scale  [pc/$\arcsec$]      &  110            & NED \\
Nuclear Spectral Class. & LINER (1.9)             & \citet{OGM2009}\\				
Morphology             & E3-4/S0             & \citet{Bellstedt2018} \\  
  \textit{i}\,[$^{\circ}$]      & 70.1   & Hyperleda \\
PA$_{\rm phot}$ & 112.7 & Hyperleda\\
R$_{eff}$ [$\arcsec$]     & 21.9 & \citet{Forbes2017} \\
M$_{\rm BH}$\,[M$_{\sun}$]            & 3.4\,(0.9)\,$\times$\,10$^{8}$& \citet{Beifiori2012}\\
PA$_{\rm jet}$\,[$^{\circ}$]  &  70 & \citet{Kadler2004a} \\
SFR [M$_{\odot}$/yr] & 0.09  & \citet{Falocco2020}\\
\hline
\end{tabular}
\label{T_properties}}
\end{center} 
\tiny{Notes. --- \lq V$_{\rm sys}$\rq, \lq D\rq \ and  \lq Scale\rq: systemic velocity,  distance, and scale, respectively, from the Local Group.  \lq Morphology\rq: Hubble classification.  \lq \textit{i} \rq \ is the inclination  angle defined as the inclination between line of sight and polar axis of the galaxy. It is determined from the axis ratio of the isophote in the B-band using a correction for intrinsic thickness based on the morphological type. \lq PA$_{\rm phot}$\rq: is the position angle of the major axis of the isophote 25 mag/arcsec$^{2}$ in the B-band measured north-eastwards (see \citealt{Paturel1997} and references therein). \lq R$_{eff}$\rq \ is the effective radius from Spitzer data. The black hole mass (M$_{\rm BH}$) is derived  from a keplerian disc model assuming an inclination of 33$^{\circ}$(81$^{\circ}$) and a distance of 18.11\,Mpc.  \lq PA$_{\rm jet}$\rq \ is the position angle of the  jet from VLBI data (covering only the central region of NGC\,1052). As the PA depends on the different components of the jet, varying from 60$^{\circ}$ to 80$^{\circ}$, in this work we consider the average value of 70$^{\circ}$. See \citet{Kadler2004a} and references therein for further details. \lq SFR\rq : upper limit to the star formation rate from FIR luminosity (2\,$\times$\,10$^{42}$ ergs$^{-1}$) as measured by \citet{Falocco2020}.} 
\end{table}

\begin{figure*}
\centering
\includegraphics[width=.95\textwidth]{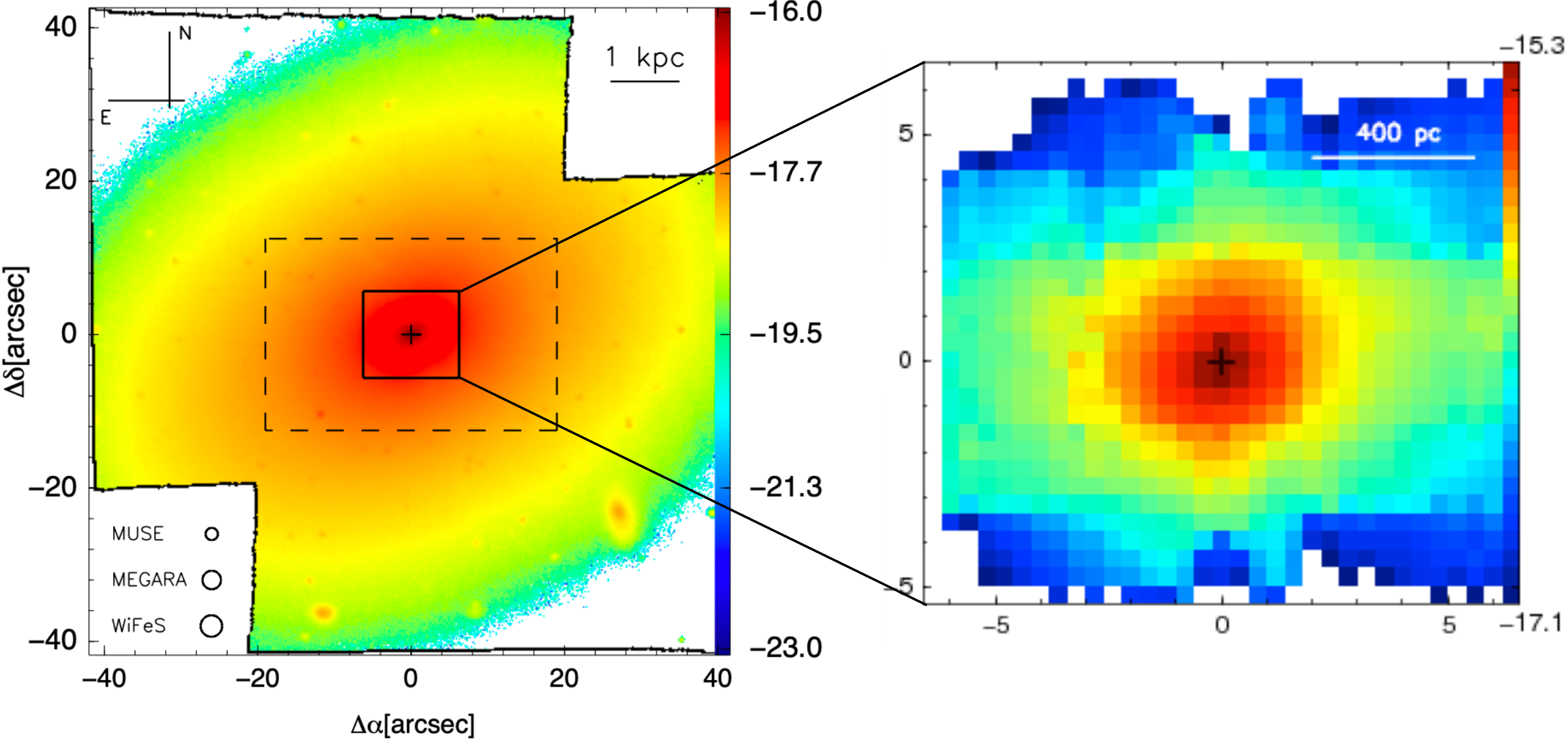}
\caption{Optical continuum images  computed from MUSE (left) and MEGARA (right) in units of erg\,s$^{-1}$/cm$^{2}$ (logarithmic scale). To obtain these images we considered a 60\,\AA-wide continuum band (i.e. from 6105\,$-$\,6165\,\AA). The cross is the photometric center, and the sizes of the different PSFs for MEGARA and MUSE data are indicated in the bottom left part of the figure (see also Sect.\,\ref{S_datared}). As reference we show the field of view (dashed rectangle) and average seeing (1$\farcs$4, bottom circle) for the WiFeS datacube analysed in D15. The black bar at the upper right represents 1\,kpc ($\sim$\,9$\arcsec$) at the redshift of NGC\,1052 (see Table\,\ref{T_properties}). Similarly, the white bar at the upper right, right panel,  represents 400\,pc ($\sim$\,3$\farcs$6).} 
\label{Fig_MM_continuum}
\end{figure*} 

\section{Observations and data reduction}
\label{S_datared}

In this section we describe MUSE and MEGARA data and their data reduction process, see Sections \ref{S_datared_MUSE} and  \ref{S_datared_MEGARA}, respectively.

\subsection{MUSE observations and data reduction}
\label{S_datared_MUSE}

\noindent The data were gathered on September 5$^{\rm th}$ 2019 with the Multi-Unit Spectroscopic Explorer (MUSE, \citealt{Bacon2010, Bacon2014}), mounted at the UT4 of the Very Large Telescope at the Paranal Observatory in Chile as part of programme 0103.B-0837(B) (PI: L.\,Hern{\'a}ndez-Garc{\'i}a). \\
\noindent They were acquired in the wide field mode configuration with the nominal setting (i.e. no extended wavelength coverage), covering the spatial extent of 1\,arcmin$^{2}$ with 0.2$\arcsec$\,pix$^{-1}$ sampling. The MUSE data has a wavelength coverage of 4800\,-\,9300\,\AA, with a mean spectral resolution of R\,$\sim$\,3000 at 1.25\,\AA \ spectral sampling. During the observations the average DIMM seeing was 0$\farcs$62 (varying between 0$\farcs$48 and 0$\farcs$85); mean airmass was 1.06. \\
In total, we obtained eight exposures with a total integration time of 93\,min.
Including overheads, the observations took two hours, i.e. two observing blocks. Each one consists in four dithered exposures of 697\,s. The relative offsets in RA(DEC) were 10$\arcsec$, 0$\farcs$5, -21$\farcs$5 and 0$\farcs$5 (11$\arcsec$, 0$\farcs$5, -21$\farcs$5 and 0$\farcs$5) with respect to the position of NGC\,1052 (Table\,\ref{T_properties}). The dither pattern also involves a 90$^{\circ}$ rotation for a better reconstruction of the final cube, i.e. an homogeneous quality across the field of view.\\
\noindent The eight pointings constitute a mosaic covering a contiguous area of 80$\arcsec$\,$\times$\,80$\arcsec$, i.e, 8.8\,kpc\,$\times$\,8.8\,kpc at the adopted spatial scale (110 pc/$\arcsec$, Table\,\ref{T_properties}). The radius of the covered area is about 3.5 times the effective radius of NGC\,1052 (i.e. 21$\farcs$9, Table\,\ref{T_properties}). 
\noindent The data reduction was performed with the  MUSE pipeline (version 2.8.1) via \texttt{EsoRex} (version 3.13.2). It performs the basic reduction steps, that is bias subtraction, flat fielding, wavelength calibration and illumination correction, as well as the combination of individual exposures in order to create the final mosaic. For flux calibration we used the spectrophotometric standard star Feige\,110 (spectral type: DOp), observed  before the science frames. Since we did not apply any telluric correction, some residuals remain in the region between $7110-7310$\,\AA. In this spectral window only the HeI$\lambda$7065.3, [Ar\,III]$\lambda$7135.80 and [Fe\,II]$\lambda$7155 lines are detected, but they are not crucial for our analysis. The sky-subtraction was performed in the latest step of the processing of MUSE observations using the sky-background  obtained from the outermost spaxels in each science exposure (no dedicated on-sky exposures were gathered). We perform the astrometry calibration using the astrometric catalogue distributed with the pipeline. \\
The final cube has dimensions of 418\,$\times$\,422\,$\times$\,3682. The total number of spectra is 176\,396, of these 28\,508 (16\,$\%$) are not useful, as they correspond to artefacts from the creation of the mosaic, i.e. empty spaxels located at  bottom-left and top-right corners, and at the edges of the field of view.\\
\noindent The radius of the point spread function (PSF) of the MUSE observations (0$\farcs$4, see Fig.\,\ref{Fig_MM_continuum}) was estimated from the full width at half-maximum (FWHM) of the 2D-profile brightness distribution of the standard star used for flux calibration. Throughout the paper, in order to avoid any possible PSF contamination in the kinematic measurements, we will conservatively consider as \lq nuclear region\rq \ a circular area of radius equal to the width at 5 per cent intensity of the PSF radial profile, i.e. 0$\farcs$8. This area does not coincide with any peculiar feature (e.g. dust lanes) visible in the MUSE continuum image shown in the left panel of Fig.\,\ref{Fig_MM_continuum}. The \lq nuclear region\rq \ is marked (with a circle) in the spectral maps computed from the MUSE datacubes (see Fig.\,\ref{Fig_MM_continuum} but also Sect.\,\ref{S_data_analysis} and Appendix\,\ref{Appendix_B}).\\
\noindent  We obtained the instrumental profile by measuring the single (not blended) OH$\lambda$\,7993.332 sky-line \citep{Osterbrock1996, Bai2017}. We measured it in the fully-reduced data cube of the standard star Feige\,110 (see above) by selecting a region of size 50\,$\times$\,50 spaxels free from stellar emission. On average, the central wavelength and the width of the OH sky-line are 7993.335\,$\pm$\,0.114\,\AA \ and  1.19\,$\pm$\,0.13\,\AA, respectively. This instrumental profile correction has been further checked with the 5577\,\AA \ sky-line. In this case, the value of the average instrumental resolution is consistent with that from the OH line (i.e. 1.2\,\AA).

\subsection{MEGARA observations and data reduction}
\label{S_datared_MEGARA}

\noindent The data were taken on December 28$^{\rm th}$ 2019 with the MEGARA instrument (see \citealt{GdP2016, Carrasco2018}), located in the Cassegrain focus of GTC using the Large Compact Bundle IFU mode (GTC94-19B, PI: S.\,Cazzoli). The 567 fibres that constitute the MEGARA IFU (100 $\mu$m in core size) are arranged on a square microlens-array that projects on the sky a field of  12$\farcs$5 x 11$\farcs$3.
Each microlens is a hexagon inscribed in a circle with diameter of 0$\farcs$62 projected in the sky. A total of 56 ancillary fibres (organised in eight fibre bundles), located at a distance of 1.75–2.0 arcmin from the centre of the IFU field of view, deliver simultaneous sky observations.\\
We made use of two low-resolution Volume Phase Holographic gratings (LR-VPHs) that provides a R\,$\sim$\,6000 in the central wavelengths of the selected bands: LR-V has a wavelength coverage $5140-6170$\,\AA\,and LR-R $6100-7300$\,\AA. \\
\noindent We obtained six exposures with a integration time of 900s per VPH in two observing blocks, leading to a total observing time of four hours. The mean signal-to-noise ratio (S/N) in the spectra continuum was 25 for the LR-R and 30 for the LR-V datacube.
\noindent The data reduction was done using the MEGARA Data Reduction Pipeline \citep{MegaraDRP2020, MegaraDRP2020ACM} available as a package inside \textsc{Python} (version 0.9.3). We performed the standard procedures: bias subtraction, flat-field correction, wavelength calibration and flux calibration using the star HR\,4963. Each fibre was traced individually at the beginning of the data reduction and, within the pipeline, we applied additional corrections for the possible differences of each fibre with respect to the whole image, including an illumination correction based on individual fibre flats. For this correction, we used \textsc{iraf} to smooth the sensitivity curve as, in the case of the LR-R VPH, some structure due to the lamp emission are present (see MEGARA cookbook). The pipeline also performs the individual exposures combination to generate the final cube (one per VPH), which can be transformed into a standard IFS cube from raw stacked spectra format by means of a regularisation grid to obtain 0$\farcs$4 square spaxels \citep[see][]{Cazzoli2020}.
The PSF of the MEGARA data was measured as in Sect.\,\ref{S_datared_MUSE} with the star HR\,4963, giving a FWHM of 1$\farcs$2 (see Fig.\,\ref{Fig_MM_continuum}, left).\\
\noindent Considering the wavelength ranges of the VPHs and the emission lines present in NGC\,1052 spectra, we decided to combine the two cubes into a single datacube to optimise the stellar modelling and subtraction (increasing the range of line-free continuum, see Sect.\,\ref{S_stellar_cont}). More specifically, the need of combining the two MEGARA cubes to reliably model the stellar continuum is twofold. First, the spectral range of the MEGARA LR-V cube covers only the MgI stellar feature whereas none are present in the LR-R one. Second, for the latter (red) cube, the stellar continuum emission is limited by the presence of the broad emission features and a telluric band (see Sect.\,\ref{S_datared_MEGARA}). For the cube combination, we scaled the fluxes for every spaxel to have the continuum at the same level in the common wavelength range of both VPHs ($6100-6170$\,\AA). The combined datacube was used in the whole analysis.

\section{Data analysis}
\label{S_data_analysis}

In this section we summarise the identification and subtraction of background sources in the MUSE field of view (Sect.\,\ref{S_bkg_sources}), as well as we describe the stellar continuum modelling (Sect.\,\ref{S_stellar_cont}) and line fitting for MUSE and MEGARA cubes (Sect.\,\ref{S_line_mod}).

\subsection{Background sources in the MUSE field of view}
\label{S_bkg_sources}

\noindent We visually inspected the white light image generated in the last step of the data reduction of MUSE data, i.e. the mosaic creation (see Sect.\,\ref{S_datared}) and the continuum image in Fig.\,\ref{Fig_MM_continuum}. We note that there are a number of sources (both point-like and extended) some of which may not be part of the NGC\,1052 galaxy. In Appendix\,\ref{Appendix_A}, we summarise the procedure to identify putative background sources. \\
We found two background galaxies at redshifts $\sim$\,0.03 and $\sim$\,0.022. Only the former is identified in NED as:  SDSSCGB$\_$67616.02. Both of these two galaxies were masked out from the final MUSE datacube used for the analysis. 

\subsection{Stellar continuum modelling}
\label{S_stellar_cont}
\begin{figure*}
\centering
\includegraphics[width=1.\textwidth]{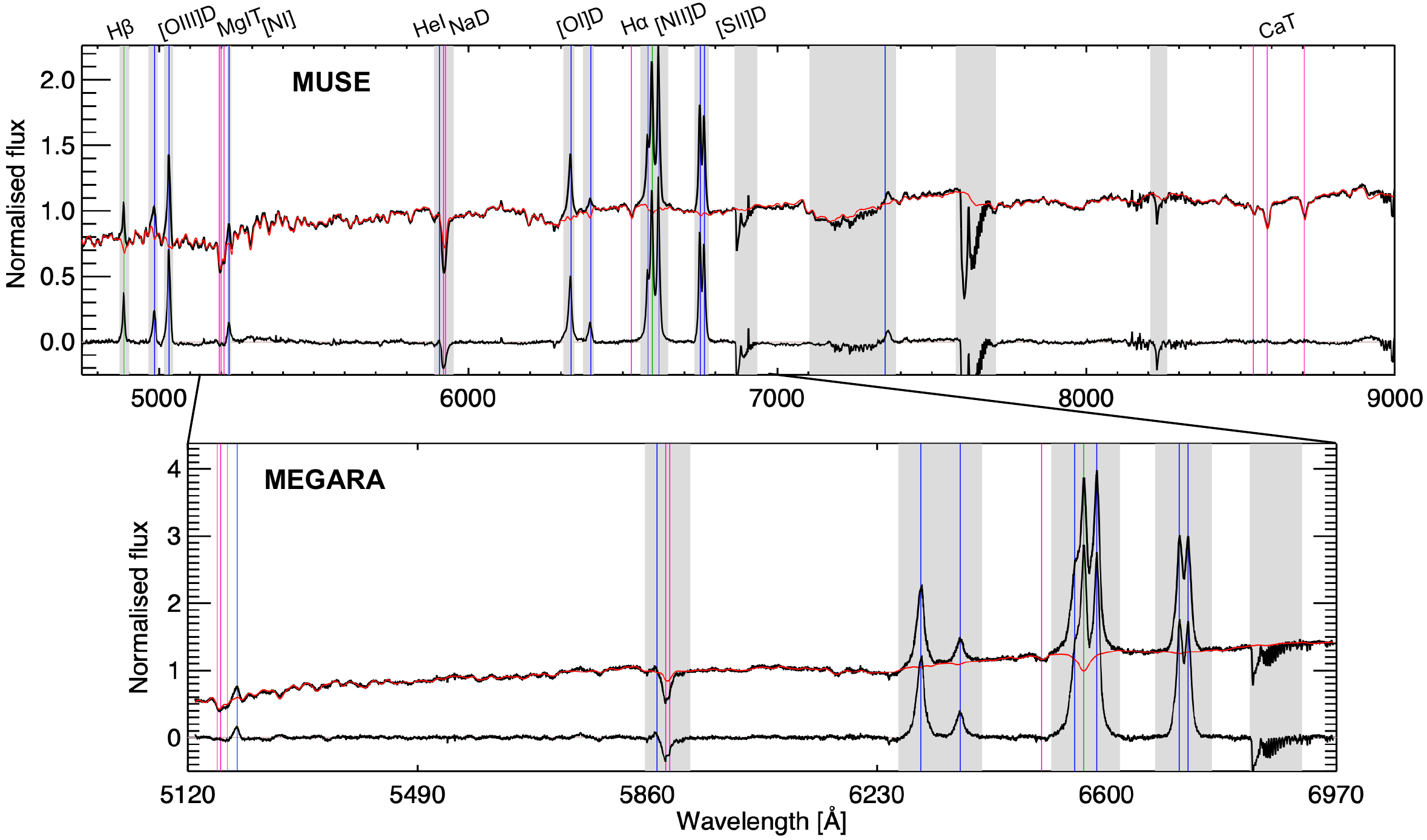}
\caption{Example of stellar continuum modelling and its subtraction for high S/N nuclear spectra from both MUSE (top panel) and MEGARA (bottom panel) data. The red line indicates the modelled stellar spectrum that matches the observed continuum, obtained applying the \texttt{pPXF} (Sect.\,\ref{S_stellar_cont}). The wavelength regions blocked for the modelling are shown in grey. Spectral features are labelled at the top, and Balmer lines, forbidden lines and absorption lines are marked in green, blue and pink, respectively. Note that, in case of MEGARA, we combined the cubes in LR-V and LR-R bands, which have a 70\,\AA \ overlap around 6130\,\AA \ (see Sect.\,\ref{S_datared_MEGARA}).}
\label{Fig_ppxf_model}
\end{figure*}

\begin{figure*}
\centering
\includegraphics[width=1\textwidth]{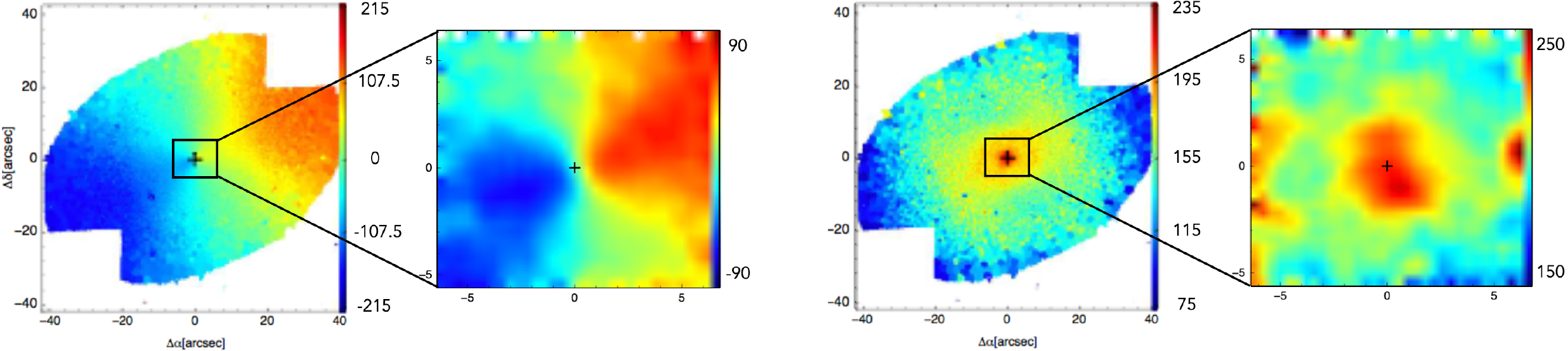}
\caption{NGC\,1052 stellar kinematics maps from our \texttt{pPXF} analysis (Sect.\,\ref{S_stellar_cont}).
These maps, i.e. velocity (left) and velocity dispersion (right), are displayed in units of \nskms. In both panels, the large scale kinematics is obtained from MUSE data whereas the insets show the smoothed \texttt{pPXF}-maps from MEGARA datacube. The cross marks the photometric center as in Fig.\,\ref{Fig_MM_continuum}.} 
\label{Fig_stellar_kin}
\end{figure*}

\noindent For the stellar continuum modelling we used the penalised PiXel-Fitting code (\texttt{pPXF}) by \citet{Cappellari2003} (see also \citealt{Cappellari2017} and references therein) for both MEGARA and MUSE, in different coding environments. We used the \texttt{pPXF} code within the GIST pipeline (see below) for MUSE and within \textsc{python} for MEGARA. \\
\noindent For MUSE, we used the GIST pipeline (v.\,3) by \citet{Bittner2019}\footnote{\url{http://ascl.net/1907.025}} as a comprehensive tool both to spatially bin the spectra in order to increase the S/N in the continuum and to model the stellar contribution to the observed spectra.
The MUSE spectra were shifted to rest-frame based on the initial guess of the systemic redshift from NED, i.e. z\,$=$\,0.005 (Table\,\ref{T_properties}).
Then the data were spatially binned using the 2D Voronoi binning technique by \citet{Cappellari2003}, that creates bins in low S/N regions, preserving the spatial resolution of those above a minimum S/N threshold. The S/N has been calculated in the line-free wavelength band between 5350 and 5800\,\AA. All spaxels with a continuum S/N\,$<$\,3 were discarded to avoid noisy spectra in the Voronoi bins. We found that a minimum S/N threshold of 30 results in reliable measurements of stellar kinematics in NGC\,1052 as well as an optimum spatial resolution. Specifically,  in general, cells are not larger than 60 spaxels (2.4\,arcsec$^{2}$ in area), hence stellar properties are likely to be homogeneous within a Voronoi-cell. \\ 
For MEGARA data the Voronoi binning was not necessary to achieve a proper stellar continuum modelling as in the spaxels with the lowest S/N ($<$\,15), that constitute $\sim$12$\%$ of the total, the resulting velocity and velocity dispersion are consistent with the rest of the cube with higher S/N.\\
\noindent To accurately  measure spectral line properties (wavelength, width and flux), it is necessary to account for stellar absorption, which primarily affects the Balmer emission lines and the NaD absorption doublet. For MUSE we limited the wavelength range used for the fit to $4800-9000$\,\AA , which contains spectral features from H$\beta$ to CaT, and excluded the region of the auroral [S\,III]$\lambda$9069 line\footnote{This line is noisy and only barely detected in a region of radius of $\sim$\,1$\arcsec$, hence no spatially resolved analysis will be done.}. For MEGARA the total wavelength range was from 5150-7000\,\AA\,covering the main spectral features in both LR-V and LR-R bands.  For both data sets, we masked the spectral regions (emission lines and atmospheric/telluric absorptions) affected by emission from the interstellar medium (ISM). Additionally, we excluded the NaD absorption that is not properly matched by the stellar templates owing to the impact of interstellar absorption. 
\noindent For MUSE, we used the Indo-U.S. stellar library \citep{Valdes2004} as in \citet{Cazzoli2014, Cazzoli2016, Cazzoli2018}. Briefly, in this library there are 885 stars selected to provide a broad coverage of the atmospheric parameters (effective temperature, surface gravity and metallicity). The stellar spectra have a continuous spectral coverage from 3460 to 9464 \AA, at a resolution of $\sim$\,1\,\AA \ FWHM  \citep{Valdes2004}. For MEGARA we used the RGD synthetic stellar library \citep{GonzalezDelgado2005,Martins2005}, since it covers the whole spectral range for the combined datacubes and the spectral resolution is consistent with that from our spectra. The library consisted on 413 stars selected with a metallicity of Z\,=\,0.02, ranging from 4000 to 7000 \AA\,, and covering a wide range of surface gravities and temperatures \citep[see][and references therein]{GonzalezDelgado2005}.\\
\noindent Finally, we set up \texttt{pPXF} using four moments of the line of sight velocity distribution (LOSVD), i.e. V, $\sigma$, h3 and h4, for both MUSE and MEGARA. The additive and multiplicative polynomials were set to 4-4 (0-12) for MUSE (MEGARA) in order to, respectively, minimise template mismatch, and match the overall spectral shape of the data so that the fit is insensitive to reddening by dust \citep[see][and references therein]{Westfall2019,Perna2020}.\\ An example of the \texttt{pPXF} modelling is shown in Fig.\,\ref{Fig_ppxf_model} for both MUSE (top panel) and MEGARA (bottom panel) data.
\noindent The results of the \texttt{pPXF} fits, i.e. the stellar kinematics maps of the first two moments of the LOSVD, are shown in Fig.\,\ref{Fig_stellar_kin}, and discussed in Sect.\,\ref{S_stellar_kin}. A detailed study of higher order moments of the stellar LOSVD (h3 and h4) is beyond the aim of the paper, hence the corresponding maps are not displayed.\\
\noindent Through the analysis we will consider formal uncertainties provided by the \texttt{pPXF} tool. These are in good agreement with those from MC-simulations performed on MUSE data. Specifically, differences are generally lower than 5\,\kms and 7\,\kms for velocity and velocity dispersion, respectively.\\
\noindent Motivated by the typical small sizes of the Voronoi cells in the MUSE data, we made the simplifying assumption that the stellar populations and kinematics do not change radically within one Voronoi bin. For each spaxel, the stellar spectrum of the corresponding bin is normalised and then subtracted to the one observed to obtain a datacube consisting exclusively of ISM absorption and emission features. For MEGARA data the stellar subtraction is performed on spaxel-by-spaxel basis.\\
In what follows, we will refer to this data cube (data\,$-$\,stellar model) as the \lq ISM-cube\rq.

\begin{figure*}
\centering
\includegraphics[width=0.975\textwidth]{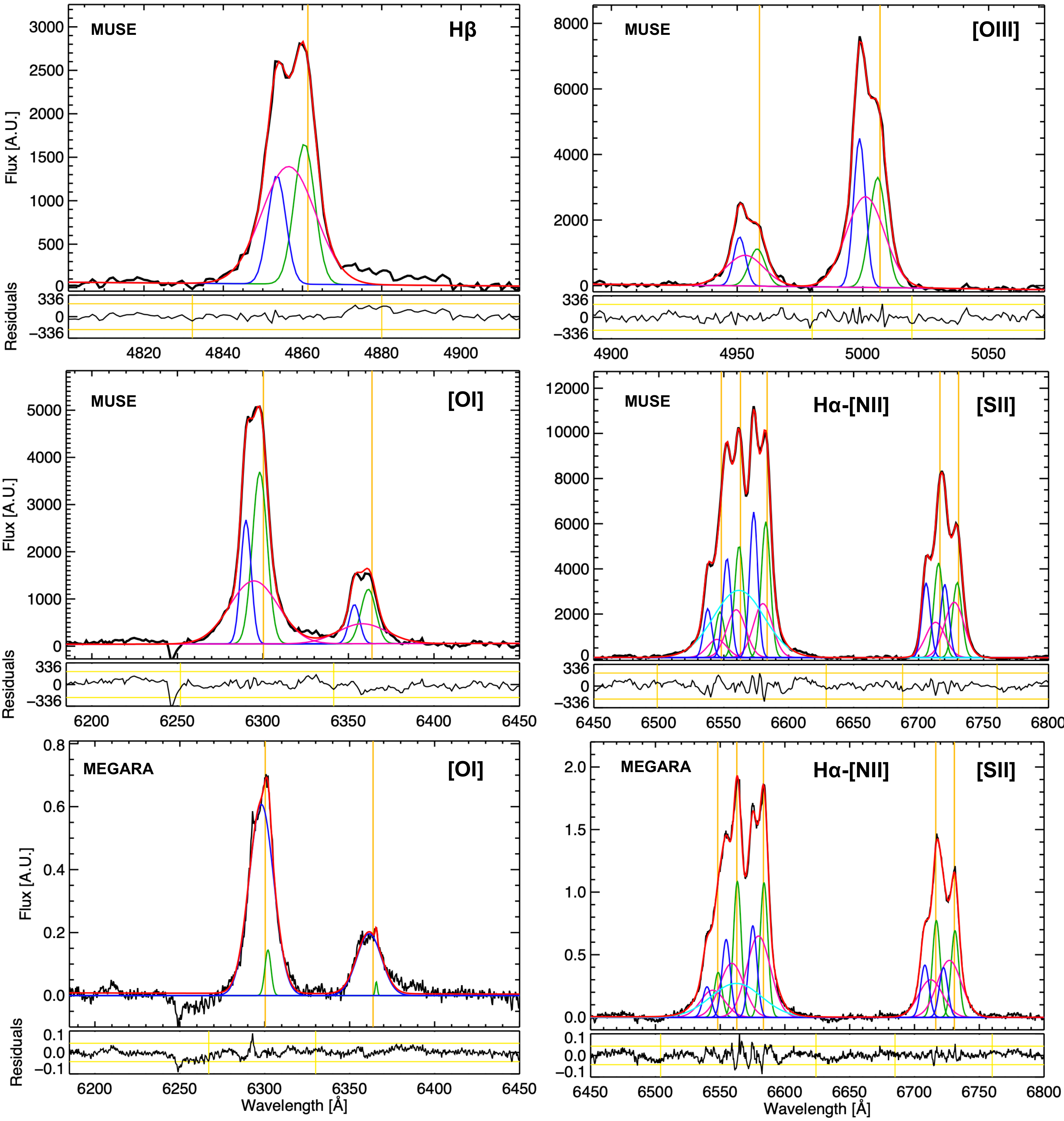}
\caption{Examples of emission line spectra (black) after stellar subtraction (Sect.\,\ref{S_stellar_cont}) and their modelling from the central region of both MUSE (R\,=\,0$\farcs$7, i.e. 77\,pc) and MEGARA data (R\,=\,0$\farcs$9, i.e. 100\,pc), see top-left labels. As reference, orange vertical lines mark the systemic wavelengths of the emission lines which are labelled on top-right. For each panel the modelled line profile (red line) and the components (with different colours) are shown. Specifically, green, blue, pink Gaussian curves indicate: primary, secondary, and third components used to model the profiles. In cyan is marked the broad H$\alpha$ component from the BLR. Residuals from the fit are shown in the small lower panels, in which yellow horizontal lines indicate the $\pm$\,2.5\,$\varepsilon_{\rm c}$ (Sect.\,\ref{S_em_lin_mod}). Vertical yellow lines mark the wavelength range considered for calculating $\varepsilon_{\rm fit}$ for each line (Sect.\,\ref{S_em_lin_mod}). Note that the high residuals redward to H$\beta$, cannot be fitted with a BLR-component (velocities and widths would be inconsistent with those of the broad H$\alpha$ component), and are likely due to some residuals from stellar subtraction (Sect.\,\ref{S_stellar_cont}).}
\label{Fig_spectral_model}
\end{figure*}

\subsection{Line modelling}
\label{S_line_mod}

\noindent  From the ISM-cube, we produce line-maps by modelling the spectral lines with multiple Gaussian functions. To achieve that, we applied a Levenberg-Marquardt least-squares fitting routine under both Interactive Data Analysis (IDL) and \textsc{Python} environments, using \textsc{mpfitexpr} by \citet{Markwardt2009} and \textsc{lmfit}, respectively (see Sections \ref{S_em_lin_mod} and \ref{S_abs_lin_mod}). We imposed the intensity ratios between the [O\,III]$\lambda$4959,5007 (only for MUSE), [O\,I]$\lambda$6300,6363 and [N\,II]$\lambda$6548,6584 to be 2.99, 3.13 and 2.99 \citep{Osterbrock2006}. The ratio between the equivalent widths (EW) of the two lines of the NaD$\lambda\lambda$5890,5896 absorption, R$_{\rm NaD}$\,=\,EW$_{5890}$/EW$_{5896}$, is restricted to vary from 1 (optically thick limit) to 2 (optically thin absorbing gas), according to \citet{Spitzer1978}.

\subsubsection{Emission Line modelling}
\label{S_em_lin_mod}

\begin{figure*}
\centering 
\includegraphics[width=1\textwidth]{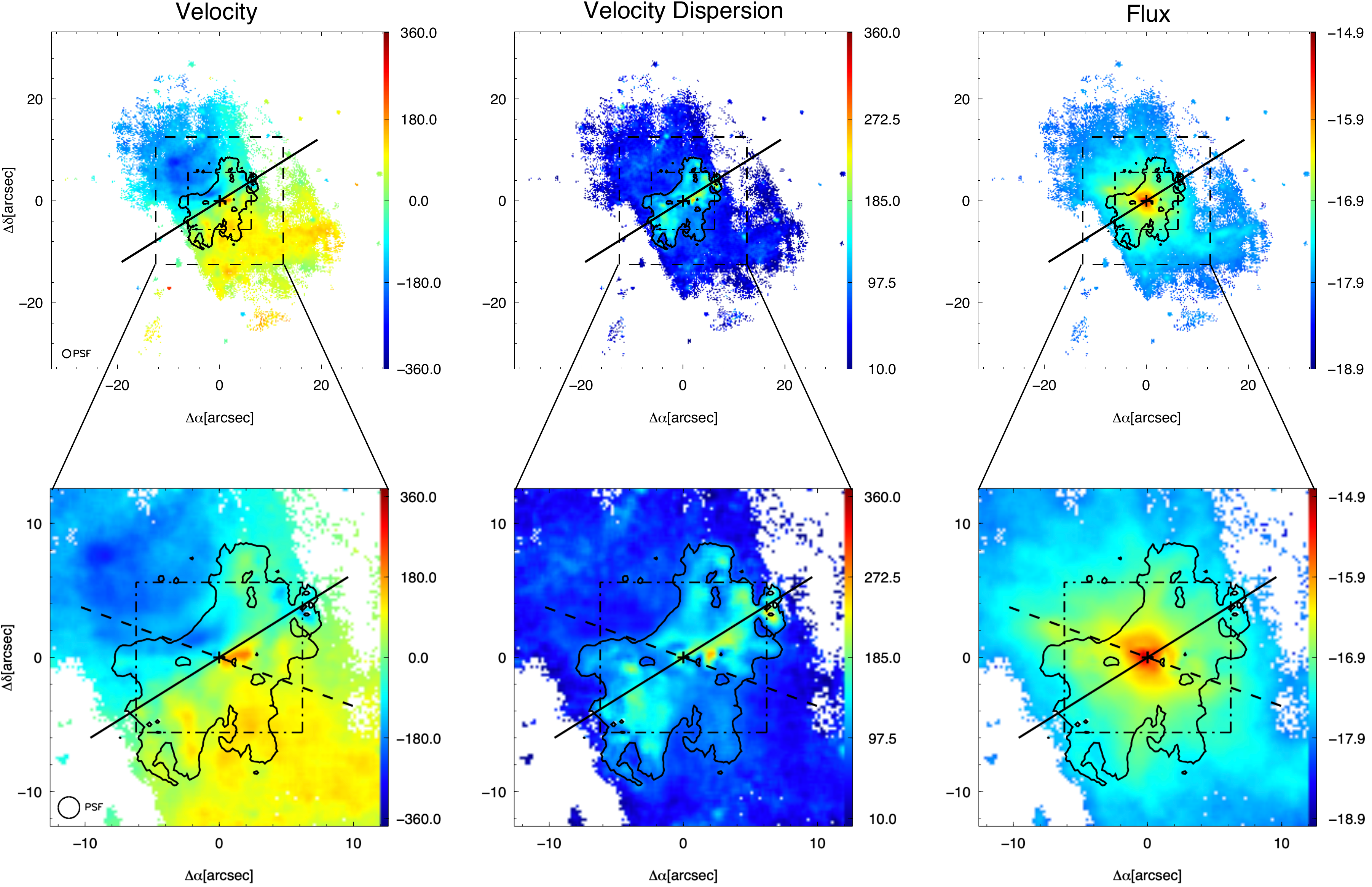}
\caption{Example of emission line maps produced from the fitting of the [O\,III]$\lambda$5007  line  using the MUSE ISM-cube (Sect.\,\ref{S_em_lin_mod}). In both panels from left to right: velocity field (\nskms), velocity dispersion (\nskms) and flux intensity (erg\,s$^{-1}$\,cm$^{-1}$, log scale) maps for the primary component.  The black solid line indicates the major axis of the stellar rotation (Table\,\ref{T_kinematics}). The dot-dashed square indicates the MEGARA field of view. The contours indicate the central region at high velocity dispersion (see text for details, e.g. Sect.\,\ref{S_result_primary_MUSE_butterfly}).
\textit{Top Panel}: The maps cover a smaller field of view with respect to original MUSE mosaic (80$\arcsec$\,$\times$\,80$\arcsec$, Sect.\,\ref{S_datared_MUSE}) to highlight weak features.  The dashed square indicate the selected zoomed view displayed in the bottom panels of this figure and for Figures from \ref{M_OIII_primary_zoom} to \ref{M_BPT_primary_zoom} of the Appendix\,\ref{Appendix_B}. 
\textit{Bottom Panel}:  zoomed area. The dashed line indicates the orientation of the radio-jet (Table\,\ref{T_properties}).}
\label{Fig_combo_OIII}
\end{figure*}

\begin{figure*}
\centering 
\includegraphics[trim = 1.8cm 15.5cm 1.8cm 5.55cm,  clip=true, width=1.\textwidth]{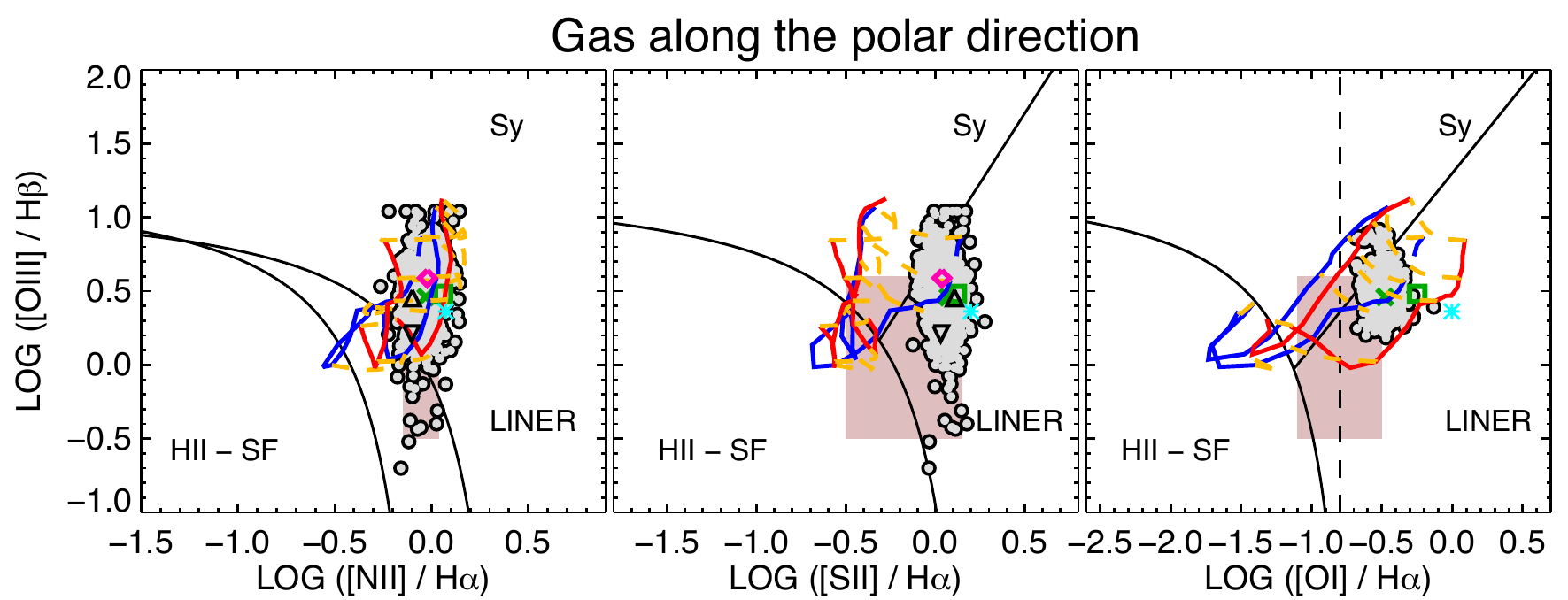}
\includegraphics[trim = 1.8cm 15.5cm 1.8cm 5.55cm,  clip=true, width=1.\textwidth]{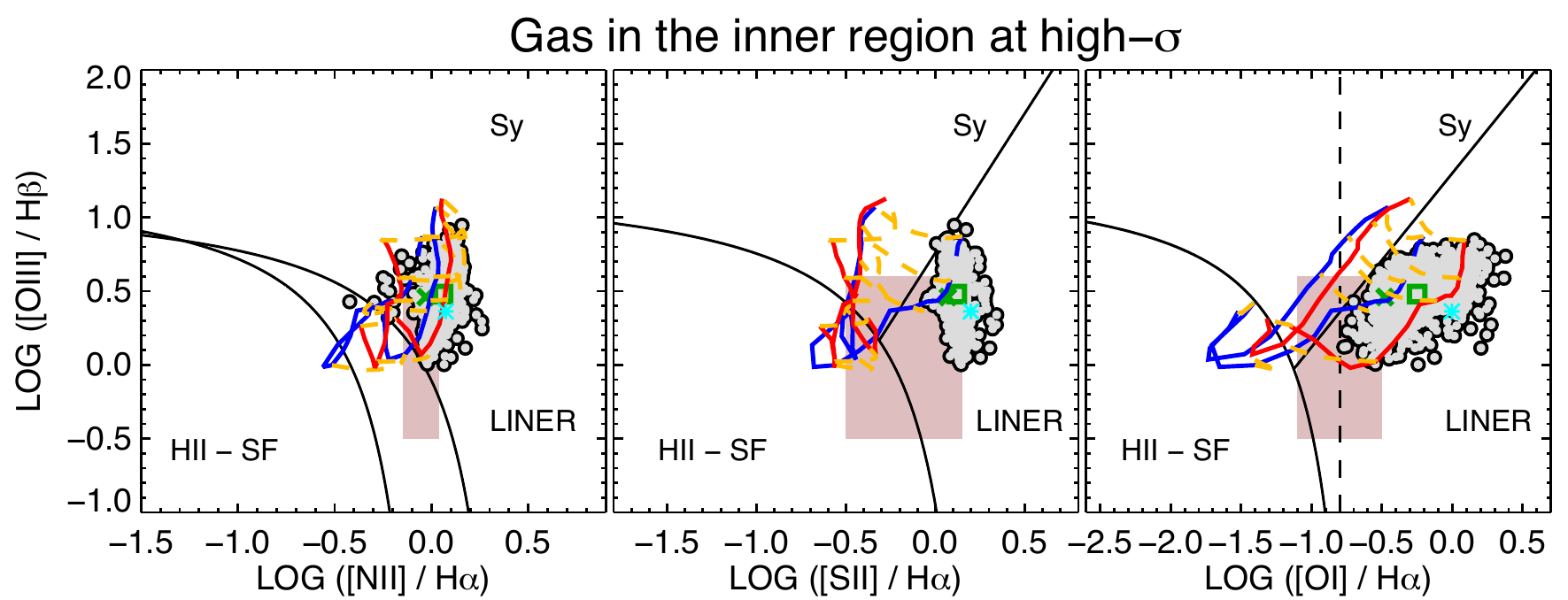}
\caption{Optical standard BPT diagrams for the primary component for the gas distributed in the polar direction and that in the central region at high-$\sigma$  (top and bottom panels, respectively) obtained from MUSE data. Grey circles mark the data points presented in this paper.  Black lines in all diagrams represent the dividing curves between H\,II star-forming regions, Seyferts, and LINERs from \citet{Kewley2006} and \citet{Kauffmann2003}. Pink boxes show the predictions of photoionisation models by pAGB stars for Z\,=\,Z$_{\sun}$, a burst age of 13 Gyr \citep{Binette1994} and ionisation parameter values (log\,U) between -3 and -4. Log\,U is typically -3.5 in LINERs \citep{Netzer2015}. The predictions of shock-ionisation models are overlaid in each diagram. Specifically, following \citet{Cazzoli2018} we considered shock+precursor grids from \citet{Groves2004}  with Z\,=\,Z$_{\odot}$ and for different n$_{e}$. Blue and red curves correspond to models with n$_{e}$\,=\,1\,cm$^{-3}$ and n$_{e}$\,=\,100\,cm$^{-3}$, respectively (see also Sect.\,\ref{S_result_primary_MUSE}). We plotted the values corresponding to the minimum and maximum preshock magnetic field allowed in each model. Also, we consider only shock-velocities from 100 to 500 km\,s$^{-1}$ (yellow dashed lines) as larger $\sigma$ are not observed for the primary component (Sect.\,\ref{S_result_primary_MUSE}). The dividing line between weak-[O\,I] and strong-[O\,I] LINERs \citep{Filippenko1992} is marked in black with a dashed line (right panels). In all diagrams, green symbols indicate the average values calculated in the polar (cross) and central (square) regions; as reference the cyan star is the typical value in the nucleus (average within the PSF region). In the top panels, the pink diamond, black triangle and black upside down triangle are the average BPT-values for the faint features: the arm, the east and south-east clumps, respectively (see Sect.\,\ref{S_result_lowSB}). These features are not detected in [O\,I] hence no symbols are displayed in the corresponding diagnostic diagrams.}
\label{Fig_BPT_primary}
\end{figure*}

\begin{figure*}
\centering 
\includegraphics[trim = 1.8cm 15.5cm 1.8cm 5.55cm,  clip=true, width=1.\textwidth]{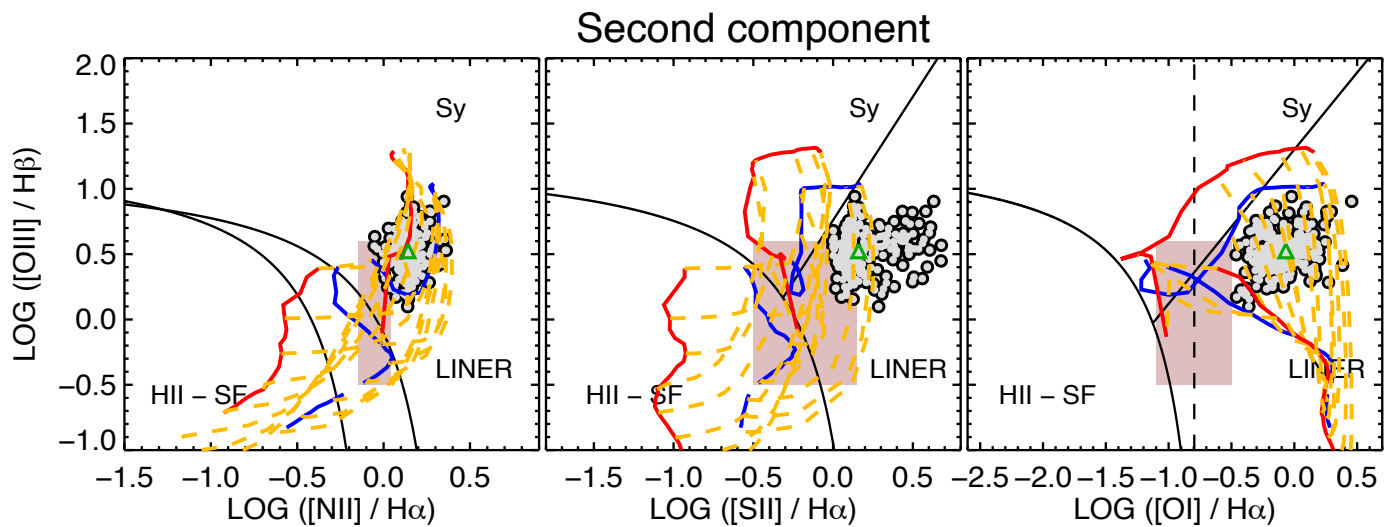}
\caption{The same as Fig.\,\ref{Fig_BPT_primary} but for the secondary component. Here, we considered shock models (no precursor), the blue and red curves correspond to models with n$_{e}$\,=\,100\,cm$^{-3}$ and n$_{e}$\,=\,1000\,cm$^{-3}$, respectively (see also Sect.\,\ref{S_result_secondary}). The green triangle indicates the average value of the line ratios distribution.}
\label{Fig_BPT_secondary}
\end{figure*}

\noindent We derive the kinematics of the ISM properties by modelling all the spectral lines available in the cubes. To perform the fitting, and hence discriminate between line models and number of components, we followed the approach proposed by \citet{Cazzoli2018}. Specifically, for both MUSE and MEGARA data, we tested the \lq [S\,II]-model\rq \ and \lq [O\,I]-model\rq, for which we first fitted in the spectrum only [S\,II] and [O\,I] lines (depending on the model) and then used them as reference to tie all the other narrow lines, so they share the same width and velocity shift. Additionally, we tested the \lq mix-models\rq, that consist on using [S\,II] and [O\,I] simultaneously as reference for [N\,II] and narrow H$\alpha$ respectively or, alternatively, using [O\,I] for narrow H$\alpha$ and [N\,II], with [S\,II] lines behaving otherwise. For MUSE data only (see Sect.\,\ref{S_datared_MEGARA}), the best fit to the H$\alpha$ ([S\,II]) line is applied to the H$\beta$ ([O\,III]) line.\\
\noindent However, none of these models provided a good fit for the whole set of lines. In the MEGARA field of view the independent fitting of [O\,I] and [S\,II] lines produced differences of $\sim$\,100\,\kms for the velocity measurements, although the line widths resulted similar, i.e. differences $\leq$\,50\,\nskms. For MUSE data, on the one hand, we found that at large spatial scales (R\,$>$\,10$\arcsec$) the kinematics of these lines are similar within 75\,\kms (mostly) when they are fitted independently. Although large discrepancies ($>$\,100\,\nskms) arise in the central region (inside the MEGARA field of view; R\,$<$\,10$\arcsec$ oriented in E-W direction), with a peculiar \lq butterfly\rq\ shape (see Sect.\,\ref{S_main_results}). A similar behaviour has been found comparing [O\,III] and [S\,II] kinematics. Moreover, the S/N of the [O\,I] (H$\beta$) drops steeply in the NW-SE direction, complicating the tying with H$\alpha$-[N\,II] (H$\alpha$) in both MUSE and MEGARA data. Taking all this into account, we decided to fit H$\beta$, [O\,III], [O\,I] and [S\,II] independently and use the latter as a template for the H$\alpha$-[N\,II] blend. Finally, as NGC\,1052 is a type 1.9 LINER (Table\,\ref{T_properties}),  we added a broad AGN component (from the unresolved BLR) with width \,$>$\,600\,\kms (1400\,\kms in FWHM) only in H$\alpha$ forcing its spatial distribution to be the same of the PSF.  Figure\,\ref{Fig_spectral_model} shows examples of the Gaussian fits of the whole set of emission lines for both MUSE (four upper panels) and MEGARA data (two lower panels). \\
\noindent The emission lines present complex profiles with broad wings and double peaks\footnote{Note that double peaks were already detected by DH19a for  NGC\,1052 (their Fig.\,3) and in other LINERs, e.g. NGC\,5077 \citep{Raimundo2021}.} (Fig.\,\ref{Fig_spectral_model}) suggesting the presence of more than one kinematic component, especially within the innermost 10$\arcsec$ of radius. In order to prevent overfit, we first fitted all emission lines with one Gaussian component, and then more components were added based on the parameter: $\varepsilon_{\rm line}$. This parameter is defined as the standard deviation of the residuals under the emission lines, after a component is added. In the cases in which $\varepsilon_{\rm line}$\,$>$\,2.5\,$\times\,\varepsilon_{\rm cont}$ (standard deviation of the line-free continuum), another Gaussian component is added. This criterion has been already successfully applied to optical spectra of active galaxies both from long-slit \citep{Cazzoli2018, LHG2019, LHM2020} and IFS \citep{Cazzoli2020}.\\ 
Overall, we allowed for a maximum of three Gaussians per line plus the BLR-component in H$\alpha$ (Fig.\,\ref{Fig_spectral_model}). This provides a good trade-off between obtaining a statistical good fit to the spectra (i.e. residuals are of the same order as the noise without any peculiar structures as spikes or bumps) and   the number of components used having a reasonable a physical explanation.\\
\noindent For each emission line and component found, we ended up with the following information: central wavelength, width, and flux intensity along with their respective fitting uncertainties. These are the formal 1-sigma uncertainty weighted with the square root of $\chi^{2}$, as in \citet{Cazzoli2020}.\\ 
\noindent Taking into account both their central velocities and line widths, we identify a \lq primary\rq, a \lq secondary\rq \ and a \lq third\rq \ component. More specifically, the primary component can be mapped over the whole galaxy line-emitting region ($\sim$\,39$\arcsec$, i.e. 4.3\,kpc), with clear blue/red velocities with generally the lowest widths (it is also clearly detected by D15b). The third component is not spatially resolved (is extended within a radius of $\leq$\,2$\arcsec$, i.e PSF size) being generally the broadest. The secondary component has intermediate properties: it is spatially resolved, being mapped up to R\,$<$\,5$\arcsec$ (i.e. 550\,pc), with extreme velocities (up to $\sim$\,660\,\nskms). Additionally, in order to discriminate between components (especially primary and secondary ones)  we considered spatial continuity  of both flux and kinematic values. For the former a visual inspection was already satisfactory to prevent wild variations, for the latter we avoid sharp variations of the kinematics between adjacent spaxels. Specifically, we imposed that the values of the velocity fields vary smoothly (differences are less than 200\,\nskms) and that the secondary component is broader than the narrow one. Differences in line-widths are of $\sim$\,160\,-\,180\,\kms on average, for the brightest lines as [O\,III] and H$\alpha$-[N\,II]. A minor number of spaxels ($<$\,40) constitute an exception to this general behaviour of velocity dispersion but these are mainly located either within the PSF or at the largest radii where the secondary component is detected.\\
\noindent  For each of these components we created velocity, velocity dispersion and flux maps. These are shown in figures in Appendix\,\ref{Appendix_B} (from Fig.\,\ref{M_OIII_primary_zoom} to Fig.\,\ref{M_SII_second_zoom} and Fig.\,\ref{M_SII_primary_megara} to Fig.\,\ref{M_SII_second_megara} for MUSE and MEGARA, respectively). An example of these maps is shown in Fig.\,\ref{Fig_combo_OIII} for the [O\,III] line for MUSE data. In this figure we display both the large and small scales mapped by our IFS data. As the large-scale emission is similar among emission lines, the maps in Appendix\,\ref{Appendix_B} show only the central region (R\,$\sim$\,10$\arcsec$) where the largest differences are observed. We refer to Sect.\,\ref{S_main_results} for details.\\
To obtain velocity dispersion, for each spectrum (i.e. on spaxel-by-spaxel basis), the effect of instrumental dispersion (i.e. $\sigma_{\rm\,INS}$, see Sect.\,\ref{S_datared}) was corrected for by subtracting it in quadrature from the observed line dispersion ($\sigma_{\rm\,obs}$) i.e. $\sigma_{\rm\,line}$\,=\,$\sqrt{\sigma_{\rm\,obs}^{2}\,-\,\sigma_{\rm\,INS}^{2}}$. \\
We use the [S\,II] ratio (i.e. [S\,II]$\lambda$6716/[S\,II]$\lambda$6731, e.g. Fig.\,\ref{M_SII_primary_zoom}, right panel) to estimate the electron density (n$_{e}$) in accordance with the relation of \citet{Sanders2016}. 
\noindent  To investigate the ionising mechanisms across the field of view, for each component used to model emission features (forbidden and narrow Balmer lines), the maps of the four line ratios used in standard BPTs diagnostic diagrams \citep{Baldwin1981} were also generated. The maps  are presented in Appendix\,\ref{Appendix_B} (Figures from \ref{M_BPT_primary_zoom} to \ref{M_BPT_second_megara}) and the diagnostic diagrams  in Figures \ref{Fig_BPT_primary} and \ref{Fig_BPT_secondary}. For the two spatially resolved components (namely primary and secondary), the typical values of kinematics and line ratios  are summarised in Tables \ref{T_ism_result_primary} and \ref{Table_ism_S_result_secondary}.




\begin{figure}
\centering 
\includegraphics[trim = 0.cm 0cm 8.5cm 0cm, width=.495\textwidth, clip=true]{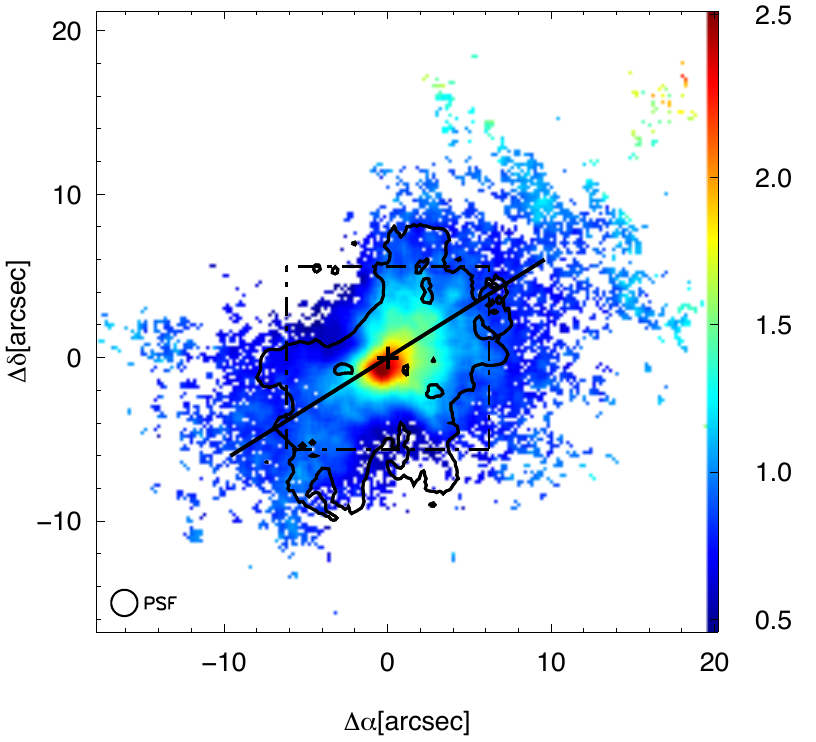}
\caption{NaD EW map in \AA-units from MUSE cube. The dot dashed square indicate the MEGARA field of view. The black solid line indicate the major axis of the stellar rotation (Table\,\ref{T_kinematics}). The contours indicate the region with enhancement velocity dispersion of emission lines (see Sect.\,\ref{S_result_primary_MUSE_butterfly} and e.g. Fig.\,\ref{Fig_combo_OIII}). }
\label{Fig_EW_NaD_abs}
\end{figure} 


\begin{figure}
\centering
\includegraphics[width=0.495\textwidth]{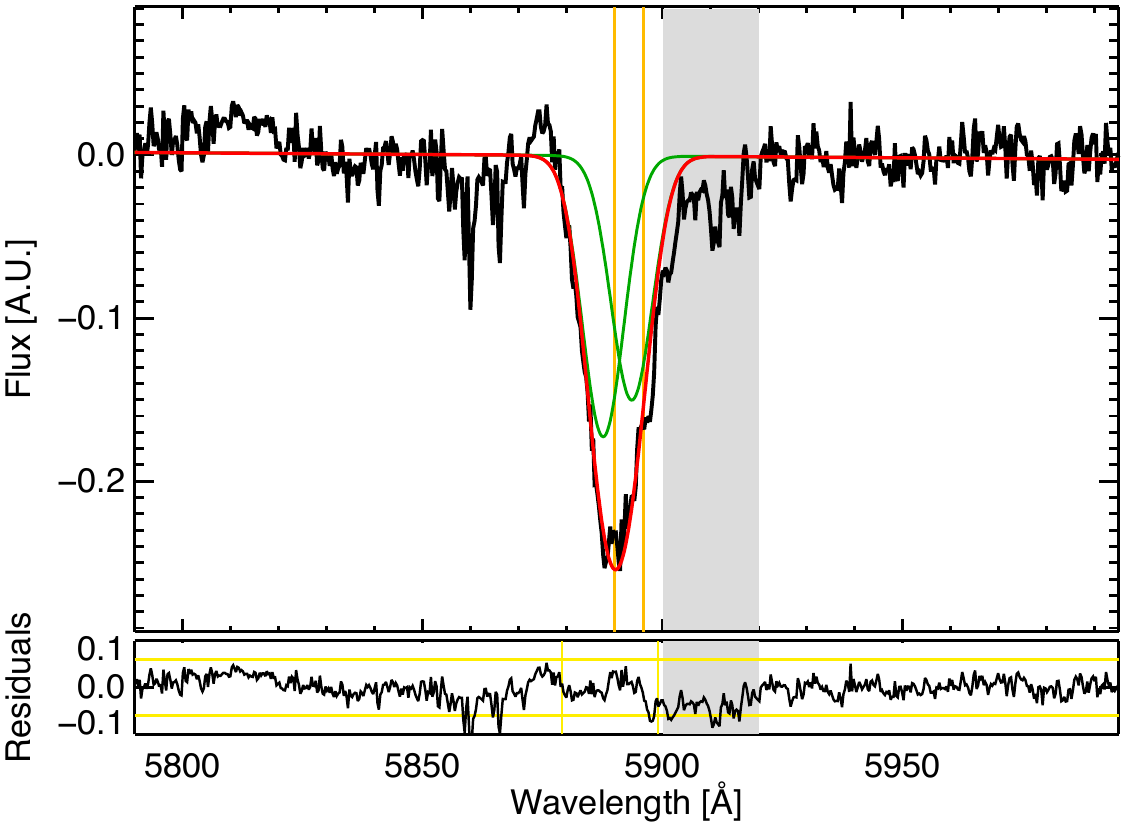}
\caption{Example of absorption line spectra (black) after stellar subtraction (Sect.\,\ref{S_stellar_cont}) and their modelling from the central region of MEGARA data (R\,=\,1$\farcs$45, i.e. 160\,pc). The grey band indicate the spectral band blocked during the fitting due to residuals from stellar subtraction (see Sect.\,\ref{S_abs_lin_mod}). Orange vertical lines and red and green curves are as Fig.\,\ref{Fig_spectral_model}, as well as both vertical and horizzontal yellow lines.}
\label{Fig_spectral_model_NaD}
\end{figure}


\begin{figure*}
\centering
\includegraphics[width=1.\textwidth]{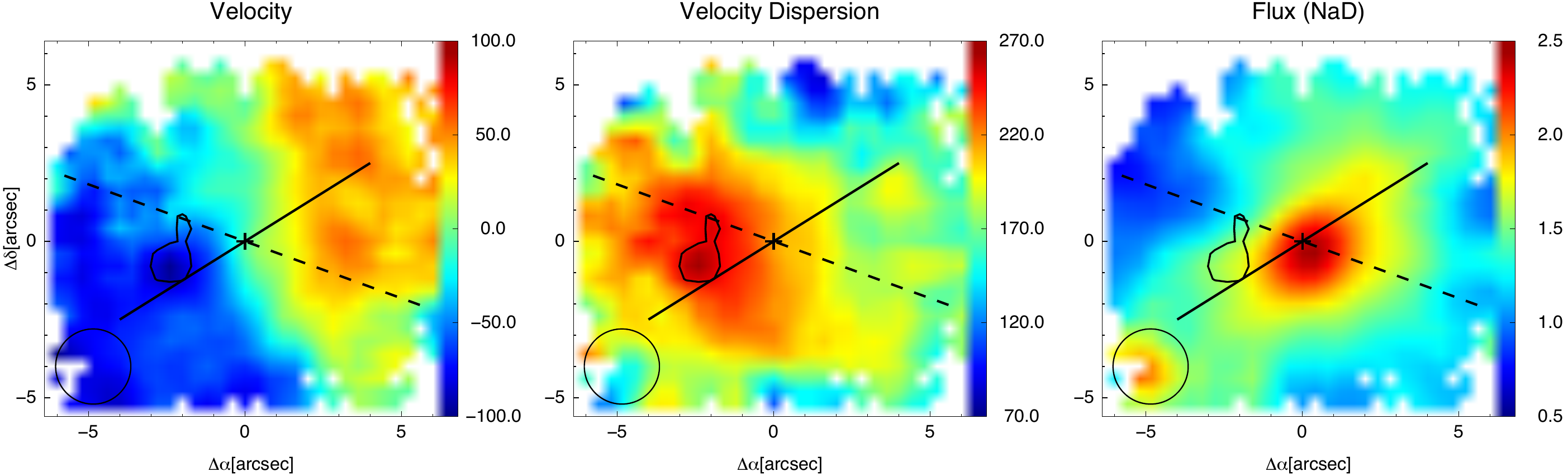}
\caption{The neutral gas velocity field (\nskms), velocity dispersion (\nskms) and flux intensity (mJy) maps for the single kinematic component used to model NaD. Black lines are as in Fig.\,\ref{M_SII_second_megara}. Specifically,  the black solid line indicate the major axis of the stellar rotation (Table\,\ref{T_kinematics}). The dashed lines indicates the orientation of the radio-jet  (Table\,\ref{T_properties}). The contours indicate the region at high velocity dispersion (Sect.\,\ref{S_outflow_kin_neutral} for details). }
\label{M_NaD_megara}
\end{figure*}

\subsubsection{Sodium doublet modelling}
\label{S_abs_lin_mod}

The wavelength coverage of  our MUSE and MEGARA data sets allow us to probe the NaD absorption doublet. This feature originates both in the cold-neutral ISM of galaxies and in the atmospheres of old stars (e.g. K-type giants). We modelled the doublet in the ISM-cubes (after the stellar subtraction, Sect.\,\ref{S_stellar_cont}) to obtain the neutral gas kinematics and hence to infer whether the cold neutral gas is either participating to the ordinary disc rotation or entraining in non-rotational motions such as outflows (see e.g. \citealt{Cazzoli2014,Cazzoli2016}).\\
For MUSE data the NaD is detected at S/N\,$>$\,3 up to R\,$\sim$\,25$\farcs$7 (2.8 kpc); however, most of the absorption (95 percent of the spaxels at S/N\,$>$\,3) is concentrated within the inner $\sim$\,16$\arcsec$ (1.8 kpc). The NaD equivalent width (EW) map is presented in Fig.\,\ref{Fig_EW_NaD_abs}. The values range from 0.4 to 3.3 \AA \ (1.1\,\AA, on average).\\
\noindent We prefer to model the NaD doublet on spaxel-by-spaxel basis in MEGARA data, as it has generally higher S/N with respect to that of MUSE data. We consider one kinematic component (a Gaussian-function each line), and we masked the wavelength range between 5900 and 5920 \AA \ due to some residuals from the stellar subtraction. \\
To infer the presence of a second component we inspect the map of the residuals (i.e. $\varepsilon$$_{\rm line}$/$\varepsilon$$_{\rm cont}$), as done for emission lines (Sect.\,\ref{S_em_lin_mod}). However, the values are in range 0.7 and 2 (1.2 on average) hence there is not a strong indication of the need of multiple components to fit the doublet.\\
Figure\,\ref{Fig_spectral_model_NaD} shows an example of the modelling of the NaD doublet absorption, and in Fig.\,\ref{M_NaD_megara} we present the corresponding kinematic and absorbed-flux maps. The results for the NaD absorption doublet are presented in Sect.\,\ref{S_results_NaD} and discussed in Sect.\,\ref{S_outflow_kin_neutral}.

\section{Main observational results}
\label{S_main_results}

In Sect.\,\ref{S_stellar_kin} we  present the results from the \texttt{pPXF} stellar kinematics analysis of both MUSE and MEGARA data.\\
\noindent The emission lines detected in both MUSE and MEGARA ISM-cubes are [S\,II], H$\alpha$-[N\,II] and [O\,I], whereas H$\beta$ and [O\,III] are covered by MUSE data only (see Sect.\ref{S_datared}). In both data sets, the maximum number of kinematic components used to model forbidden lines and narrow H$\alpha$ is three (Sect.\,\ref{S_em_lin_mod}). These components have different kinematics and spatial distribution supporting that they are distinct components. In Sect.\,\ref{ISM_kinematics_MUSE} we present the spatial distributions of kinematics and ISM properties (e.g. line ratios and electron density), measured for each of the three components in MUSE data. The comparison between MUSE and MEGARA results is presented in Sect.\,\ref{S_MUSE_vs_MEGARA}. An additional broad H$\alpha$ component originated in the BLR of the AGN has been used to model spectra  within the nuclear region (Sect.\,\ref{S_em_lin_mod}). Its properties are presented in Sect.\,\ref{S_BLR_detection} for both data sets.\\
Finally, Sect.\,\ref{S_results_NaD} summarises the main results from the modelling of the NaD absorption (Sect.\,\ref{S_abs_lin_mod}).

\subsection{Stellar Kinematics}
\label{S_stellar_kin}

\begin{figure}
\centering
\includegraphics[width=.495\textwidth]{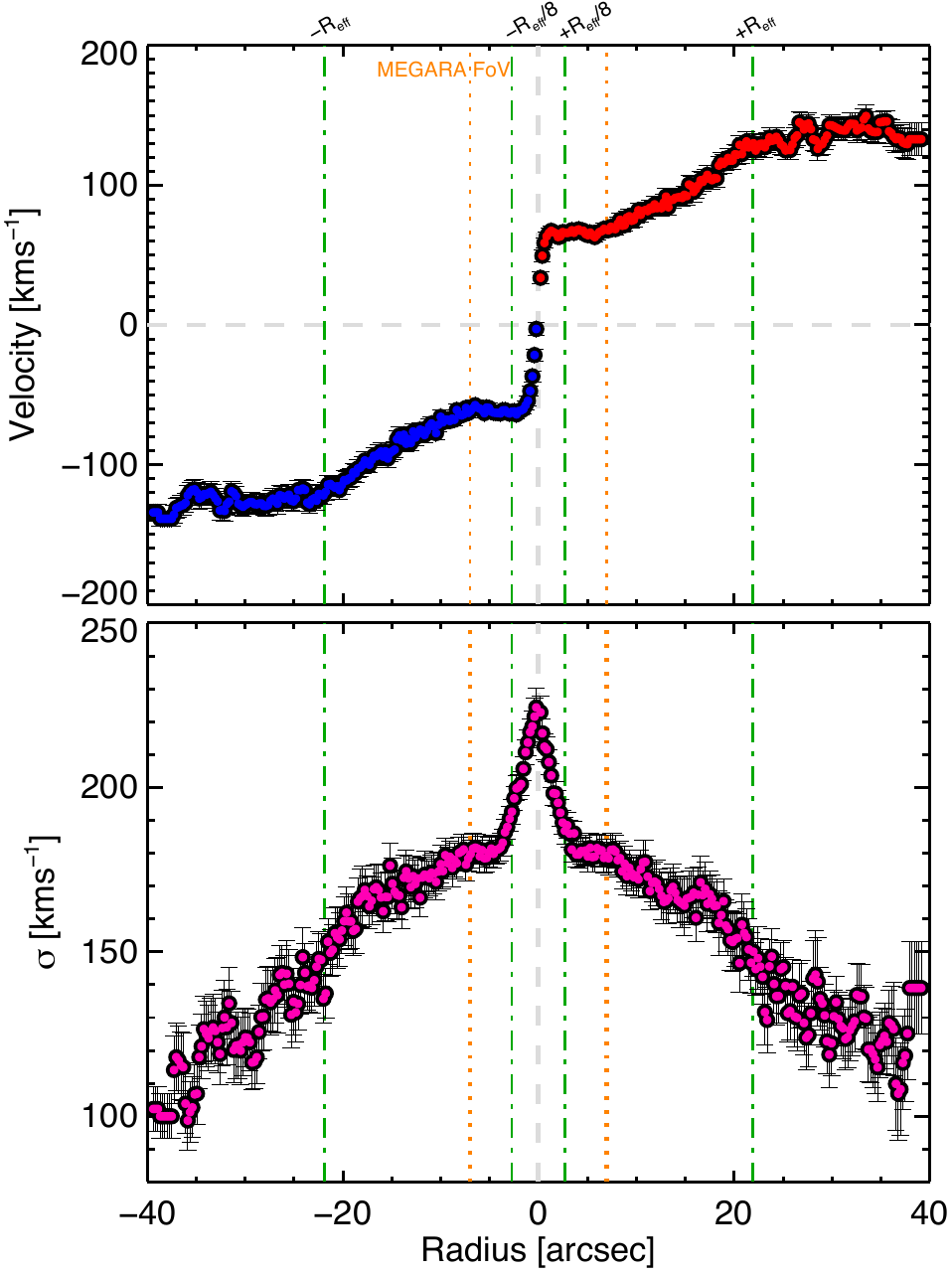}
\caption{Position-Velocity (P-V, top) and Position-Velocity Dispersion (P-$\sigma$, bottom) curves of the stellar component of NGC\,1052 from MUSE data  (Sect.\,\ref{S_stellar_kin}). Both curves were obtained considering a pseudo-slit of 1$\arcsec$-width aligned according to the major axis of the rotation  (i.e. 112$^{\circ}$, Table\,\ref{T_kinematics}). Velocities are centred to the kinematic center, and the radius is calculated as the distance from the photometric center. In the top panel, blue and red symbols indicate the approaching (negative velocities) and receding sides (positive velocities) of the rotation, respectively. Green lines mark the R$_{\rm eff}$ (21$\farcs$9, i.e. 2.4\,kpc, Table\,\ref{T_properties}) and R$_{\rm eff}$/8 (2$\farcs$75, i.e. 303\,pc,  Sect.\,\ref{S_stellar_kin}), as labelled on the top. Grey dashed lines show zero-points for position and velocity, as reference. The field of view of MEGARA observations is marked with orange dotted lines.  Note that the the typical uncertainty (extracted from the uncertainties estimated with \texttt{pPXF}) on the velocity and velocity dispersion measurements are generally $\leq$\,12\,\kms and $\leq$\,14\,\nskms, respectively.}  
\label{Fig_PVPS}
\end{figure} 


\begin{table}
\caption[kinsummary]{Stellar kinematic properties of NGC\,1052 from MUSE and MEGARA.}
\begin{center}
\begin{tabular}{l  c  c  c  c  }
\hline \hline
IFU$_{\rm FoV}$ & $\Delta$V  &  PA & $\sigma_{\rm c}$ &  $\sigma$  \\
                & \nskms  &   \degr & \nskms  &  \nskms   \\
\hline
MEGARA           & 78\,$\pm$\,3   &  112\,$\pm$\,6  & 215\,$\pm$\,13 & 201\,$\pm$\,16   \\
MUSE$_{\rm MEGARA}$ & 75\,$\pm$\,9   &  122\,$\pm$\,5  & --- & 180 $\pm$ 6   \\
MUSE             &  167\,$\pm$\,19 &  122\,$\pm$\,10 & 201\,$\pm$\,10 & 145\,$\pm$\,22   \\
\hline
\end{tabular}
\label{T_kinematics}
\end{center} 
\tiny{Notes. --- \lq $\Delta$V\rq: observed velocity amplitude; PA: position angle of the major kinematic axis; \lq $\sigma_{c}$\rq \ and \lq $\sigma$\rq \ are the central velocity dispersion (at R\,$<$\,R$_{eff}$/8, i.e. 303\,pc) and the mean velocity dispersion, respectively (see Sect.\,\ref{S_stellar_kin}). For velocity dispersion measurement the quoted uncertainties are 1 standard deviation. The \lq MUSE$_{\rm MEGARA}$\rq \ line indicate that the values are measured using MUSE data but over the field of view (FoV) of MEGARA.} 
\end{table}

\noindent As explained in Sect.\,\ref{S_stellar_cont}, we used \texttt{pPXF} to fit the stellar continuum of the spectra for both MEGARA and MUSE datacubes. The  maps of the stellar kinematics (velocity and velocity dispersion) for both data sets are shown in Fig.\,\ref{Fig_stellar_kin} and the main properties are summarised in Table\,\ref{T_kinematics}.\\
\noindent The stellar velocity field (Fig.\,\ref{Fig_stellar_kin}, left panels) shows the typical spider-pattern consistent with a rotating disc, at both large and small spatial scales mapped by our IFS data. The peak-to-peak velocity ($\Delta$V, Table\,\ref{T_kinematics}) from MUSE (MEGARA) data is 167\,$\pm$\,19\,\kms (78\,$\pm$\,3\,\nskms) at a galactocentric distance of 40$\arcsec$ (4$\arcsec$), that corresponds to 4.4 kpc (0.4 kpc). The $\Delta$V from MUSE map within the MEGARA footprint, 75\,$\pm$\,9\,\kms (Table\,\ref{T_kinematics}), is consistent with that from MEGARA cube.\\
\noindent The stellar major kinematic axis, estimated at the largest scales for  MUSE and MEGARA data are (122\,$\pm$\,10)$^{\circ}$ and (112\,$\pm$\,6)$^{\circ}$ measured north-eastwards, respectively (Table\,\ref{T_kinematics}). Both measurements indicate this axis is aligned with the photometric major axis (112.7$^{\circ}$, Table\,\ref{T_properties}). \\
\noindent Overall, the stellar velocity dispersion varies from 75 to 235\,\kms for MUSE and from 100 to 250\,\kms for MEGARA
(Fig.\,\ref{Fig_stellar_kin}, right panels). As expected in the case of a rotating disc, the stars exhibit a centrally peaked velocity dispersion map, with a maximum value of 233\,$\pm$\,6\,\kms and 241\,$\pm$\,4 \kms as measured from MUSE and MEGARA maps, respectively, being in positional agreement within uncertainties with the nucleus (considered as the photometric center, i.e. the cross in all maps). \\
\noindent Following \citet{Cappellari2013} for the ATLAS$^{\rm 3D}$ legacy project, the central velocity dispersion ($\sigma_{\rm c}$) is calculated at a distance corresponding to R$_{eff}$/8, which is R\,$<$\,2$\farcs$75 (303 pc) for NGC\,1052. The value for the central velocity dispersion is 201\,$\pm$\,10\,\kms (215\,$\pm$\,13\,\nskms) whereas the extra-nuclear mean velocity dispersion is 145\,$\pm$\,22\,\kms (201\,$\pm$\,16\,\nskms) for MUSE (MEGARA) data (see Table\,\ref{T_kinematics}). The mean velocity dispersion from MUSE data within the MEGARA footprint is 180\,$\pm$\,6 \kms (Table\,\ref{T_kinematics}), hence consistent within uncertainties with that measured directly from the MEGARA velocity dispersion map.\\
Besides the main point-symmetric disc-like pattern, in MUSE data towards the north-east and south-west and up to R\,$\sim$\,30$\arcsec$ (i.e. $\sim$\,3.3\,kpc) we observe a smooth local enhancement of the  velocity dispersion values. This enhancement is of about $150-180$\,\kms (hence above the average, Table\,\ref{T_kinematics}) but does not match features in either the continuum or ISM maps (Fig.\ref{Fig_MM_continuum} and Appendix\,\ref{Appendix_B}),  and it is not an artefact from cross-talk effects.\\
\noindent Higher velocity dispersion ($\sim$\,220\,\nskms) with respect to the mean values that seems to be present only in MEGARA at R\,$\sim$\,5$\arcsec$ prominent only to the east and to the west. Given its the position, this feature  it is likely caused by the lower S/N of the spaxels near the edges (see Sect.\,\ref{S_stellar_cont}). \\
\noindent We obtained position-velocity and position-dispersion diagrams, i.e. the \lq P-V\rq \ and \lq P-$\sigma$\rq \ diagrams in Fig.\,\ref{Fig_PVPS}, in a 1$\arcsec$-width pseudo-slit along their major axis of rotation listed in Table\,\ref{T_kinematics}. We checked that in the (central) region mapped by both data sets kinematics and curves are in agreement within uncertainties (Table\,\ref{T_kinematics}). However, as MEGARA observations cover only the innermost region (see Fig.\,\ref{Fig_MM_continuum} and Sect.\,\ref{S_datared}), in this work we will consider the kinematics from MUSE cube as reference for the stellar component.\\
The large scale rotation curve (Fig.\,\ref{Fig_PVPS}, top) is characterised by two plateau. The first flattening is at a galactocentric distance of  $\sim$\,2$\arcsec$, i.e. 220\,pc  with velocities of $\sim$\,70\,\nskms. At large distances, between 10$\arcsec$-20$\arcsec$, the curve rise slowly reaching values up to 140\,\nskms, and then finally flattens at 30$\arcsec$.\\
\noindent The velocity dispersion profile shows a sharp peak within the innermost 3$\arcsec$ (i.e. 330 pc) without an exponential decline up to the largest distances mapped by MUSE (Fig.\,\ref{Fig_PVPS}, bottom). 

\subsection{Kinematics and fluxes of the different ISM components detected by MUSE}
\label{ISM_kinematics_MUSE}

\begin{sidewaystable}
\caption{Summary of measurements for the primary component from MUSE and MEGARA.}    
\label{T_ism_result_primary}      
\centering                          
\begin{tabular}{l  c c c c c c c c c }        
\hline\hline      
 & \multicolumn{2}{c}{whole FoV} & \multicolumn{3}{c}{Polar Emission} & \multicolumn{3}{c}{Central region (high-$\sigma$)}\\ 
Line &  $\sigma$ & BPT  & $\sigma$ &  $\Delta$V  & BPT  &   $\sigma$  & $\Delta$V & BPT \\     
\hline   
& \nskms  &   &  \nskms  & \nskms &  & \nskms  & \nskms &  & \\
\hline   
H$\beta$                         & 60\,(52)\,$\pm$\,51    & --                          &  47\,(47)\,$\pm$\,25    &247\,$\pm$\,13 &  --                           &    128\,(118)\,$\pm$\,34  & 358\,$\pm$\,51 & -- \\
$[$O\,III$]$                     &  66\,(62)\,$\pm$\,39   & 0.47\,(0.46)\,$\pm$\,0.16   &   54\,(57)\,$\pm$\,21   & 251\,$\pm$\,3  & 0.46\,(0.45)\,$\pm$\,0.16    & 121\,(114)\,$\pm$\,27  & 215\,$\pm$\,6 &  0.48\,(0.48)\,$\pm$\,0.15 \\ 
$[$O\,I$]$                       & 204\,(142)\,$\pm$\,151 & -0.36\,(-0.44)\,$\pm$\,0.21 &   115\,(110)\,$\pm$\,32 & 207\,$\pm$\,11 &  -0.48\,(-0.48)\,$\pm$\,0.07 &  351\,(358)\,$\pm$\,106 & 231\,$\pm$\,34 & -0.25\,(-0.30)\,$\pm$\,0.22\\ 
H$\alpha$-$[$N\,II$]$            &  66\,(54)\,$\pm$\,47   & -0.02\,(-0.03)\,$\pm$\,0.07 &   50\,(49)\,$\pm$\,17    & 190\,$\pm$\,3 & -0.03\,(-0.03)\,$\pm$\,0.06  &  149\,(134)\,$\pm$\,52  & 295\,$\pm$\,6 &  0.06\,(0.05)\,$\pm$\,0.05 \\ 
$[$S\,II$]$                      &  58\,(48)\,$\pm$\,46   & 0.08\,(-0.08)\,$\pm$\,0.06  &   44\,(44)\,$\pm$\,21    & 200\,$\pm$\,16 & 0.07\,(0.07)\,$\pm$\,0.06   &  143\,(130)\,$\pm$\,44  &  260\,$\pm$\,15 &  0.12\,(0.12)\,$\pm$\,0.04 \\
\hline    
$[$O\,I$]$                       &  157\,(121)\,$\pm$\,115 &  -0.84\,(-0.81)\,$\pm$\,0.34  &   101(94)\,$\pm$\,33 &520\,$\pm$\,117  & -0.63(-0.66)\,$\pm$\,0.19 & 282(276)\,$\pm$\,66  & 175\,$\pm$\,92  &  -0.53(-0.59)\,$\pm$\,0.23  \\
H$\alpha$-$[$N\,II$]^{\dagger}$  &  --                          & 0.02(0.03)\,$\pm$\,0.04       & --  & --  & 0.03(0.03)\,$\pm$\,0.04 & -- & -- & 0.01(0.01)\,$\pm$\,0.03 \\
$[$S\,II$]$                      & 154\,(138)\,$\pm$\,69        & 0.17(0.17)\,$\pm$\,0.06       &  78(78)\,$\pm$\,7  &192\,$\pm$\,80  &  0.14(0.15)\,$\pm$\,0.06  &   170(155)\,$\pm$\,65 & 259\,$\pm$\,97 & 0.17(0.18)\,$\pm$\,0.06 \\  
\hline  
\end{tabular}

\tiny{Notes. --- \lq $\Delta$V\rq: observed velocity amplitude; average velocity dispersion and value of the average line ratio used for standard BPTs in Fig.\,\ref{Fig_BPT_primary} in log units. These latter are reported in correspondence of the numerator of the standard line ratios. The values are reported for the different spatial scales labelled on the top, except for the \lq whole field of view (FoV)\rq \ for which we did not report $\Delta$V as it coincides with that of polar emission (indeed the most extreme velocity values are seen at large galactocentric distance). For velocity dispersion and line-ratios measurements the quoted uncertainties are 1 standard deviation. $^{\dagger}$ [S\,II] and H$\alpha$-[N\,II] lines were fixed to have the same kinematics; only the line ratios differ.}

\end{sidewaystable}

\noindent  As mentioned at the end of Sect.\,\ref{S_line_mod}, Tables \ref{T_ism_result_primary} and \ref{Table_ism_S_result_secondary} summarise the most important properties of the two spatially resolved components (primary and secondary). Figures \ref{Fig_BPT_primary} and \ref{Fig_BPT_secondary} show the location of the line ratios for the narrow and secondary emission line components onto standard \lq BPT diagrams\rq \ \citep{Baldwin1981}. Note that, a direct comparison of gas and stellar motions for the primary component is presented in  Fig.\,\ref{P_kin} that includes the P-V and P-$\sigma$ along the three major axes, i.e. major and minor axes of the host galaxy, and the radio jet.\\
In the following we describe the overall results for each component.

\subsubsection{Overall properties of the primary component}
\label{S_result_primary_MUSE}

The primary component is the narrowest among the three detected (Sect.\,\ref{S_line_mod}), with $\sigma$\,$\leq$\,66\,\kms on average (except for [O\,I] which is of 204\,\nskms). Exceptions to this general behaviour are few spaxels ($<$\,65) mostly within the PSF area (circle in all maps in Appendix\,\ref{Appendix_B}, see also Sect.\,\ref{S_data_analysis}). The velocities are generally $\mid$V$\mid$\,$<$\,350\,\nskms, except for H$\beta$, which are up to 450\,\kms (these extreme values are observed only towards the north-west). \\
\noindent  The kinematic maps (both velocity and velocity dispersion) lack of any symmetry typical of a rotation dominated system (e.g. left and central panels of Fig.\,\ref{Fig_combo_OIII}). A clear distinguishable feature in the velocity dispersion map is the $\sigma$-enhancement crossing the galaxy from east to west (along the major axis of rotation) with a \lq butterfly-shape\rq \ (contours in Fig.\,\ref{Fig_combo_OIII} and Figures\,\ref{M_OIII_primary_zoom} to \ref{M_SII_primary_zoom}). The gas here present complex motions that differ markedly from gas elsewhere. \\
\noindent For the identification of this region with high-$\sigma$, we consider as reference the average velocity dispersion in two square regions of side 15$\arcsec$ (1.65\,kpc) in the outer part of the maps lacking of any peculiar $\sigma$ feature. Specifically, at a distance of 15$\arcsec$ from the photometric center towards the north-east and south-west. In the case of [O\,I], the box size and distance are 5$\arcsec$ and 8$\arcsec$ (550 and 880 pc), respectively, due to the decrease in S/N already visible at a radius of 10$\arcsec$ (1.1\,kpc).\\
The final threshold (i.e. 2$\sigma$ above the average velocity dispersion) is 90\,\kms for all the emission lines but [O\,I], for which is 180\,\nskms. Hereafter, we consider as polar\footnote{Throughout this paper, the polar direction (NE-SW) correspond to that of the minor photometric kinematic axis. It is not related to the direction of the AGN ionisation cones.} emission all the spaxels with velocity dispersion below those thresholds  (Sect.\,\ref{S_result_primary_MUSE_polar}). These are mostly distributed along the minor axis of rotation, i.e. NE-SW direction. The properties of the intriguing feature  with high-$\sigma$ in the central region of NGC\,1052 will be described separately from that of emitting gas organised along the  polar direction (Sect.\,\ref{S_result_primary_MUSE_butterfly}). \\
\noindent Maps of line fluxes (Fig.\,\ref{Fig_combo_OIII} and Figures\,\ref{M_OIII_primary_zoom} to \ref{M_SII_primary_zoom}, right panels) show a similar general morphology which is very different from the smooth continuum flux (Fig.\,\ref{Fig_MM_continuum}). More specifically, the gas emission within the inner 3$\arcsec$ resembles a mini-spiral while it appears extended along the NE-SW direction with some filaments and irregularities especially relevant up to R\,$\sim$\,10$\arcsec$ (mostly within the central region at high-$\sigma$). However, flux maps do not show any \lq butterfly-morphology\rq \ matching that of  the innermost region at high velocity dispersion.  Outside the inner 10$\arcsec$\,$\times$\,9$\arcsec$  (i.e. 1.1\,kpc\,$\times$\,1.0\,kpc, see Sect.\,\ref{S_result_primary_MUSE_butterfly}), the flux maps do not reveal any peculiar morphology (e.g. filaments or clumps). Taken all this into account, we prefer to describe the morphology of line-fluxes only in this section and not separately for the polar and central region (Sections \ref{S_result_primary_MUSE_polar} and \ref{S_result_primary_MUSE_butterfly}). \\
\noindent At all scales line ratios from standard BPT-diagnostic indicate LINER-like ionisation (see Table\,\ref{T_ism_result_primary} for typical values and Fig.\,\ref{M_BPT_primary_zoom}). These line ratios will be discussed in Sect.\,\ref{S_ionisation_structure} together with the weak-[O\,I] and strong-[O\,I]  LINERs classification by \citet{Filippenko1992} as in  \citet{Cazzoli2018}.\\
The [S\,II] lines ratio varies from 1.2 to 1.7 (Fig.\,\ref{M_SII_primary_zoom}) excluding extreme values (i.e. the 5 per\,cent at each end of the line-ratio distribution). This ratio is 1.47\,$\pm$\,0.2 on average, indicating a gas with relatively low density (n$_{e}$\,$<$\,100\,cm$^{-3}$).

\subsubsection{Polar emission on kpc scale}
\label{S_result_primary_MUSE_polar}

\noindent The velocity field of the primary component for all the lines show a similar overall  pattern (see  Fig.\,\ref{Fig_combo_OIII} for [O\,III]), with well defined blue and red sides oriented along the minor axis of rotation (polar direction, i.e. NE-SW). Despite that, the velocities do not show a rotating disc features (spider diagram) in any emission line (Fig.\,\ref{Fig_combo_OIII} and Appendix\,\ref{Appendix_B}).\\
The region with negative velocities extends from the photometric center towards the north-east up to 30$\arcsec$, i.e. 3.3\,kpc (12$\arcsec$, i.e. 1.3\,kpc for [O\,I], Figures \ref{Fig_combo_OIII} and \ref{M_OI_primary_zoom}, left) with an opening angle of $105^{\circ}$ as measured from the velocity maps of [O\,III]. \noindent The most blueshifted value of the observed velocity field is $\sim$\,250\,\nskms, located at a distance of $\sim$\,11$\farcs$5, i.e. 1.3\,kpc as measured from [O\,III] line (Fig.\,\ref{Fig_combo_OIII}, top-left). Similar negative velocities (within uncertainties) are seen for all the other emission lines.  The unique exception is [O\,I], for which the maximum blueshifted velocity is of about -250\,kms at a radius of 7$\farcs$5 (825\,pc) in the NE-direction (e.g. Fig.\,\ref{Fig_combo_OIII}, top-left). \\
It is worth to note that these blueshifted velocities do not decrease smoothly up to its minimum. Instead, the maps show three concentric arcs which do not cross each other (see e.g. Fig.\,\ref{Fig_combo_OIII}). These arcs are not symmetric since they are absent where positive velocities are observed (see e.g. Figures \ref{Fig_combo_OIII} and \ref{P_kin}) i.e. towards the south-west and up to 25$\arcsec$, corresponding to 2.75\,kpc (15$\arcsec$, i.e. 1.65\,kpc for [O\,I]). We checked the possibility that  extinction due to the galaxy dusty stellar disc might cause this asymmetry. By comparing the velocity maps of the ionised gas and that of the ratio between H$\alpha$ and H$\beta$ fluxes we did not find evident dusty structures at the location of the arcs. Hence, we excluded this possibility. \\
\noindent The average velocity dispersion is typically of about 50\,\kms varying between 44\,$\pm$\,21 and 54\,$\pm$\,21 \kms for [S\,II] and [O\,III], respectively (Table\,\ref{T_ism_result_primary}). The [O\,I] emission represents the exception, with a velocity dispersion of 115\,$\pm$\,32\,\kms on average (Table\,\ref{T_ism_result_primary} and Fig.\,\ref{M_OI_second_zoom}).\\
\noindent The [N\,II]/H$\alpha$, [S\,II]/H$\alpha$, [O\,I]/H$\alpha$ line ratios for the large scale gas distribution are rather homogeneous (Fig.\,\ref{M_BPT_primary_zoom}), see values in Table\,\ref{T_ism_result_primary} and the discussion in Sect.\,\ref{S_ionisation_structure}. The typical standard deviation of the values in the maps is 0.08 in log units; the scatter for the [O\,III]/H$\beta$  ratio is larger, i.e. about $\sim$\,0.2 (Fig.\,\ref{M_BPT_primary_zoom}, left). Note that, low log\,[O\,III]/H$\beta$ ratio values (i.e. $<$\,0.1) corresponding to both log\,[N\,II]/H$\alpha$ and log\,[S\,II]/H$\alpha$ of about $\sim$\,$-$0.1\,-\,0.0, are sparsely observed at large distances from the nucleus (R\,$>$\,10$\arcsec$) and towards the north-east and the south where faint clumpy features are detected (see Sect.\,\ref{S_result_lowSB}).

\subsubsection{High-$\sigma$ feature in the central region of NGC\,1052}
\label{S_result_primary_MUSE_butterfly}

\noindent For all emission lines, the region of higher velocity dispersion with $\sigma$\,$>$\,90\,\kms ($\sigma$\,$>$\,180\,\kms for [O\,I], see e.g. Figures \ref{M_OIII_primary_zoom} and \ref{M_OI_primary_zoom} and Sect.\,\ref{S_result_primary_MUSE}) is located in the innermost parts of the maps, 10$\arcsec$\,$\times$\,9$\arcsec$ (i.e. 1.1\,kpc\,$\times$\,1.0\,kpc, Table\,\ref{T_ism_result_primary}, contours in Fig.\,\ref{Fig_combo_OIII} and Figures \ref{M_OIII_primary_zoom} to  \ref{M_SII_primary_zoom}). It is mostly aligned with the major axis of the stellar rotation, with a PA of $\sim$\,124$^{\circ}$ and opening angle of $\sim$\,70$^{\circ}$ measured from [O\,III] line (Fig.\,\ref{Fig_combo_OIII}).  This region is partially mapped also with MEGARA data (see Section\,\ref{S_MUSE_vs_MEGARA} and Figures \ref{M_SII_primary_megara} and \ref{M_OI_primary_megara}). \\
\noindent The line-emitting gas is spatially resolved with MUSE into streams of filamentary strands with  a tail (clearly visible especially in [O\,III] line-maps, Fig.\,\ref{M_OIII_primary_zoom}) departing from the photometric center towards the south, with velocities up to 150\,\nskms.
\noindent In this central region, the velocity of the narrow component does not closely match the motion of the large scale gas in the polar direction. \\
Similar patterns in kinematics maps are seen for all emission lines (Figures \,\ref{M_OIII_primary_zoom}, \ref{M_Ha_primary_zoom},  \ref{M_SII_primary_zoom} and \ref{P_kin}) except [O\,I] (Figures \ref{M_OI_primary_zoom} and \ref{P_kin}) for which we summarise the main results separately. \\
\noindent For Balmer features, [O\,III], [N\,II] and [S\,II] lines large blueshifted (redshifted) velocities up to -290 (260) \nskms, are detected towards the east and west of the center of the \lq butterfly\rq \ region. The southern tail has redshifted velocities from 100 to 180 \kms generally, with a typical velocity dispersion that varies from 90 to 110 \nskms. The $\sigma$-map show non-symmetric, clumpy structures in the west strands. Such clumpiness is particularly evident in the H$\alpha$-[N\,II] velocity dispersion map (Fig.\,\ref{M_Ha_primary_zoom}  central panel). \\
\noindent For the [O\,I] line, the morphology of the high-$\sigma$ region is characterised by two well defined regions with a triangular projected area (contours in Fig.\,\ref{M_OI_primary_zoom}). The apex of the east projected triangle is at 2\arcsec \ from the photometric center whereas that of the west one is at the photometric center.\\
The velocity distribution is skewed to negative (blueshifted) velocities (60\% of the spaxels in this region). The main difference of [O\,I] kinematics with respect to the common patterns of all other lines is seen to the east. Specifically, at this location in the velocity map, two thick strands are clearly visible, both at negative velocities $\sim$\,200 and $\sim$\,160 \kms in the northern and southern directions, respectively (Fig.\,\ref{M_OI_primary_zoom}, left panel). For other emission lines, at the same spatial location, the velocities are negative and positive, hence partially kinematically distinct to what found for [O\,I].\\
The values of the [O\,I] $\sigma$-map increase gradually from the photometric center both to the east and to the west from $\sim$\,200\,\kms up to $\sim$\,500\,\kms (Fig.\,\ref{M_OI_primary_zoom}, central panel). The highest values are seen in correspondence to the most extreme velocities (e.g. the two strands towards the east). \\
\noindent Apart from the flux features summarised in Sect.\,\ref{S_result_primary_MUSE}, in the innermost 10$\arcsec$ the maps do not reveal any peculiar morphology (e.g. clumps or filament) but only a gradual decrease towards the external part of this region.\\
\noindent At the location of enhanced-$\sigma$, line ratios indicate LINER-like emission (Fig.\,\ref{M_BPT_primary_zoom}).
More specifically, the [O\,III]/H$\beta$ line ratio is typically  $>$\,0.1 in log units (on average 0.46\,$\pm$\,0.16, Table\,\ref{T_ism_result_primary}); except for an elongated region from the east to the south-west crossing the photometric center. At this location line the log\,[O\,III]/H$\beta$ varies between 0.005 and 0.3. This peculiar structures do not match any feature of any other map for the narrow component. However, it overlaps with the location of the secondary component. Any putative link between the properties of these two components will be discussed in Sect.\,\ref{S_outflow_jet}.\\
\noindent The main feature of the [N\,II]/H$\alpha$ ratio map (Fig.\,\ref{M_BPT_primary_zoom} second panel) is the presence of two clumps of similar size (diameter is 1$\farcs$2, i.e. 130\,pc). On the one hand, one clump is located within the PSF region (Sect.\,\ref{S_datared}) with log\,[N\,II]/H$\alpha$\,$\sim$\,0.2. On the other hand, the other clump with log\,[N\,II]/H$\alpha$\,$\sim$\,$-$0.3 is located 2$\farcs$6 (290\,pc) westward to the photometric center. This clump is embedded in an area with a local enhancement of the [N\,II]/H$\alpha$ ratio. Specifically, this region emerges from the photometric center and extends for 8$\arcsec$ towards the west, and partially matches the region where velocity dispersion is higher (about 250\,-\,350\,\nskms) with respect to the \lq butterfly\rq \ average, i.e. 149\,$\pm$\,52\,\kms (Table\,\ref{T_ism_result_primary}). Local [N\,II]/H$\alpha$ ratios are also enhanced at a distance of 7$\arcsec$ to the north and to the west.\\
Similarly, two clumps with log\,[S\,II]/H$\alpha$\,$\sim$\,0.03 (hence lower than the average, i.e. 0.07\,$\pm$\,0.06, Table\,\ref{T_ism_result_primary}) are detected to the north of the photometric center at R\,$\sim$\,1$\farcs$5 (Fig.\,\ref{M_BPT_primary_zoom}, right).\\
\noindent The observed values of the log\,[O\,I]/H$\alpha$ vary between $-$0.69 and 0.25 ($-$0.48\,$\pm$\,0.07 on average, Table\,\ref{T_ism_result_primary}). The morphology of this line ratio closely match the one seen in the [O\,I] kinematic maps (with well defined  strands) at the same position (Fig.\,\ref{M_BPT_primary_zoom} third panel).\\
For a detailed discussion of ionisation mechanisms from BPTs we refer to Sect.\,\ref{S_ionisation_structure}.

\subsubsection{Properties of the secondary component}
\label{S_result_secondary}

\begin{table}

\caption{Summary of measurements for the second component from MUSE and MEGARA.}    
\centering                          
\begin{tabular}{l c c c}        
\hline\hline      
Line & $\sigma$  & $\Delta$V & BPT  \\    
\hline   
& \nskms  & \nskms  &          \\   
\hline   
H$\beta$               & 313\,(316)\,$\pm$\,128 &  637\,$\pm$\,59  &  --  \\ 
$[$O\,III$]$           &  267\,(277)\,$\pm$\,44 &  582\,$\pm$\,12  &   0.52\,(0.53)\,$\pm$\,0.14\\ 
$[$O\,I$]$             &  637\,(704)\,$\pm$\,167 & 371\,$\pm$\,51  &  -0.07\,(-0.07)\,$\pm$\,0.18\\ 
H$\alpha$-$[$N\,II$]$  &  281\,(277)\,$\pm$\,105 & 569\,$\pm$\,12 & 0.14\,(0.14)\,$\pm$\,0.08\\ 
$[$S\,II$]$            &  260\,(256)\,$\pm$\,96 &  571\,$\pm$\,14 & 0.16\,(0.11)\,$\pm$\,0.14\\ 
\hline    
H$\alpha$-$[$N\,II$]$  & --  &  -- &  0.05\,(0.05)\,$\pm$\,0.09 \\ 
$[$S\,II$]$            &   445\,(434)\,$\pm$\,106 & 430\,$\pm$\,175  & 0.28\,(0.28)\,$\pm$\,0.09\\

\hline  
\end{tabular}
\label{Table_ism_S_result_secondary}
\tiny{Notes. -- The same as Table\,\ref{T_ism_result_primary} but for the secondary component. For MEGARA data: [S\,II] and H$\alpha$-[N\,II] lines were fixed to have the same kinematics (Sect.\,\ref{S_em_lin_mod}); we do not report measurements for [O\,I] due to its low S/N.}
\end{table}

For MUSE data, the spatial distribution of the secondary component has a bipolar shape extended up to 7\farcs2, that corresponds to 790\,pc  (Figures from \ref{M_OIII_second_zoom} to \ref{M_SII_second_zoom}); its properties are summarised in Table\,\ref{Table_ism_S_result_secondary}. This emission is aligned with the radio jet (PA\,=\,70$^{\circ}$, Table\,\ref{T_properties}) with a PA of $\sim$\,75$^{\circ}$, thought not center but slightly more extended to the south to the photometric center.  The  morphology is almost symmetric with respect to the photometric center with a redshifted region towards the west of the nucleus, and a blueshifted region towards the east. Overall, the velocity distribution is large, with velocities ranging from -680 to 730 \kms (Table\,\ref{Table_ism_S_result_secondary}). The line profile is broad, generally with $\sigma$\,$>$\,150\,\nskms. The average values of the $\sigma$-maps are within 260 and 320 \kms for all emission lines, except for [O\,I] which is 637\,$\pm$\,167 \kms (Table\,\ref{Table_ism_S_result_secondary}, Fig.\,\ref{M_OI_second_zoom}). Despite these high values, there is a $\sigma$-decrement ($\sigma$\,$\sim$\,80\,\nskms) that mostly corresponds to the PSF region. This feature is more evident in the H$\beta$, [O\,III] and [O\,I] maps with respect to the same maps for [S\,II] and H$\alpha$-[N\,II]. The unique feature of the flux maps outside the PSF region is a shallow elongation towards the south-west (Figures from \ref{M_OIII_second_zoom}  to \ref{M_SII_second_zoom}, right panels).\\
The average value for the [S\,II] line ratio is 1.2\,$\pm$\,0.5 (Fig.\,\ref{M_SII_second_zoom}) indicating a gas with relatively high density (100\,$<$\,n$_{e}$\,$<$\,1000\,cm$^{-3}$).
The values of the standard BPT line ratios (see Table\,\ref{Table_ism_S_result_secondary} for average values, and Fig.\,\ref{M_BPT_second_zoom}) indicate the LINER-like AGN-photoionisation as the dominant mechanism for the gas of this component (see Fig.\,\ref{Fig_BPT_secondary}). We refer to Sect.\,\ref{S_ionisation_structure} for further discussion.

\subsubsection{Faint features}
\label{S_result_lowSB}

\noindent All emission line maps from MUSE (e.g. [O\,III], Fig.\,\ref{Fig_combo_OIII}, top panels), except [O\,I] due to the lower S/N (Fig.\,\ref{M_OI_primary_zoom}), show two peculiar faint features with typical fluxes of about 3\,$\times$\,10$^{-18}$\,erg\,s$^{-1}$\,cm$^{-2}$ with kinematics (velocity and velocity dispersion) consistent with that observed in the polar direction (Sect.\,\ref{S_result_primary_MUSE_polar}). \\
On the one hand, towards the west, a stream is well visible in [O\,III] (Fig.\,\ref{Fig_combo_OIII}, top), and H$\alpha$-[N\,II]; whereas it is weakly or barely detected in [S\,II] and H$\beta$ maps. It is extended for 18$\arcsec$ (2\,kpc) as measured from H$\alpha$-[N\,II] maps considering only the detached region to the west. The same measurement in the [O\,III] map (Fig.\,\ref{Fig_combo_OIII}) is more difficult due to the fact that the stream is connected to the main body of NGC\,1052, and no peculiar feature in kinematic and flux maps allow us to disentangle the stream from the body of the galaxy.\\
This stream is found to have nearly systemic velocities  (i.e. $\pm$\,60\,\nskms) and low velocity dispersion ($<$\,50\,\nskms, generally). A small clump of radius 0$\farcs$4 (45\,pc) is detected at high-$\sigma$ ($>$\,100\,\nskms) in [O\,III] only. \\
\noindent On the other hand, towards the south and south-east, there are two detached clumps. Both clumps show redshifted velocities, but the one to the south shows the most extreme kinematics. Specifically, at this location observed velocities vary from 80 to 150 \kms (130\,$\pm$\,16\,\nskms, on average) whereas, towards the south-east, the velocity maps show values between 65 and 115 \kms (95\,$\pm$\,7\,\nskms, on average). Among these two clumps the differences in velocity dispersion are mild. The average values are 45\,$\pm$\,13\,\kms and 28\,$\pm$\,9\,\kms for the south and south-east clumps, respectively.\\
The location of the line ratios for all these faint features onto the standard BPT diagrams (Fig.\,\ref{Fig_BPT_primary} top panels, black and pink symbols) are generally consistent with those observed in AGNs (LINER-like) considering the dividing curves proposed by \citet{Kewley2006} and \citet{Kauffmann2003}. This result excludes star formation as dominant ionisation mechanism in these clumps. 
\subsection{Main kinematic properties of the third spatially unresolved component}
\label{S_result_third}

For MUSE data this component is generally the broadest one ($\sigma$\,$>$\,400\,\nskms) for H$\beta$ and Oxygen lines. For [S\,II] and H$\alpha$-[N\,II] the average line widths are 134\,$\pm$\,45 and 217\,$\pm$\,104 \nskms, respectively. Its velocity distribution is skewed to blueshifted velocities (typically within -600 and 200 \nskms). \\
In none of earlier works but D19b, the detection of a broad (FWHM\,$\sim$\,1380\,\nskms) and blueshifted  (V\,$\sim$\,490\,\nskms) unresolved component in narrow lines has been reported. D19b found such a broad component in [O\,III] only, whereas with our current MUSE data we detect it in all emission lines. \\
The FWHM of the [O\,III] line is 1053\,$\pm$\,84\,\nskms, on average, hence lower than the measurements by D19b. Although this discrepancy, considering such a large FWHM of the [O\,III] and the AGN-like BPT-ratios measured for this third component, it could probe either an unresolved AGN component as proposed by DH119b or a more recent AGN-driven outflow, which is very central and therefore unresolved.  \\
However, as mentioned in Sect.\,\ref{S_em_lin_mod}, this component is found only in the central region affected by the PSF (Sect.\,\ref{S_datared}), hence no  spatially resolved analysis can be done. 

\subsection{Comparison between MUSE and MEGARA results}
\label{S_MUSE_vs_MEGARA}

\noindent Similarly to the case of MUSE data, with MEGARA we map three different kinematic components in narrow lines and the BLR  emission in H$\alpha$. Among the detected emission lines in MEGARA ISM cube (Sect.\,\ref{S_main_results}), [O\,I] has the lowest S/N. Hence we focus on the results from the modelling of [S\,II] and H$\alpha$-[N\,II]. These lines were tied to share the same kinematics (Sect.\,\ref{S_em_lin_mod}).\\
The field of view of MEGARA data is almost completely coincident with the region at high-$\sigma$, with a minor fraction of few spaxels ($\sim$14\%) corresponding to the polar emission. Hence, we focus the comparison between the results from the MUSE and MEGARA ISM cubes on the \lq butterfly region\rq. However, we summarised the properties of the polar emission from MEGARA in Table\,\ref{T_ism_result_primary} for sake of completeness.\\
\noindent For the primary component, the velocity maps for both [S\,II] and [O\,I] lines from MEGARA data set (Figures \ref{M_SII_primary_megara} and \ref{M_OI_primary_megara}, left panels) show a rotation pattern, with larger redshifted velocities in the [O\,I] (systematically $\sim$\,100\,\kms larger). For both lines, there is a velocity decrement at  R\,$\sim$\,5$\arcsec$ north-westward from the photometric center which continues spatially up to $\sim$\,770\,pc, as seen from MUSE maps (e.g. Fig.\,\ref{M_OIII_primary_zoom}), at larger distances. This  decrement is spatially coincident with the high-$\sigma$ region, and divides the two strands seen in the \lq butterfly\rq \ region defined by the MUSE maps (see Sect.\,\ref{S_result_primary_MUSE_butterfly}). Additionally, the velocity map of the [S\,II] line (Fig.\,\ref{M_SII_primary_megara}, left) clearly shows an arc at almost rest frame velocities at approximately 3$\arcsec$ northward of the photometric center, that is also seen in  MUSE maps (see Sect.\,\ref{S_result_primary_MUSE_polar}; Fig.\,\ref{M_SII_primary_zoom}).\\
\noindent The velocity dispersion shows an average value of the [S\,II] lines of 154\,$\pm$\,38\,\nskms, broadly consistent within uncertainties with that of MUSE in the same innermost region (Table\,\ref{T_ism_result_primary}). The [S\,II] and [O\,I] lines share the same structure (Figures \ref{M_SII_primary_megara} and \ref{M_OI_primary_megara}), with increasing values in the west and east regions of the photometric center (also mentioned before for MUSE; see Sect.\,\ref{S_result_primary_MUSE_butterfly}). The photometric center has lower values ($\sim$\,100\,\nskms) than the east and west parts of the map (generally $>$\,200\,\nskms), that emerge in a biconical shape (defining the \lq wings\rq \ of the \lq butterfly\rq) from the center in a similar way as in MUSE maps (e.g. Figures \ref{M_OI_primary_zoom} and \ref{M_SII_primary_zoom}).\\
\noindent The flux maps for the narrow component of all the emission lines in MEGARA data are not centrally peaked, but show instead a spiral-like shape with high fluxes (right panels in Figures \ref{M_OI_primary_megara} and \ref{M_SII_primary_megara}). It does not correspond to any peculiar feature in the kinematic maps (velocity or velocity dispersion). This structure is also present in MUSE maps limited to the region of MEGARA field of view, being the only noticeable feature in the maps (as mentioned in Sect.\,\ref{S_result_primary_MUSE}).\\
\noindent The limited spectral coverage of MEGARA data allow us to estimate the [S\,II]/H$\alpha$, [N\,II]/H$\alpha$ and [O\,I]/H$\alpha$ line ratios (see Sect.\,\ref{S_main_results}). 
The [S\,II]/H$\alpha$ ([N\,II]/H$\alpha$) ratio in log for the primary component ranges between -0.17 to 0.44 (-0.19 to 0.18), with an average value of 0.17\,$\pm$\,0.06 (0.02\,$\pm$\,0.04). As for the [O\,I]/H$\alpha$, the values range from -1.6 to 0.3, on average -0.84\,$\pm$\,0.34 in the complete MEGARA field of view (see Table\,\ref{T_ism_result_primary} and Fig.\,\ref{M_BPT_primary_megara}). In the maps of this latter ratio (Fig.\,\ref{M_BPT_primary_megara} center), a clump is present near the photometric center, within the PSF, that is spatially coincident with an enhanced region of this ratio also in MUSE maps. Table\,\ref{T_ism_result_primary} shows that the ratios are consistent within the uncertainties independently of the high-$\sigma$\,/\,polar emission splitting. We have also estimated the electronic density using the [S\,II] line-ratio (Fig.\,\ref{M_SII_primary_megara}, right) which indicates a low density regime, as for MUSE data (see Sect.\,\ref{S_result_primary_MUSE}). The density maps of this component are homogeneous, with small deviations only in the outer parts of the field of view (with lower S/N).\\ 
\noindent For the second component detected in MEGARA data (Fig.\,\ref{M_SII_second_megara}), it has the same spatial extension as in MUSE, accounting for the differences in the spatial resolution of the two data sets. For both [S\,II] and [O\,I] velocity maps, the same structure is seen, with a clear velocity distribution ranging up to an absolute value of $\sim$\,400\,\kms for both lines. For this component, the velocities of both lines are in well agreement, also with MUSE data (see Table\,\ref{Table_ism_S_result_secondary}). As for the velocity dispersion, this component is the broadest of all the components detected in MEGARA data (excluding the broad H$\alpha$ in Sect.\,\ref{S_BLR_detection}). The values are consistent for all lines, although the [O\,I] measured in MEGARA differs considerably with that from MUSE (average of 359\,$\pm$\,64 vs. 627\,$\pm$\,167 \nskms), probably due to the lower S/N of this line in the MEGARA data. Therefore, we cannot ensure a proper determination of the properties of the secondary component with the [O\,I] lines.\\
The flux maps of all the lines show a centrally-peaked distribution, with no peculiar features. However, as in MUSE, the line ratios present elongated substructures both east and south-west from the photometric center in both [S\,II]/H$\alpha$ and [N\,II]/H$\alpha$, that do not correspond to any kinematic feature (Fig.\,\ref{M_BPT_second_megara}). The mean values of these ratios are summarised in Table\,\ref{Table_ism_S_result_secondary}. For both MUSE and MEGARA data sets, the [S\,II] flux ratio of the second component (Fig.\,\ref{M_SII_second_megara}) indicates a gas with high density, i.e. n$_{e}$\,$\sim$\,1000\,cm$^{-3}$.\\
\noindent  As already mentioned, MEGARA also identified a third spatially unresolved kinematic component in the emission lines. However, differing from MUSE data, this component is detected only in [S\,II]. Its main kinematic properties are velocities ranging between -365 and 221 \kms (mean error 72\,\kms), and an average velocity dispersion of 127\,$\pm$\,47 \nskms. This results are in broad agreement within uncertainties with that obtained with MUSE data for the [S\,II] lines (see Sect.\,\ref{S_result_third}).

\subsection{BLR component}
\label{S_BLR_detection}

The broad H$\alpha$ component from the spatially unresolved BLR of NGC\,1052 is observed only within the PSF radius (i.e. 0$\farcs$8 and 1$\farcs$2 for MUSE and MEGARA respectively, Sect.\,\ref{S_datared}) in both data sets. For this component we obtained, on average, velocities near rest frame, i.e. -38 \kms (-60\,\nskms) as measured from MUSE (MEGARA) data. Overall, the average velocity dispersion is 1031\,$\pm$\,141 \kms and 998\,$\pm$\,200  \kms (2427 and 2350 \kms in FWHM) for MUSE and MEGARA data, respectively.\\
Finally, note that our final modelling of the H$\beta$ line does not require a broad component confirming the type 1.9 AGN classification of the active nucleus in NGC\,1052 (see Table\,\ref{T_properties}).\\
The FWHM of this AGN component is compared to that of previous works in Sect.\,\ref{S_BLR_properties}.

\subsection{NaD Absorption}
\label{S_results_NaD}
Figure\,\ref{Fig_EW_NaD_abs} shows the equivalent width map of the NaD absorption corresponding to spaxels with S/N\,$\geq$\,5 in the MUSE ISM-cube. Its overall spatial distribution has an intriguing morphology similar to that of the central, butterfly-like, region at high-$\sigma$ described in Sect.\,\ref{S_result_primary_MUSE_butterfly}. It is oriented in the SE-NW direction with the north-west side more prominent (EWs generally $>$\,1.5\,\AA).\\
Our kinematic maps obtained from MEGARA data indicate a complex neutral gas kinematics (Fig.\,\ref{M_NaD_megara}). Specifically, on the one hand the velocity map shows the blue/red pattern of a rotating disc (velocities are from -96 to 57 \nskms) but with a flat gradient ($\Delta$V  is 77\,$\pm$\,12\,\nskms, Fig.\,\ref{M_NaD_megara}, left). On the other hand, the peak of velocity dispersion map is off-centred (Fig.\,\ref{M_NaD_megara} center). It peaks at 2$\farcs$5 (277 pc) eastwards with a value of 263\,$\pm$\,10\,\nskms. Moreover, large velocity dispersion values (i.e. $>$\,220\,\nskms, larger than the central velocity dispersion of the stars, $\sigma_{\rm c}$ in Table\,\ref{T_kinematics}) are observed up to 4$\farcs$8 (530\,pc) towards the north-east. These large values do not have any counterparts in either velocity or flux maps (Fig.\,\ref{M_NaD_megara}, left and right).\\
The maps of the ratio between the NaD fluxes indicate that the gas is optically thick (R$_{\rm NaD}$\,=\,1.3\,$\pm$\,0.1, on average) similarly to what was estimated for the nuclear spectrum analysed in \citet{Cazzoli2018} (R$_{\rm NaD}$\,=\,1.0), so far the only study of the NaD-absorption in NGC\,1052.

\section{Discussion}
\label{S_discussion}

The results obtained with MUSE data are in  general agreement with those from MEGARA cube at higher spectral resolution (Sections \ref{S_MUSE_vs_MEGARA} and \ref{S_BLR_detection}). \\
In Sect.\,\ref{S_disc_kin}, we discuss the stellar kinematics and dynamics using the full data set, whereas the discussion in Sect.\,\ref{S_ISM_prop} is mostly based on the results from MUSE data only in order to exploit its capabilities (spectral range, spatial sampling and field of view, Sect.\,\ref{S_datared}). Sections \ref{S_outflow_kin_ion} and \ref{S_outflow_kin_neutral} are dedicated to explore the kinematics and energetics of the multi-phase (ionised and neutral gas)  outflow. Finally, in Sect.\,\ref{S_BLR_properties} we will compare the FWHM of the unresolved BLR component with previous measurements.\\
Note that the estimation of the black hole mass based on the stellar kinematics and the broad H$\alpha$ components is discussed in Sect.\,\ref{S_disc_kin} and Sect.\,\ref{S_BLR_properties}, respectively.

\subsection{Kinematics and dynamics of the stellar disc}
\label{S_disc_kin}

\noindent As mentioned in Sect.\,\ref{S_stellar_kin}, the stellar component of NGC\,1052 shows features of rotational motions at both on small (MEGARA) and large (MUSE) scales. These include a spider pattern in the velocity field and a centrally peaked velocity dispersion map (Fig.\,\ref{Fig_stellar_kin}). Besides, the fact that the kinematic major axis coincides with the photometric major axis further confirms the presence of rotation-dominated kinematics.\\
\noindent NGC\,1052 is classified as oblate galaxy of E3-4/S0-type (\citealt{Bellstedt2018}, Table\,\ref{T_properties}). Its stellar kinematic properties (e.g. large velocity amplitude, Table\,\ref{T_kinematics}  and Fig.\,\ref{Fig_PVPS}, bottom) suggest that NGC\,1052 is more likely a lenticular-S0 galaxy (see  \citealt{Cappellari2016} for a review). The motivation is twofold. First, the lack of the exponential decline of the P-$\sigma$ curve (Fig.\,\ref{Fig_PVPS}, bottom) that indicates the presence of relevant random motions. Second, the combination of a large velocity amplitude and a symmetric velocity field (Table\,\ref{T_kinematics}, Fig.\,\ref{Fig_PVPS}, top) suggesting that NGC\,1052 has a prominent rotating disc. \\
\noindent The rotational support of the stellar disc can be  drawn from the observed (i.e. no inclination corrected) velocity-to-velocity dispersion (V/$\sigma$) ratio\footnote{Some authors (e.g. \citealt{Perna2022} and references therein) to calculate the dynamical ratio use the inclination-corrected velocity. For NGC\,1052, such a correction does not strongly affect the V/$\sigma$ ratio, i.e. it would be 1.23 instead of 1.16, hence $\sim$\,1.2 in both cases.}, calculated as the ratio between the amplitude and the mean velocity dispersion across the disc. For MUSE (MEGARA) the dynamical ratio is $\sim$\,1.2 (0.8) indicating a strong random motion component, hence a dynamical hot disc. \\
\noindent The results from the analysis of the stellar kinematics from present IFS data are generally  in agreement with those from previous works by D15 and DH19a with optical IFS from WiFEs and GMOS/GEMINI, respectively, although these are limited in either spectral range or in field of view,  and in spatial sampling (see Sect.\,\ref{S_introduction}). For both these past works, the stellar velocity field shows clearly a smooth rotation. Although, a 1:1 comparison is not possible as no velocity amplitude measurements are given by the authors. The velocity dispersion shows a central cusp ($\sim$\,200\,\kms  and $\sim$\,250\,\kms as measured by D15 and DH19a, respectively). This is qualitatively consistent  with the shape of the P-$\sigma$ curve (Fig.\,\ref{Fig_PVPS}, bottom). Finally, our results are broadly consistent with those by \citet{Bellstedt2018} obtained with DEIMOS/Keck, i.e. a rotational velocity and central velocity dispersion of $\sim$\,120\,\kms and $\sim$\,200\,\nskms, respectively.\\
\noindent The large scale rotation curve (Fig.\,\ref{Fig_PVPS}, top) is characterised by two plateau. The first flattening is at a galactocentric distance of  $\sim$\,2$\arcsec$ (i.e. 220\,pc)  with velocities of $\sim$\,70\,\nskms. At large distances, between 10$\arcsec$-20$\arcsec$, the curve rise slowly reaching values up to 140\,kms, and then finally flattens at 30$\arcsec$. \\
Thanks to our measurement of the stellar dynamics, we can provide an estimate of the black hole mass (M$_{\rm BH}$). From the central velocity dispersion of stars measured in MUSE data (201\,$\pm$\,10\,\nskms, Table\,\ref{T_kinematics}), the Eq.\,8 by \citet{Bluck2020} (see also  \citealt{Saglia2016}) yields M$_{\rm BH}$ of  2\,$\pm$\,0.5 $\times$\,10${^8}$ M$_{\sun}$. This value is in good agreement with the previous estimates by \citet{Beifiori2012} listed in Table\,\ref{T_properties}. Note that the use of other prescriptions can return different black hole masses (see e.g. \citealt{Ho2008} and reference therein) as briefly discussed in Sect.\,\ref{S_BLR_properties}.

\subsection{Multi-phase ISM properties}
\label{S_ISM_prop}
\begin{figure*}
\centering 
\includegraphics[width=.995\textwidth]{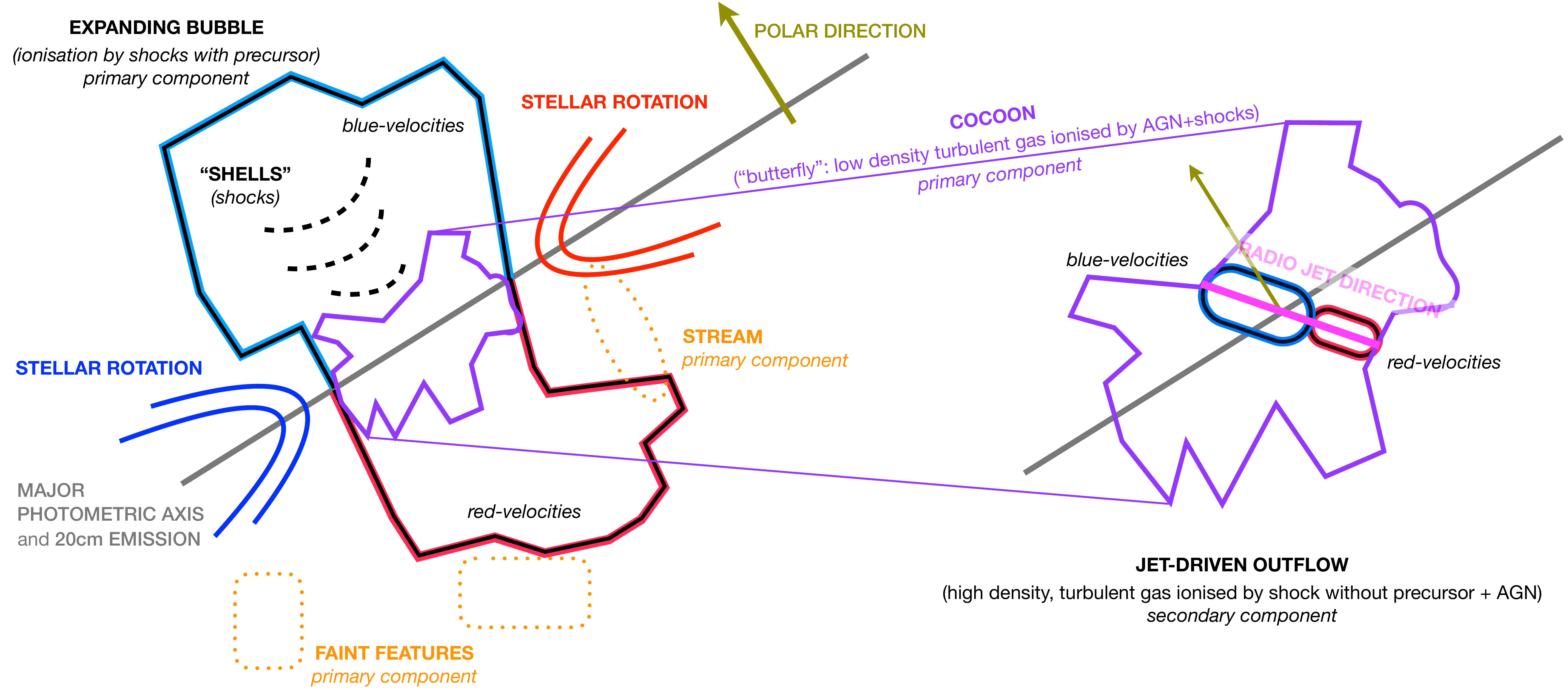}
\caption{Cartoon illustrating the proposed scenario for the stellar component and the ionised ISM for  NGC\,1052  (see text for details).}
\label{Fig_cartoon}
\end{figure*}

\noindent Early-type galaxies were traditionally thought to be uniform stellar systems with
little or no gas and dust \citep{FalconBarroso2005}.  The spatial distribution and kinematics of the ionised gas in NGC\,1052 challenges this view, as ISM and stars seem completely decoupled indicating a complex interplay between the two galaxy components. The proposed scenario is summarised in the cartoon shown in Fig.\,\ref{Fig_cartoon}.\\
In what follows we mostly focus on spatially resolved components (i.e. primary and second for emission lines and that for the NaD absorption). Note that, a third component is needed to reproduce lines profiles in all forbidden lines and narrow H$\alpha$ (see Sect.\,\ref{S_em_lin_mod}). The presence of this component has been previously reported by \citet{Dahmer2019b} but only in [O\,III] with FWHM\,$\sim$\,1380\,\nskms. These authors propose that is tracing the interaction between the jet and the ISM-environment. Despite the fact that we were able to map such a component for all emission lines (from H$\beta$ to [S\,II]), it is spatially unresolved (see Sect.\,\ref{ISM_kinematics_MUSE}).  Due to this limitation we are not investigating this component further. However, its general properties are summarised in Sect.\,\ref{S_result_third}.

\subsubsection{The intriguing ISM kinematics in NGC\,1052} 
\label{S_kinematics}

For NGC\,1052, the presence of non rotational motions  such as an AGN-driven outflow has been suggested in many previous works on the basis of \textit{HST} imaging \citep{Pogge2000, Walsh2008} as well as 1D and IFS spectroscopy \citep{Sugai2005, Dopita2015, Cazzoli2018,  Dahmer2019b}, mostly in the optical band.\\
Generally, the detection of outflows is widely based on the comparison between the observed velocity field and line width distribution and what expected in a case of a rotating disc (see e.g. \citealt{Veilleux2020} and reference therein). However, for NGC\,1052 the deviations from the disc-like behaviour (i.e. outflow signatures) in the kinematics maps of the two spatially resolved components are ambiguous. \\
On the one hand, for the primary ISM-component it is observed a clear velocity gradient in the perpendicular direction with respect to the stars (SE-NW direction, e.g. Figures \ref{Fig_combo_OIII} and \ref{P_kin}). This feature can be explained in terms of either large buoyant bubbles or a polar disc\footnote{We discard the scenario in which the polar gas arise from the AGN's narrow line region (NLR). Indeed, by means of the relation between the X-ray luminosity and size of the NLR for LINERs by \citet{Masegosa2011} (see their Fig.\,4), for NGC\,1052 the NLR physical size would be $\sim$\,600\,pc. Hence, the NLR is much less extended than the polar emission detected at a distance $>$\,3\,kpc.}. Such a bipolar velocity field is not perfectly symmetrical (see Sect.\,\ref{S_result_primary_MUSE_polar} and Figures \ref{Fig_combo_OIII} and \ref{P_kin}) indicating that the putative disc would be either a perturbed rotator or a complex kinematic object according to the classification by \citealt{Flores2006}. \\
On the other hand, the velocity dispersion map is not centrally peaked as expected for rotating discs (e.g. Figures \ref{Fig_combo_OIII} and \ref{P_kin}). Instead, a $\sigma$-enhancement\footnote{Such line-width enhancement cannot be explained in terms of beam smearing because the scale on which we observe it is much larger than the spatial resolution of the observations.} $>$\,90\,\kms is present at a galactocentric distance lower than 10$\arcsec$ with a peculiar \lq butterfly\rq \ shape (Sect.\,\ref{S_result_primary_MUSE_butterfly} and contours in Fig.\,\ref{Fig_combo_OIII} and those in the Appendix, from \ref{M_OIII_primary_zoom}  to \ref{M_SII_primary_zoom}, and Fig.\,\ref{P_kin}). At this location, the maximum velocity gradient is oriented nearly along the stellar major axis of rotation (black solid line in Figures in Appendix\,\ref{Appendix_B}). 
The morphology and the kinematics of this \lq butterfly\rq \ feature are suggestive of the presence of two bubbles outside the plane of the galaxy similarly to the well known superwind in NGC\,3079 (an optically thin bubble with blue and red sides from the front and back volumes, e.g. \citealt{Veilleux1994}). Indeed, if two bubbles (or biconical outflows) are moving away in the polar direction, high velocity dispersion is expected along the major axis of rotation due to the overlap of the blue and red clouds to the line-of-sight. The observed  \lq butterfly\rq \ feature may represent this effect. \\
\noindent  This twofold behaviour of the ionised gas on different spatial scales might indicate that the gas probed by the primary ISM-component is tracing two different substructures that are possibly related. Neither of them are likely probing a rotating disc due to the irregularities in the kinematics and the significance of shocks in ionising the gas (as discussed in Sect.\,\ref{S_ionisation_structure}).\\
\noindent For the second ISM-component the blue-to red velocity gradient is mostly aligned with the radio jet (70$^{\circ}$, Table\,\ref{T_properties}) with large widths (Table\,\ref{Table_ism_S_result_secondary}) extended mostly within 5$\arcsec$ from the photometric center (see e.g. Fig.\,\ref{M_OIII_second_zoom}).\\
As mentioned in Sect.\,\ref{S_results_NaD} the spatial distribution of the NaD absorption  has a morphology  similar that of the central region at high-$\sigma$ described with a prominent north-west side (Fig.\,\ref{Fig_EW_NaD_abs}). However, the kinematics maps do not show clear evidence of a neutral gas outflow (Fig.\,\ref{M_NaD_megara}; Sections \ref{S_results_NaD} and \ref{S_outflow_kin_neutral}). \\
Hence, by using the kinematics only we cannot claim the robust detection of a multi-phase outflow. In the next section we explore the ionisation structure and the possible connection with the radio jet in order to pinpoint the location of the outflow and hence study the kinematic, energetic and power source.

\subsubsection{Line Ratios and Ionisation structure}
\label{S_ionisation_structure}

\noindent We use the observed spatially resolved narrow emission line fluxes and line ratios to investigate the excitation mechanisms at work in NGC\,1052 by means of the standard diagnostic diagrams by \citet{Baldwin1981}, also known as BPT diagrams (Figures\,\ref{Fig_BPT_primary} and \ref{Fig_BPT_secondary}). \\
For MUSE data, H$\beta$ and [O\,I] lines are the weakest among the detected. Therefore, they constrain the spatial regions where the BPT analysis can be carried out (see maps in Appendix\,\ref{Appendix_B}). For MEGARA data the main limitation is that the observed spectra lack both H$\beta$ and [O\,III], preventing to exploit BPT-diagnostics.\\
\noindent In this section we mainly compare our results about ionisation mechanism with those in D15 due to the similarities in spatial and wavelength coverage. Such a comparison would be more difficult with the results by DH19b, as these authors present the analysis of [N\,II]/H$\alpha$ ratio, complemented by NIR BPT-diagrams, in the central region of NGC\,1052 (i.e. 3$\farcs$5\,$\times$\,5$\farcs$0). However, their general findings, i.e. LINER-like line-ratios throughout the whole GMOS/GEMINI field of view and a combination of shocks and photoionisation mechanisms in act in NGC\,1052, are in broad agreement with our results (see Sect.\,\ref{S_main_results} and below). \\
\noindent For the primary (narrowest) component, we exclude the pAGB or H\,II-ionisation scenarios in favour of a mixture of AGN-photoionisation and shocks-excitation as the dominant mechanisms of ionisation. \\
On the one hand, the large majority of the line ratios lies above the empirical dividing curves between H\,II- and AGN- like ionisation by \citet{Kewley2006} and \citet{Kauffmann2003}. These line ratios are not fully reproduced by pAGBs models  by \citet{Binette1994}. Furthermore, the observed [O\,I]/H$\alpha$ ratios indicate that NGC\,1052 is a strong-[O\,I] object (i.e. genuine AGN) according to the criterion for dividing weak-[O\,I] and strong-[O\,I] LINERs, proposed by \citealt{Filippenko1992}, that is [O\,I]/H$\alpha$\,$>$\,0.16. Hence, these findings indicate the need of an ionisation mechanism more energetic than star formation or pAGB-stars such as AGN photoionisation. Note that only for a small number of spaxels (50, i.e. $<$\,1\,per\,cent of the map), the AGN scenario is disfavoured as the log\,([O\,III]/H$\beta$) ratio is $<$\,0.3 and log\,([N\,II]/H$\alpha$) is $<$\,0.2. However, these spaxels are sparsely distributed at large distances (R\,$>$\,20$\arcsec$, i.e. 2.2\,kpc), where faint gas-clumps are detected (see Sect.\,\ref{S_result_lowSB}). \\
On the other hand, shocks models with a photoionising precursor (grids in Fig.\,\ref{Fig_BPT_primary}) are able to reproduce the large majority of the observed line ratios in the [N\,II]/H$\alpha$ and [O\,I]/H$\alpha$ diagrams, and only partially in the [S\,II]/H$\alpha$ diagram. \\
The match between data-points and shocks models is more accurate for the gas distributed along the polar direction (Fig.\,\ref{Fig_BPT_primary}, top) than for that within the central region at high-$\sigma$ (Fig.\,\ref{Fig_BPT_primary}, bottom).\\
\noindent The same two dominant sources of ionisation (AGN and shocks) acting in NGC\,1052 were identified by D15. These authors propose that part of the ionised line-emitting gas is photoionised by the AGN with a central region (R\,$<$\,1$\arcsec$) that appears shock excited. Emission lines have been modelled with a dusty plasma having a three times solar abundance and via double-shock model. The latter combines an accretion shock with velocities of about 150\,\kms and a cocoon shock at higher velocities i.e. 200\,$-$\,300\,\nskms. Such a model explains the high densities observed ($\sim$\,10$^{4}$\,$-$\,10$^{6}$\,cm$^{-3}$) in WiFes data and provides a good fit to the observed emission-line spectrum. The proposed physical scenario establishes the existence of a higher ionisation cone and a large-scale bipolar outflow (energised by the jet) and a turbulent flow along the major axis of the galaxy. However, the model by D15 only marginally fits our measurements, as explained below.\\
On the one hand, our 2D mapping of primary component reveals a central region at high velocity dispersion consistent with the accretion shocks proposed by D15, i.e. $\sigma$\,$\sim$\,120\,$-$\,150\,\kms, (except for [O\,I] for which  $\sigma$\,$\sim$\,350\,\nskms; Table\,\ref{T_ism_result_primary}). Generally the high velocity dispersion region seen in WiFES data match that of our IFS data. However, thanks to the high sensitivity and spatial resolution of MUSE we can map this region on a larger area and spatially resolve substructures in flux and kinematics. \\
\noindent On the other hand, the discrepancy is threefold. \noindent First, we do not find any indication of such extremely high density. We rather measure two regimes of gas densities both at lower densities (i.e. n$_{e}$\,$<$\,10$^{4}$\,cm$^{-3}$) for the primary and second components, as mentioned in Sections \ref{S_result_secondary} and \ref{S_result_primary_MUSE}. Second, by using the line-flux maps from MUSE, in the central region  we measured a metallicity of 8.18\,$\pm$\,2.059 (the solar value is 8.69; \citealt{Asplund2009}) following \citet{PerezDiaz2021}, so using the HII-CHI-MISTRY tool by  \citet{PerezMontero2014}). 
Hence  there are no hints of extreme metallicities as adopted by D15. Third, velocities consistent with the cocoon shock velocities (200$-$300\,\nskms) in the model by D15 are observed for the secondary component (Table\,\ref{Table_ism_S_result_secondary}, except for [O\,I]). This is a separated component with respect to that distributed on kpc-scales being extended up to 7\farcs2, that corresponds to 790\,pc (hence not only in the central R\,$<$1$\arcsec$, Sect.\,\ref{S_result_secondary}) and oriented similarly to the radio jet.\\
\noindent  For the secondary component, shock models (without precursor) are able to reproduce satisfactorily the observed [N\,II]/H$\alpha$ and [O\,I]/H$\alpha$ ratios. Nearly half of the data points in the [S\,II]/H$\alpha$ diagram are too high to be modelled either with shocks or pAGBs models (Fig.\,\ref{Fig_BPT_secondary}, grids and pink boxes, respectively).\\
\noindent  Hence, taking into account all this, we conclude that the emission lines ionisation in NGC\,1052 cannot be explained by one mechanism alone, as proposed by D15 and DH19b. The ionisation in the central region (R\,$<$\,10$\arcsec$) is a mixture of AGN-photoionisation and shock-ionisation, while at larger galactocentric distances the shock mechanism is dominating.
Finally, note that we used different shocks models (with and without precursor) to reproduce line ratios of the two spatially resolved components. The gas probed by the primary component is \lq self-ionizing\rq \ including both shocks and precursor. For the secondary component, the gas is collisionally ionised by the shock (i.e. no precursor) likely as a consequence of the passage of the radio jet,  given the alignment between the axis of the radio jet and the secondary component.\\
Note that, although the [O\,I] kinematic-properties are different from those of other lines (e.g. H$\alpha$), they are consistent with the current scenario (see e.g. Fig.\,\ref{Fig_cartoon}). Indeed, [O\,I] is highly sensitive to shocks (especially shock-heating), and it is being enhanced in the region where the butterfly-feature is observed.

\subsubsection{Connection between ISM and radio jet}
\label{S_outflow_jet}

\noindent NGC\,1052 is a radio-loud AGN (L$_{\rm 1-100\,GHz}$\,$\sim$\,4.4\,$\times$\,10$^{40}$\,ergs$^{-1}$, \citealt{Wrobel1984}) with a twin radio jet strongly interacting with its environment \citep{Kadler2004a,Mukherjee2018}. The jet has been detected in numerous observational studies (\citealt{Falocco2020} and references therein) associated with a X-ray emitting region (spatially coincident with radio emission at 1.5 GHz, \citealt{Kadler2004b}).\\
Radio jets can produce gaseous outflows impacting the host-galaxy on sub-kpc and kpc-scales (e.g.  \citealt{Harrison2014}, \citealt{LHG2018}, \citealt{Jarvis2019}, \citealt{Molyneux2019} and \citealt{Venturi2021}) with different observational features. On the one hand, powerful AGNs show high velocity dispersion along the full extent of the radio emission (e.g. \citealt{Oosterloo2019}). On the other hand, an enhancement of the emission-line velocity width is found to be perpendicular to the direction of the AGN ionisation cones and jets (e.g. \citealt{Venturi2021}).\\
In NGC\,1052 the footprints of the interaction between the jet and the gas in the galaxy disc are probed by both primary and secondary components. Specifically, the alignment between the radio emission and ionised gas in the inner $\sim$\,1-1.2\,kpc is indicative of the radio jet interacting with the ISM (PAs are about 70$^{\circ}$, Table\,\ref{T_properties} and Sect.\,\ref{S_result_secondary}). Such an interaction can trigger an outflow as well as induce large turbulence and kpc-scale bow-shocks, i.e. a jet-induced outflow acting on both sub-kpc and kpc scales. In this scenario, the outflow is probed by the second component, whereas the primary component is tracing both the cocoon of gas surrounding the expanding jet-induced outflow (with enhanced turbulence) and the large scale gas (extended up to $\sim$\,2-3\,kpc) expanding perpendicular to the axis of the jet. The shells seen in the blue part of the velocity field of the primary component along the polar direction could indicate shock waves propagating in a smooth medium. These could be absent in the red-side of the polar emission at positive velocities due ISM anisotropy.\\
The proposed scenario for NGC\,1052 is similar to that presented by \citet{Morganti2021} for PKS\,0023--26 (a far IR bright source hosting a young powerful radio source) on the basis of the results from ALMA CO\,(2$-$1) and 1.7\,mm continuum data. In PKS\,0023--26, the highly perturbed gas tends to follow the edge of the radio emission on sub-kpc scales, whereas the relatively mild expansion of the cocoon, created by the interaction between jet and ISM, is pushing the gas aside. For NGC\,1052 the strong coupling between radio jets and the ISM is limited to the innermost 7\farcs2, corresponding to 790\,pc (Sect.\,\ref{S_result_secondary}), with large buoyant bubbles extended up to 30$\arcsec$, that corresponds to 3.3\,kpc (Sect.\,\ref{S_result_primary_MUSE_polar}). As for PKS\,0023--26, the cocoon is not reaching extreme velocities but it is injecting turbulence into the ISM, triggering the creation of the bubbles along the polar direction. With the present data set we cannot infer the presence of cavities devoid of dense gas at larger radii due to the maintenance phase of outflow, neither any relation between radio lobes and the ISM, as the former are absent for the jet in NGC\,1052.\\ Although the comparison between PKS 0023--26 and our results for NGC\,1052 is illustrative, it has to be taken with caution as we are tracing different gas phases within the jet-ISM interaction.

\subsection{Ionised gas outflow kinematics and energetic}
\label{S_outflow_kin_ion}

On the basis of the morphology and kinematics of the different components, and taking into account that shocks are a crucial mechanism of ionisation in NGC\,1052, we claim the detection of an ionised gas outflow. It is probed by the secondary component, with a bipolar morphology and velocity dispersion $>$\,150\,\nskms. The outflow is strongly interacting with the surrounding ISM mapped by the primary component. Such an interplay is suggested by both the high-$\sigma$ ($>$\,90\,\nskms) region with a peculiar \lq butterfly like\rq \ morphology and the presence of two kpc-scale buoyant bubbles. \\
In this section, we summarise the main properties (kinematics and energetic) of the outflow as well as its power source (i.e. verify the jet-driven scenario proposed in Sect.\,\ref{S_outflow_jet}).\\
We assume a simple outflow model, with inclination-corrected velocities and distances, that considers the outflow oriented perpendicular to the plane of the disc.\\ 
We estimated the total mass of the emitting ionised hydrogen gas following \citet{Venturi2021} (see also \citealt{Carniani2015},  \citealt{Cresci2017}). We calculated the H$\alpha$ luminosity corrected for extinction (L$_{\rm H\alpha}$), considering the corresponding distance (i.e. 22.6\,Mpc, Table\,\ref{T_properties}) and  using the attenuation law  by \citet{Calzetti2000} for galactic diffuse ISM (R$_{V}$ = 3.12) and an intrinsic ratio (H$\alpha$/H$\beta$)\,=\,2.86 (for an electron temperature of 10$^{4}$\,K, \citealt{Osterbrock2006}). The intrinsic H$\alpha$ luminosity is converted in mass of the ionised gas with the Eq.\,1 in \citet{Venturi2021} using the median value of the electron density (i.e. 360\,cm$^{-3}$). We obtain a total L$_{\rm H\alpha}$ of (1.8\,$\pm$\,0.7) $\times$\,10$^{40}$ erg\,s$^{-1}$ and a total mass of ionised gas in the outflow of M$_{\rm OF, ion}$ = (1.6\,$\pm$\,0.6) $\times$\,10$^{5}$ M$_{\sun}$ with our data. \\
\noindent The mass outflow rate is $\dot{\rm M}_{\rm OF, ion}$\,= (0.4\,$\pm$\,0.2) M$_{\sun}$yr$^{-1}$, estimated with 3\,$\times$\,V$_{\rm OF, ion}$/R$_{\rm OF, ion}$\,$\times$\,M$_{\rm OF, ion}$ as in \citet{Cresci2015}. Note that, in this estimation, we assumed V$_{\rm OF, ion}$ to be the maximum of the velocity field from the map of the secondary component ($\sim$\,655\,\nskms). \\
\noindent We also estimated the kinetic energy and power of the outflowing ionised gas, i.e. E$_{\rm OF, ion}$\,=\,0.5\,$\times$\,$\sigma_{\rm OF, ion}^{2}$\,M$_{\rm OF, ion}$ and $\dot{\rm E}_{\rm OF, ion}$\,=\,0.5\,$\times$\,$\dot{\rm M}$\,$\times$\,(V$_{\rm OF, ion}^{2}$+\,3$\sigma_{\rm OF, ion}^{2}$), using an average velocity dispersion of H$\alpha\sim$\,280\,\nskms. We obtained E$_{\rm OF, ion}$ = (1.3\,$\pm$\,0.9) $\times$ 10$^{53}$ erg and $\dot{\rm E}_{\rm OF, ion}$= (8.8\,$\pm$\,3.5) $\times$ 10$^{40}$ erg\,s$^{-1}$.\\
\noindent An upper limit on the outflow mass and energy could be estimated by considering the whole \lq outflow-phenomenon\rq \ i.e. the  \lq outflow-core\rq \ (secondary component) plus the buoyant bubbles (primary component, excluding the faint features described in Sect.\,\ref{S_result_lowSB}). Hence, the mass of the ionised gas associated to the outflow-phenomenon (bubbles) is 1.8\,$\pm$\,1.1 $\times$\,10$^{6}$ M$_{\sun}$ (1.7\,$\pm$\,1.1 $\times$\,10$^{6}$) M$_{\sun}$, the corresponding energy is 1.9\,$\pm$\,1.5\,$\times$ 10$^{53}$ (8.1\,$\pm$\,1.1 $\times$ 10$^{52}$) erg.\\
\noindent We exclude the star-formation as a power source of the outflow since the kinetic power of the starburst associated to supernovae is low ($\sim$\,6.3\,$\times$\,10$^{40}$ erg\,s$^{-1}$, as calculated following \citealt{Veilleux2005} from the total SFR, i.e. 0.09 M$_{\odot}$yr$^{-1}$, Table\,\ref{T_properties}). In what follows we focus in disentangling between the two most likely  scenarios: AGN- vs. jet- driven outflow.\\
\noindent The energy rate is of the order of 0.01 of the bolometric luminosity of NGC\,1052 (L$_{\rm bol}$\,=\,10$^{42.91}$ erg\,s$^{-1}$ \citealt{Onori2017b}). This is in broad agreement with the results of \citet{Fiore2017} (see their Fig.\,1\,right) that showed that the average ratio $\dot{\rm E}_{\rm OF, ion}$/L$_{\rm bol}$ for AGN-driven ionised outflows is generally below 0.1.
\noindent As in \citet{Venturi2021}, in order to infer whether the jet is energetic enough to power the observed features, we compared the total kinetic energy of the jet (E$_{\rm jet}$) with the  kinetic energy of the outflow. By assuming the power and travelling time of the jet (i.e. 10$^{45}$\,erg\,s$^{-1}$ and 0.7\,Myr, respectively) used in \citet{Mukherjee2018} to simulate the observed kinematics and morphology of the ionised gas in NGC\,1052 we obtain a total energy of the jet of E$_{\rm jet}$\,=\,2.2\,$\times$\,10$^{58}$\,erg.\\
\noindent The comparisons $\dot{\rm E}_{\rm OF, ion}$ vs. L$_{\rm bol}$ and E$_{\rm jet}$ vs. E$_{\rm OF, ion}$ indicate that both the AGN and the jet in NGC\,1052 are capable to inject the required energy into the ISM to power the outflow. However, taking into account the alignment between the radio jet, the secondary component, and the cocoon with enhanced turbulence, we consider that the most likely power source of the outflow is the jet, although some contribution from the AGN is possible.

\subsection{Neutral gas outflow detection}
\label{S_outflow_kin_neutral}

\noindent As mentioned in Sect.\,\ref{S_kinematics}, the mapping of the neutral gas properties does not show evident outflow features (e.g. a broad kinematic component with significant blueshifted velocities). Hence, the identification of the neutral gas outflow and the corresponding estimates provided in this Section is exploratory and hence must be taken with caution. \\
To identify the putative neutral gas outflow we used the velocity dispersion map  (Fig.\,\ref{M_NaD_megara}, center) that shows the most clear deviations from the rotating-disc behaviour among those obtained from the NaD modelling (Sect.\,\ref{S_results_NaD}). \\
As a threshold to identify the outflowing neutral gas, we consider the 75th percentiles of the  distributions of the velocity dispersion, that is, $\sigma_{\rm thr}$\,$>$\,245\,\nskms. The selected region (with $\sigma$\,$>$\,$\sigma_{\rm thr}$) is marked with contours in the maps shown in Fig.\,\ref{M_NaD_megara}. It is extended up to a galactocentric distance of 4$\farcs$8 (530\,pc), with an elongated  morphology (oriented north-south), and a projected area of 3.8\,arcsec$^{2}$. The region is characterised by a mild kinematics with velocity and velocity dispersion of 63\,$\pm$\,21\,\kms and 251\,$\pm$\,5\,\nskms, respectively, on average. The EW is on average of  1.2\,$\pm$\,0.3\,\AA. This value is converted into column density of the wind (N$_{\rm H}$) via reddening (E$_{\rm B-V}$) following the approach by \citet{Cazzoli2014, Cazzoli2016}, already used for MEGARA data by \citet{CatalanTorrrecilla2020}. On average, the column density of the outflow is (2.8\,$\pm$\,0.7) $\times$ 10$^{21}$ cm$^{-2}$. As in \citet{CatalanTorrrecilla2020} we assumed that the outflow is organised in a series of thin shells and, to obtain the deprojected velocities, distances and solid angle,  we used a simple geometrical model of a conical outflow that emerges perpendicular to the disc.\\
Following these prescriptions, the total mass of neutral gas contained in the outflowing region is (7.1\,$\pm$\,2.8) $\times$ 10$^{6}$ M$_{\sun}$ and the outflow rate is (0.86\,$\pm$\,0.30) M$_{\sun}$yr$^{-1}$. We also derived the total energy of the neutral outflow which is (1.1\,$\pm$\,0.4)\,$\times$\,10$^{55}$\,erg.\\
\noindent Cold neutral gas outflows in LINERs and early-type galaxies (ETG), probed by the NaD absorption, have been less studied compared to e.g. ionised/molecular outflows in Seyferts or U/LIRGs (e.g. \citealt{Arribas2014, PereiraSantaella2016, Venturi2018, PereiraSantaella2020, Wylezalek2020, Perna2021, Comeron2021, Riffel2021}). 
However, there are two systematic studies of neutral gas in LINERs in large samples by \citet{Lehnert2011} and \citet{Sarzi2016}. \citet{Lehnert2011} detected neutral ISM-gas in about one-third of their sample of 691 radio-loud ETGs on the basis of SDSS data. The detected NaD profiles suggest the presence of outflows with low velocities ($\sim$\,50\,\nskms) and broad profiles ($\sim$\,500\,\nskms). On the contrary \citet{Sarzi2016} found that only  a dozen radio-AGNs (out of 103 objects) show NaD absorption from ISM, but the neutral gas never appears to be outflowing. \\
The unique study of the NaD absorption in NGC\,1052 is by \citet{Cazzoli2018}  on the basis of slit spectroscopy.
In this work, the neutral gas kinematics has been interpreted as due to rotation. However,  slit observations give only a partial description of the outflow phenomenon, hence in the case of NGC\,1052, IFS observations could have been the key for our (tentative) detection of the neutral gas outflow.

\subsection{Comparison with current and previous H$\alpha$ broad component measurements}
\label{S_BLR_properties}
The BLR component in NGC\,1052 has been observed in polarized light \citep{Barth1999} and at different wavelengths (\citealt{Onori2017a},  \citealt{Cazzoli2018}, \citealt{Dahmer2019a} and references therein).\\
\citet{Onori2017a} modelled the BLR component in both optical and near-IR bands with \textit{HST}/FOS  (R\,$\sim$\,2800) and ISAAC  (R\,$\sim$\,730)  spectra. The near-IR He\,I$\lambda$1.083$\mu$m line has been modelled with a broad Gaussian curve with width of 2455\,\nskms, a slightly smaller value (i.e. 2193\,\nskms) has been used for H$\alpha$. The broad H$\alpha$ emission was also measured by \citet{Balmaverde2014} (FWHM\,$\sim$\,2240\,\nskms), \citet{Constantin2015} (FWHM\,$\sim$\,2800\,\nskms) and \citet{Cazzoli2018}\footnote{They found evidence for the BLR only in \textit{HST}/STIS and not in ground-based CAHA/CAFOS data, due  to a less reliable fit to the H$\alpha$ emission line in ground- and space-based data sets.} (FWHM\,$\sim$\,2915\,\nskms) with \textit{HST}/STIS slit spectra, all obtaining values that are in fair agreement.\\
There exists three measurements of the width of the broad H$\alpha$ component with optical IFS. Two of them are from the present MEGARA and MUSE IFS data, that are 2427\,$\pm$\,332 and 2350\,$\pm$\,470 \nskms, respectively. These values are consistent within uncertainties but smaller from the value reported by DH19a, from their GMOS/GEMINI cube (R\,$\sim$\,1700 and final angular resolution 0$\farcs$7), that is $\sim$\,3200\,\nskms.\\
We considered as the main sources of the discrepancies the number of components used to model emission lines and the different spectral/spatial resolution of the different data sets (see e.g. \citealt{Cazzoli2020}). Another possibility for explaining the differences in the FWHM of the broad component is AGN variability (see e.g. \citealt{LHG2014}) whose study is beyond the aim of the paper.\\
\noindent The FWHM and luminosity of the broad H$\alpha$ component determined from the best-fitting model of the H$\alpha$ broad component can be converted in black hole mass using the virial relation. For NGC\,1052, we found that M$_{\rm BH}$ is $\sim$\,3\,$\times$\,10${^5}$ M$_{\sun}$ from Eq.\,3 in \citet{Koss2017}. Considering that the assumed value of luminosity is a lower limit as  we did not applied any correction for reddening, the estimate of the black hole mass from H$\alpha$ is in broad agreement with  that by \citet{Onori2017b} using the virial relation i.e. $\sim$\,4\,$\times$\,10${^6}$ M$_{\sun}$ (Table\,\ref{T_properties}). \\
\noindent However, the determination based in the broad H$\alpha$ has been explored the most for luminous type 1 AGNs (see e.g. \citealt{Greene2005} for details) hence for type 1.9  LINERs as NGC\,1052 (Table\,\ref{T_properties}), could be uncertain (e.g.  it challenging to  isolate the AGN contribution unambiguously). Therefore, we consider the M$_{\rm BH}$ from the stellar velocity dispersion,  more reliable (it is the result of coevolution between the host galaxies  and the supermassive black holes).

\section{Conclusions}
\label{S_conclusions}

\noindent On the basis of optical MUSE and MEGARA IFS data we have studied the properties of the stellar and ionised/neutral gas components in the LINER\,1.9  NGC\,1052, using as tracers both emission lines (from H$\beta$ to [S\,II]) and the NaD absorption doublet.\\
The conclusions of this study can be summarised as follows: 

\begin{enumerate}

\item \textit{Kinematics and dynamical support for the stellar component.}
The stellar velocity field is characterised by ordered large-scale rotational motions ($\Delta$V\,=\,167\,$\pm$\,19\,\nskms), although the velocity dispersion is generally high as measured from MUSE (145\,$\pm$\,22\,\nskms) and MEGARA data (201\,$\pm$\,16\,\nskms). The rotational support is, however, low. The dynamical ratio,  V/$\sigma$\,=\,1.2 (0.8) from MUSE (MEGARA) data, is indicative of a dynamically hot disc with a significant random motion component. In both data sets, the stellar major axis is well aligned with the photometric one. The kinematic and dynamics of the stellar disc of NGC\,1052 favour its classification as a S0-type. The black hole mass estimated from stellar dynamics is 2\,$\pm$\,0.5 $\times$\,10${^8}$ M$_{\sun}$.\\ 

\item \textit{Ionisation mechanisms}.  By combining the location of line ratios onto BPTs, theoretical models of shocks and pAGBs ionisation, and the weak/strong [O\,I] classification, we exclude star formation and pAGB scenarios in favour of a mixture of shock excitation and AGN activity as the main mechanisms of ionisation in NGC\,1052. The general behaviour is that the ionisation in the central region (R\,$<$\,10$\arcsec$) is a mixture of AGN-photoionisation and shocks while at larger galactocentric distances the shock excitation is dominating.\\

\item \textit{The intriguing properties of the ionised gas probed by the primary component}. The velocity field shows a large scale structure extended in the polar direction (NE-SW direction) up to $\sim$\,30$\arcsec$ ($\sim$\,3.3\,kpc) with blue and red velocities (typically $<$\,$\mid$\,250\,$\mid$\,\kms). The velocity dispersion map lacks of any symmetry typical of a rotation dominated system with a notable enhancement ($\sigma$\,$>$\,90\,\nskms) crossing the galaxy along the major axis of rotation in the central $\sim$\,10$\arcsec$ (also called \lq butterfly\rq \ region within the main text). We consider that both features are likely related to the presence of an ionised gas outflow instead of, for example, a polar disc.\\ 

\item \textit{Ionised gas outflow}. It is probed by the secondary component with a bipolar morphology, velocity dispersions $>$\,150\,\kms and velocities up to 660\,\nskms. The outflow (with mass of 1.6\,$\pm$\,0.6 $\times$\,10$^{5}$ M$_{\sun}$, and mass rate of 0.4\,$\pm$\,0.2 M$_{\sun}$yr$^{-1}$) is propagating in a cocoon of gas with enhanced turbulence (the \lq butterfly\rq\ region) and triggering the onset of kpc-scale buoyant bubbles (polar emission). Considering the energy (1.3\,$\pm$\,0.9 $\times$ 10$^{53}$ erg) and energy rate (8.8\,$\pm$\,3.5 $\times$ 10$^{40}$ erg\,s$^{-1}$) of the outflow both the AGN and the radio jet are able to launch the outflow. However, taking into account both its alignment with the jet and with the cocoon, and that the gas is collisionally ionised, we consider that the most likely power source of the outflow is the jet, although some contribution from the AGN is possible. \\

\item \textit{Neutral gas content}. The kinematics maps of the NaD absorption obtained with MEGARA data indicate optically thick neutral gas with complex kinematics. The velocity field is consistent with a slow rotating disc ($\Delta$V\,=\,77\,$\pm$\,12\,\nskms) but the velocity dispersion maps is off-centred with a peak value of 263\,$\pm$\,10 \kms observed at 2$\farcs$5 (277 pc) eastwards to the photometric center without any counterpart in the (centrally peaked) flux map. The hints of the presence of the neutral gas outflow are weak and our identification its tentative. The putative neutral gas outflow is extended to the west with projected area of 3.8 arcsec$^{2}$ with mild kinematics i.e. with velocity and velocity dispersion of 63\,$\pm$\,21\,\kms and 251\,$\pm$\,5\,\nskms, respectively. The mass, the mass rate and the energy of the neutral would be (7.1\,$\pm$\,2.8)\,$\times$\,10$^{6}$ M$_{\sun}$, (0.86\,$\pm$\,0.30)\,M$_{\sun}$yr$^{-1}$, and (1.1\,$\pm$\,0.4)\,10$^{55}$\,erg, respectively. \\

\item \textit{BLR properties}. In the nuclear region of NGC\,1052 ($\leq$\,1$\arcsec$) the broad H$\alpha$ component originated in the (unresolved) BLR of the AGN is modelled with a Gaussian component with FWHM of 2427\,$\pm$\,332 and 2350\,$\pm$\,470\,\nskms, respectively for MUSE and MEGARA data.\\

\item \textit{Unresolved component}.  It has been detected with MUSE data (barely with MEGARA) in all emission lines. This component is observed in the central region, with a spatially extension matching that of the PSF, with an average FWHM\,$\sim$\,1380\,\kms and line ratios indicating AGN-ionisation. It could probe either an unresolved AGN component as proposed by DH119b or a more recent AGN-driven outflow. However,  with the current data set it is not possible to disentangle among the two scenarios.

\end{enumerate}

\begin{acknowledgements}
The authors acknowledge  the anonymous referee for her/his instructive comments that helped to improve the presentation of this paper.\\
SC, IM, JM and LHM acknowledge financial support from the State Agency for Research of the Spanish MCIU through the \lq Center of Excellence Severo Ochoa\rq \ award to the Instituto de Astrof{\'i}sica de Andaluc{\'i}a (SEV-2017-0709). These authors are also supported by the Spanish Ministry of Economy and Competitiveness under grants no. AYA2016-76682-C3 and PID2019-106027GB-C41. LHM acknowledges financial support under the FPI grant BES-2017-082471. AGdP and ACM acknowledge the grant RTI-2018-096188-B-I00. LHG acknowledges funds by ANID – Millennium Science Initiative Program – ICN12$\_$009 awarded to the Millennium Institute of Astrophysics (MAS). FLF acknowledges support from PRIN MIUR project \lq Black Hole winds and the Baryon Life Cycle of Galaxies: the stone-guest at the galaxy evolution supper\rq, contract no. 2017PH3WAT.  CRA acknowledges financial support from the European Union's Horizon 2020 research and innovation programme under Marie Sk\l odowska-Curie grant agreement No 860744 (BiD4BESt) and from the State Research Agency (AEI-MCINN) and the Spanish MCINN under grant \lq Feeding and feedback in active galaxies\rq, with reference PID2019-106027GB-C42.
This research has made use of the NASA/IPAC Extragalactic Database (NED), which is operated by the Jet Propulsion Laboratory, California Institute of Technology, under contract with the National Aeronautics and Space Administration. We acknowledge the usage of the HyperLeda database (\url{http://leda.univ-lyon1.fr}).\\
This work has made extensive use of \texttt{IRAF} and \textsc{Python}, particularly with \textsc{astropy} \url{http://www.astropy.org} \citep{astropy:2013, astropy:2018}, \textsc{matplotlib} \citep{Hunter:2007}, \textsc{numpy} and \textsc{lmfit}.\\
This paper made use of the plotting package \textsc{jmaplot}, developed by Jes{\'u}s Ma{\'i}z-Apell{\'a}niz  available at: \url{http://jmaiz.iaa.es/software/jmaplot/ current/html/jmaplot_overview.html}. \\
This research has made use of the \texttt{Skycat} tool that combines visualization of images and access to catalogues and archive data for astronomy. In particular, \texttt{EXTRACTOR} as part of the \texttt{GAIA} (Graphical Astronomy and Image Analysis Tool) package.\\
We thanks J.\,Perea Duarte for the technical support and B. Perez for the calculation of gas metallicity.
\end{acknowledgements}

\bibliographystyle{aa} 
\bibliography{Bibliography.bib}
\begin{appendix}
\section{Background/foreground emission}
\label{Appendix_A}
\begin{figure}
\centering
\includegraphics[width=.495\textwidth]{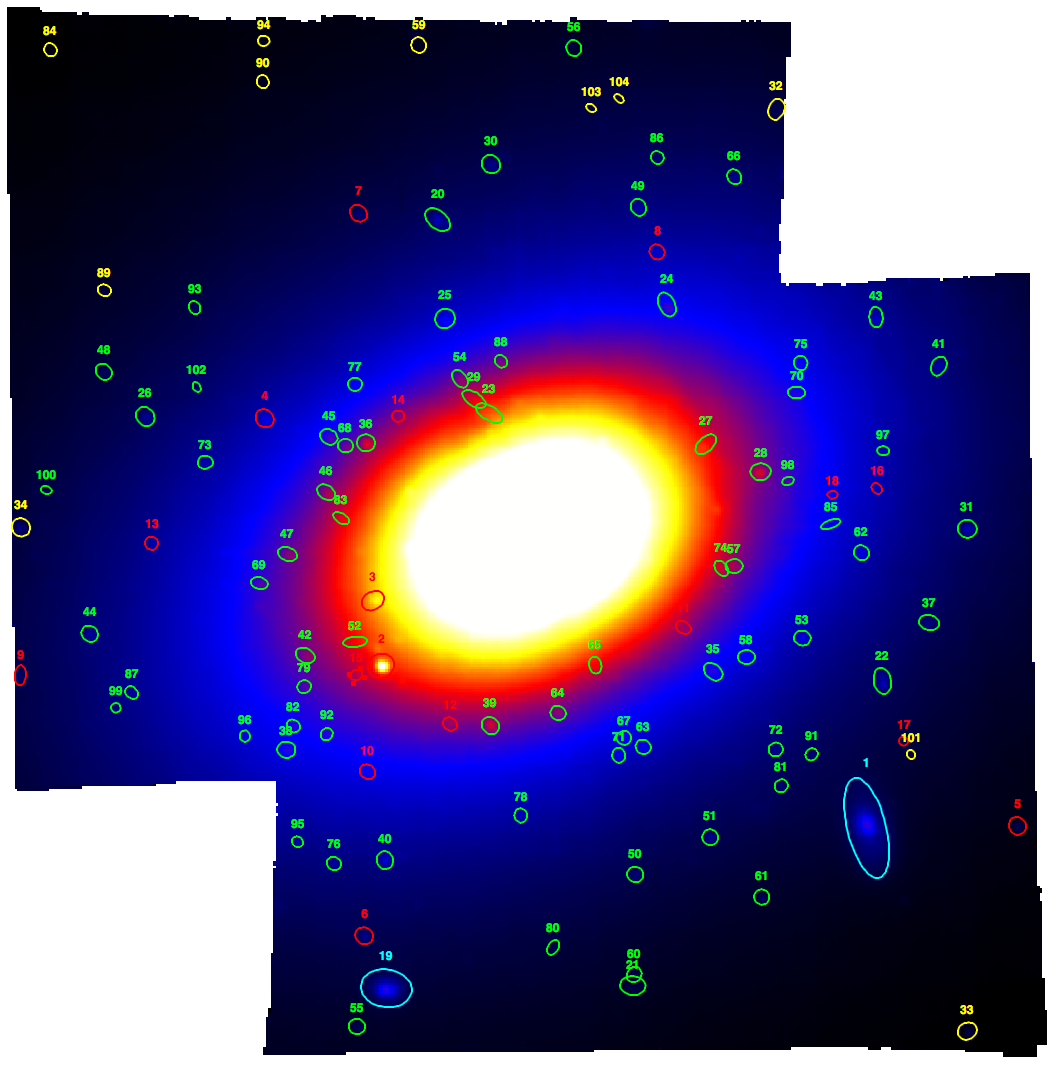}
\caption{The white light image of NGC\,1052 from MUSE data cube. The different symbols indicate the sources listed in Table\,\ref{T_ext_emission}. See text for the color coding.}
\label{Fig_white_reg}
\end{figure}

\begin{figure*}
\centering
\includegraphics[width=.75\textwidth]{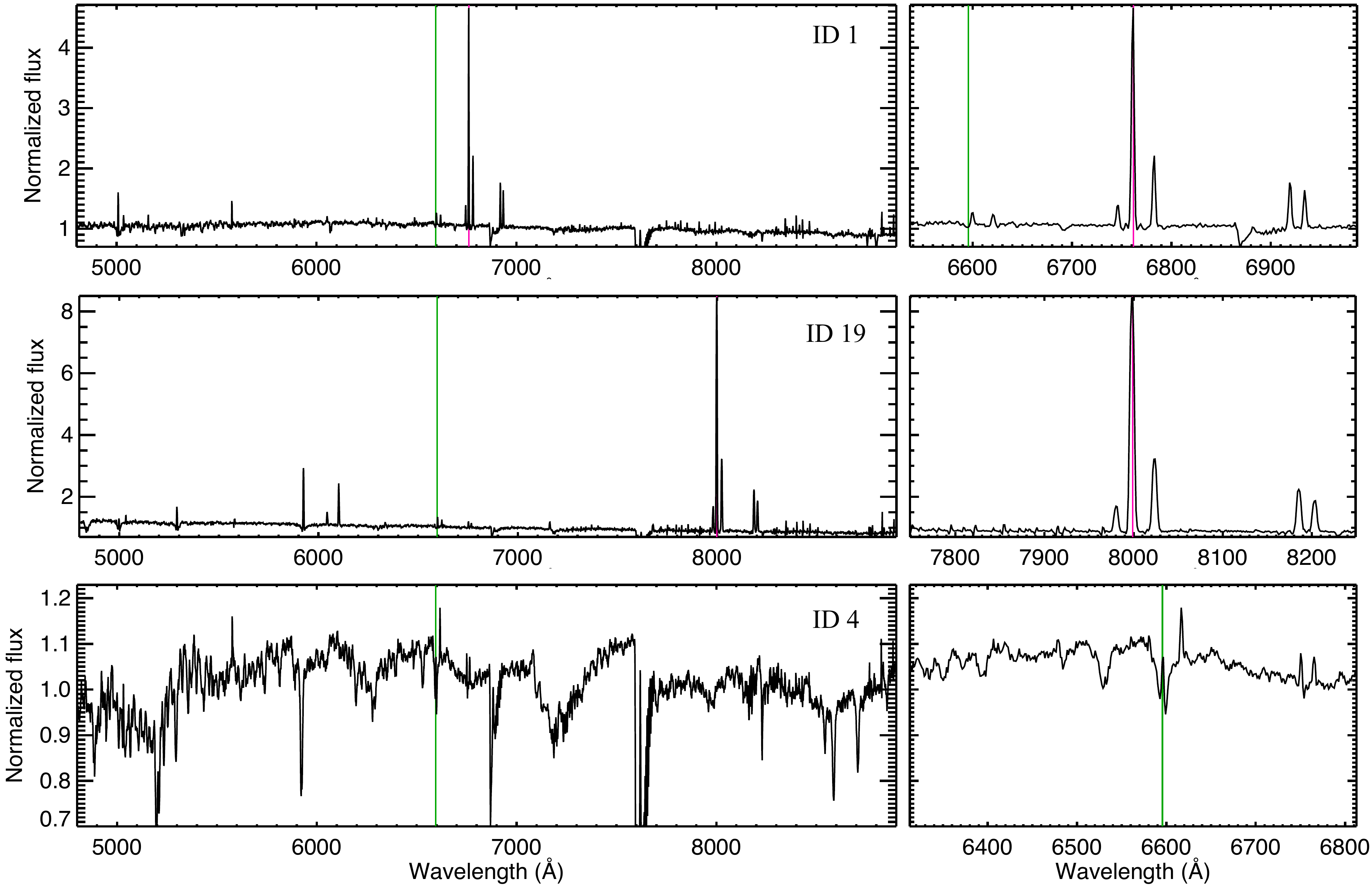}
\caption{Spectra of the objects with ID \lq1\rq , \lq19\rq\ and \lq 4\rq \ from Table\,\ref{T_ext_emission}, see also Fig.\,\ref{Fig_white_reg}. The right panel show a zoomed-in view of the spectra region around H$\alpha$. The green lines mark the wavelength of  H$\alpha$ at the redshift of NGC\,1052 (z\,=\,0.005, Table\,\ref{T_properties}). The magenta lines in the first and second panels mark the wavelength of H$\alpha$ at the corresponding redshift (see text for detail).}
\label{Fig_spec_sources}
\end{figure*}

To account for the \lq external\rq \ (background/foreground) emission from sources different from NGC\,1052, from the white-light MUSE frame shown in Fig.\,\ref{Fig_white_reg}, we create a \lq sharp-divided\rq \ image \citep[see][]{Marquez1999, Marquez2003}. The latter is obtained by dividing the original image, by a filtered version of it (generated using the \texttt{IRAF} using the command \lq median\rq \ with a box of 15 pixels). In the final  sharp-divided  image the identification of features departing from axisymmetry is facilitated (see e.g. \citealt{Cazzoli2018}, \citealt{LHM2020, LHM2022}). \\
Then we used the \lq sharp-divided\rq \ image as input for the \texttt{EXTRACTOR} tool as part of the \texttt{GAIA} (Graphical Astronomy and Image Analysis Tool, \citealt{Draper2009}) package  through the \texttt{STARLINK} software \citep{Currie2014} currently supported by the east Asian Observatory. We considered a 4$\sigma$ threshold, a minimum area of 8 pixels (in order to avoid the inclusion of noise-spikes) and MUSE parameters from the manual (i.e. detector gain: 1.1 ADU/\textit{e}$^{-}$, readout noise: 2.6\,\textit{e}$^{-}$ and saturation 65\,000\,\textit{e}$^{-}$). With these prescriptions, we generated a catalogue of putative background/foreground objects.\\ 
The catalogue is composed by 104 objects (excluding the NGC\,1052 nucleus), these are marked in Fig.\,\ref{Fig_white_reg} in different colors and listed in Table\,\ref{T_ext_emission} along with their position. \\
Of these, 18 (marked in red in Fig.\,\ref{Fig_white_reg}) are present in the NED database (within less than 3$\arcsec$ from the position measured in the MUSE field) being classified either as \lq Infrared source\rq, \lq Radio source\rq , \lq X-ray Source\rq , \lq UltraViolet Source\rq \ or \lq Star Cluster\rq. \\
We visually inspected the spectra of the remaining 85 objects. For the large majority of these spectra (74 out of 85, green symbols in Fig.\,\ref{Fig_white_reg}), the H$\alpha$ line is clearly visible, either in emission or in absorption, at the same redshift of NGC\,1052 (see Table\,\ref{T_properties}). Hence, we will consider these sources as part of the galaxy. The remaining 11 out of 86 sources are mostly located at the edge of the MUSE field of view with barely detected emission/absorption lines in their (noisy) spectra. Hence these will be considered as non-detections (yellow symbols in Fig.\,\ref{Fig_white_reg}). \\
Only in two cases, the spectra show a strong continuum and evident emission-line features, typically of emission-line galaxies. These are \lq1\rq \ and \lq19\rq \ in Table\,\ref{T_ext_emission}, and are marked in cyan in Fig.\,\ref{Fig_white_reg}.  Source \lq1\rq \ (Table\,\ref{T_ext_emission}) has a counterpart in NED with identification SDSSCGB\,67616.02, but with no redshift mentioned. In our data the H$\alpha$ emission is observed at wavelength of $6761.86$\,\AA, resulting in a redshift measurement of $0.0303$. Source \lq19\rq \ (Table\,\ref{T_ext_emission}) is not present in the NED database (within 3$\arcsec$ from its measured position). The H$\alpha$ emission peaks at 7999.09\,\AA, hence the redshift is $\sim$0.2189. The spectra of both galaxies are shown in Fig.\,\ref{Fig_spec_sources}, along with an example of the spectrum of a source having H$\alpha$ in emission/absorption at the same redshift as NGC\,1052, for comparison. \\
The emission from the two external galaxies was masked out from the final MUSE datacube used for the analysis, whereas that of sources at the same redshift of NGC\,1052 were included.\\

\newpage

\newpage

\begin{table*}
\caption[]{Coordinates of the sources marked in Fig.\,\ref{Fig_white_reg}.}
\begin{center}
\tiny{\begin{tabular}{l  c  c|l c c|l c c}
\hline \hline
ID &  RA & DEC & ID & RA & DEC\\
\hline
1 &02$^{h}$41$^{m}$02$^{s}$.928 &-08$^{d}$15$^{m}$43$^{s}$.83 & 36 &02$^{h}$41$^{m}$05$^{s}$.640 &-08$^{d}$15$^{m}$12$^{s}$.88 & 71 &02$^{h}$41$^{m}$04$^{s}$.270 &-08$^{d}$15$^{m}$37$^{s}$.98 \\
2 &02$^{h}$41$^{m}$05$^{s}$.558 &-08$^{d}$15$^{m}$30$^{s}$.79 & 37 &02$^{h}$41$^{m}$02$^{s}$.590 &-08$^{d}$15$^{m}$27$^{s}$.32 & 72 &02$^{h}$41$^{m}$03$^{s}$.421 &-08$^{d}$15$^{m}$37$^{s}$.51 \\
3 &02$^{h}$41$^{m}$05$^{s}$.601 &-08$^{d}$15$^{m}$25$^{s}$.56 & 38 &02$^{h}$41$^{m}$06$^{s}$.070 &-08$^{d}$15$^{m}$37$^{s}$.54 & 73 &02$^{h}$41$^{m}$06$^{s}$.510 &-08$^{d}$15$^{m}$14$^{s}$.44 \\
4 &02$^{h}$41$^{m}$06$^{s}$.188 &-08$^{d}$15$^{m}$10$^{s}$.90 & 39 &02$^{h}$41$^{m}$04$^{s}$.966 &-08$^{d}$15$^{m}$35$^{s}$.60 & 74 &02$^{h}$41$^{m}$03$^{s}$.715 &-08$^{d}$15$^{m}$22$^{s}$.96 \\
5 &02$^{h}$41$^{m}$02$^{s}$.112 &-08$^{d}$15$^{m}$43$^{s}$.65 & 40 &02$^{h}$41$^{m}$05$^{s}$.537 &-08$^{d}$15$^{m}$46$^{s}$.41 & 75 &02$^{h}$41$^{m}$03$^{s}$.285 &-08$^{d}$15$^{m}$06$^{s}$.45 \\
6 &02$^{h}$41$^{m}$05$^{s}$.649 &-08$^{d}$15$^{m}$52$^{s}$.50 & 41 &02$^{h}$41$^{m}$02$^{s}$.538 &-08$^{d}$15$^{m}$06$^{s}$.70 & 76 &02$^{h}$41$^{m}$05$^{s}$.813 &-08$^{d}$15$^{m}$46$^{s}$.69 \\
7 &02$^{h}$41$^{m}$05$^{s}$.680 &-08$^{d}$14$^{m}$54$^{s}$.42 & 42 &02$^{h}$41$^{m}$05$^{s}$.969 &-08$^{d}$15$^{m}$29$^{s}$.96 & 77 &02$^{h}$41$^{m}$05$^{s}$.699 &-08$^{d}$15$^{m}$08$^{s}$.16 \\
8 &02$^{h}$41$^{m}$04$^{s}$.063 &-08$^{d}$14$^{m}$57$^{s}$.53 & 43 &02$^{h}$41$^{m}$02$^{s}$.878 &-08$^{d}$15$^{m}$02$^{s}$.77 & 78 &02$^{h}$41$^{m}$04$^{s}$.801 &-08$^{d}$15$^{m}$42$^{s}$.82 \\
9 &02$^{h}$41$^{m}$07$^{s}$.527 &-08$^{d}$15$^{m}$31$^{s}$.64 & 44 &02$^{h}$41$^{m}$07$^{s}$.135 &-08$^{d}$15$^{m}$28$^{s}$.21 & 79 &02$^{h}$41$^{m}$05$^{s}$.974 &-08$^{d}$15$^{m}$32$^{s}$.48 \\
10 &02$^{h}$41$^{m}$05$^{s}$.631 &-08$^{d}$15$^{m}$39$^{s}$.31 & 45 &02$^{h}$41$^{m}$05$^{s}$.840 &-08$^{d}$15$^{m}$12$^{s}$.39 & 80 &02$^{h}$41$^{m}$04$^{s}$.625 &-08$^{d}$15$^{m}$53$^{s}$.43 \\
11 &02$^{h}$41$^{m}$03$^{s}$.922 &-08$^{d}$15$^{m}$27$^{s}$.70 & 46 &02$^{h}$41$^{m}$05$^{s}$.855 &-08$^{d}$15$^{m}$16$^{s}$.83 & 81 &02$^{h}$41$^{m}$03$^{s}$.391 &-08$^{d}$15$^{m}$40$^{s}$.44 \\
12 &02$^{h}$41$^{m}$05$^{s}$.184 &-08$^{d}$15$^{m}$35$^{s}$.46 & 47 &02$^{h}$41$^{m}$06$^{s}$.065 &-08$^{d}$15$^{m}$21$^{s}$.81 & 82 &02$^{h}$41$^{m}$06$^{s}$.033 &-08$^{d}$15$^{m}$35$^{s}$.63 \\
13 &02$^{h}$41$^{m}$06$^{s}$.801 &-08$^{d}$15$^{m}$20$^{s}$.94 & 48 &02$^{h}$41$^{m}$07$^{s}$.059 &-08$^{d}$15$^{m}$07$^{s}$.15 & 83 &02$^{h}$41$^{m}$05$^{s}$.773 &-08$^{d}$15$^{m}$18$^{s}$.93 \\
14 &02$^{h}$41$^{m}$05$^{s}$.465 &-08$^{d}$15$^{m}$10$^{s}$.73 & 49 &02$^{h}$41$^{m}$04$^{s}$.164 &-08$^{d}$14$^{m}$53$^{s}$.94 & 84 &02$^{h}$41$^{m}$07$^{s}$.348 &-08$^{d}$14$^{m}$41$^{s}$.28 \\
15 &02$^{h}$41$^{m}$05$^{s}$.694 &-08$^{d}$15$^{m}$31$^{s}$.53 & 50 &02$^{h}$41$^{m}$04$^{s}$.181 &-08$^{d}$15$^{m}$47$^{s}$.57 & 85 &02$^{h}$41$^{m}$03$^{s}$.123 &-08$^{d}$15$^{m}$19$^{s}$.40 \\
16 &02$^{h}$41$^{m}$02$^{s}$.874 &-08$^{d}$15$^{m}$16$^{s}$.55 & 51 &02$^{h}$41$^{m}$03$^{s}$.776 &-08$^{d}$15$^{m}$44$^{s}$.57 & 86 &02$^{h}$41$^{m}$04$^{s}$.062 &-08$^{d}$14$^{m}$49$^{s}$.94 \\
17 &02$^{h}$41$^{m}$02$^{s}$.729 &-08$^{d}$15$^{m}$36$^{s}$.84 & 52 &02$^{h}$41$^{m}$05$^{s}$.698 &-08$^{d}$15$^{m}$28$^{s}$.87 & 87 &02$^{h}$41$^{m}$06$^{s}$.908 &-08$^{d}$15$^{m}$32$^{s}$.94 \\
18 &02$^{h}$41$^{m}$03$^{s}$.114 &-08$^{d}$15$^{m}$17$^{s}$.05 & 53 &02$^{h}$41$^{m}$03$^{s}$.276 &-08$^{d}$15$^{m}$28$^{s}$.56  & 88 &02$^{h}$41$^{m}$04$^{s}$.909 &-08$^{d}$15$^{m}$06$^{s}$.31 \\
19 &02$^{h}$41$^{m}$05$^{s}$.529 &-08$^{d}$15$^{m}$56$^{s}$.71 & 54 &02$^{h}$41$^{m}$05$^{s}$.130 &-08$^{d}$15$^{m}$07$^{s}$.72 & 89 &02$^{h}$41$^{m}$07$^{s}$.055 &-08$^{d}$15$^{m}$00$^{s}$.61 \\
20 &02$^{h}$41$^{m}$05$^{s}$.251 &-08$^{d}$14$^{m}$54$^{s}$.93 & 55 &02$^{h}$41$^{m}$05$^{s}$.689 &-08$^{d}$15$^{m}$59$^{s}$.77 & 90 &02$^{h}$41$^{m}$06$^{s}$.197 &-08$^{d}$14$^{m}$43$^{s}$.84\\
21 &02$^{h}$41$^{m}$04$^{s}$.194 &-08$^{d}$15$^{m}$56$^{s}$.50 & 56 &02$^{h}$41$^{m}$04$^{s}$.546 &-08$^{d}$14$^{m}$40$^{s}$.81 & 91 &02$^{h}$41$^{m}$03$^{s}$.227 &-08$^{d}$15$^{m}$37$^{s}$.91\\
22 &02$^{h}$41$^{m}$02$^{s}$.843 &-08$^{d}$15$^{m}$32$^{s}$.02 & 57 &02$^{h}$41$^{m}$03$^{s}$.646 &-08$^{d}$15$^{m}$22$^{s}$.78 & 92 &02$^{h}$41$^{m}$05$^{s}$.853 &-08$^{d}$15$^{m}$36$^{s}$.28\\
23 &02$^{h}$41$^{m}$04$^{s}$.973 &-08$^{d}$15$^{m}$10$^{s}$.48 & 58 &02$^{h}$41$^{m}$03$^{s}$.579 &-08$^{d}$15$^{m}$30$^{s}$.09 & 93 &02$^{h}$41$^{m}$06$^{s}$.567 &-08$^{d}$15$^{m}$02$^{s}$.00\\
24 &02$^{h}$41$^{m}$04$^{s}$.011 &-08$^{d}$15$^{m}$01$^{s}$.75 & 59 &02$^{h}$41$^{m}$05$^{s}$.370 &-08$^{d}$14$^{m}$39$^{s}$.94 & 94 &02$^{h}$41$^{m}$06$^{s}$.193 &-08$^{d}$14$^{m}$40$^{s}$.57\\
25 &02$^{h}$41$^{m}$05$^{s}$.211 &-08$^{d}$15$^{m}$02$^{s}$.88 & 60 &02$^{h}$41$^{m}$04$^{s}$.187 &-08$^{d}$15$^{m}$55$^{s}$.61 & 95 &02$^{h}$41$^{m}$06$^{s}$.010 &-08$^{d}$15$^{m}$44$^{s}$.94 \\
26 &02$^{h}$41$^{m}$06$^{s}$.835 &-08$^{d}$15$^{m}$10$^{s}$.75 & 61 &02$^{h}$41$^{m}$03$^{s}$.496 &-08$^{d}$15$^{m}$49$^{s}$.37 &  96 &02$^{h}$41$^{m}$06$^{s}$.295 &-08$^{d}$15$^{m}$36$^{s}$.44 \\
27 &02$^{h}$41$^{m}$03$^{s}$.800 &-08$^{d}$15$^{m}$13$^{s}$.00  & 62 &02$^{h}$41$^{m}$02$^{s}$.957 &-08$^{d}$15$^{m}$21$^{s}$.70 & 97 &02$^{h}$41$^{m}$02$^{s}$.838 &-08$^{d}$15$^{m}$13$^{s}$.51 \\
28 &02$^{h}$41$^{m}$03$^{s}$.503 &-08$^{d}$15$^{m}$15$^{s}$.19 & 63 &02$^{h}$41$^{m}$04$^{s}$.139 &-08$^{d}$15$^{m}$37$^{s}$.31 & 98 &02$^{h}$41$^{m}$03$^{s}$.354 &-08$^{d}$15$^{m}$15$^{s}$.93 \\
29 &02$^{h}$41$^{m}$05$^{s}$.054 &-08$^{d}$15$^{m}$09$^{s}$.35  & 64 &02$^{h}$41$^{m}$04$^{s}$.599 &-08$^{d}$15$^{m}$34$^{s}$.58 & 99 &02$^{h}$41$^{m}$06$^{s}$.994 &-08$^{d}$15$^{m}$34$^{s}$.16 \\
30 &02$^{h}$41$^{m}$04$^{s}$.963 &-08$^{d}$14$^{m}$50$^{s}$.46 & 65 &02$^{h}$41$^{m}$04$^{s}$.398 &-08$^{d}$15$^{m}$30$^{s}$.73 & 100 &02$^{h}$41$^{m}$07$^{s}$.370 &-08$^{d}$15$^{m}$16$^{s}$.66 \\
31 &02$^{h}$41$^{m}$02$^{s}$.383 &-08$^{d}$15$^{m}$19$^{s}$.77 & 66 &02$^{h}$41$^{m}$03$^{s}$.645 &-08$^{d}$14$^{m}$51$^{s}$.49 & 101 &02$^{h}$41$^{m}$02$^{s}$.688 &-08$^{d}$15$^{m}$37$^{s}$.91 \\
32 &02$^{h}$41$^{m}$03$^{s}$.417 &-08$^{d}$14$^{m}$46$^{s}$.06 & 67 &02$^{h}$41$^{m}$04$^{s}$.243 &-08$^{d}$15$^{m}$36$^{s}$.59 &102 &02$^{h}$41$^{m}$06$^{s}$.556 &-08$^{d}$15$^{m}$08$^{s}$.38 \\
33 &02$^{h}$41$^{m}$02$^{s}$.384 &-08$^{d}$16$^{m}$00$^{s}$.15 & 68 &02$^{h}$41$^{m}$05$^{s}$.750 &-08$^{d}$15$^{m}$13$^{s}$.10 &103 &02$^{h}$41$^{m}$04$^{s}$.420 &-08$^{d}$14$^{m}$45$^{s}$.94 \\
34 &02$^{h}$41$^{m}$07$^{s}$.524 &-08$^{d}$15$^{m}$19$^{s}$.67 & 69 &02$^{h}$41$^{m}$06$^{s}$.217 &-08$^{d}$15$^{m}$24$^{s}$.14 &104 &02$^{h}$41$^{m}$04$^{s}$.269 &-08$^{d}$14$^{m}$45$^{s}$.20 \\
35 &02$^{h}$41$^{m}$03$^{s}$.759 &-08$^{d}$15$^{m}$31$^{s}$.27 & 70 &02$^{h}$41$^{m}$03$^{s}$.308 &-08$^{d}$15$^{m}$08$^{s}$.82 & && \\
\hline  
\end{tabular}
\label{T_ext_emission}}
\end{center} 
\tiny{Notes. ---  \lq ID\rq \ indicate the sources numbered in Fig.\,\ref{Fig_white_reg}. \lq RA\rq \ and \lq DEC\rq \ are the coordinates from MUSE data.}  
\end{table*}

\newpage
\section{Spectral Maps}
\label{Appendix_B}

The first part of this appendix is devoted to present the ionised gas velocity, velocity dispersion and flux maps for NGC\,1052 from MUSE and MEGARA data, for the two spatially resolved components detected (Sect.\,\ref{S_main_results}). In the second part, we show maps for the line ratios of the standard diagnostic diagrams used to pinpoint the ionisation mechanisms of the ISM gas in NGC\,1052. \\  
As for the figures in the main text, the north is at the top and east to the left in all the panels.  The center (0,0) is identified with the photometric center (see Fig.\,\ref{Fig_MM_continuum}). The black solid line indicates the major axis of the stellar rotation (i.e. 122$^{\circ}$, Table\,\ref{T_kinematics}). The dot-dashed square indicates the MEGARA field of view, as e.g. in Fig.\,\ref{Fig_EW_NaD_abs}.\\
Flux intensity maps are in units of erg\,s$^{-1}$\,cm$^{-2}$ and mJy for MUSE and MEGARA, respectively, and displayed on logarithmic-scale. Line-fluxes are not converted to a common units as we are mainly interested in the analysis of line-ratios and we do not compare directly line-fluxes of the two data sets (due to e.g. differences in spatial resolution).

\begin{figure*}
\vspace{1cm}
\centering
\includegraphics[trim = 1.1cm 6.cm 1.2cm 7.2cm,  clip=true, width=1.\textwidth]{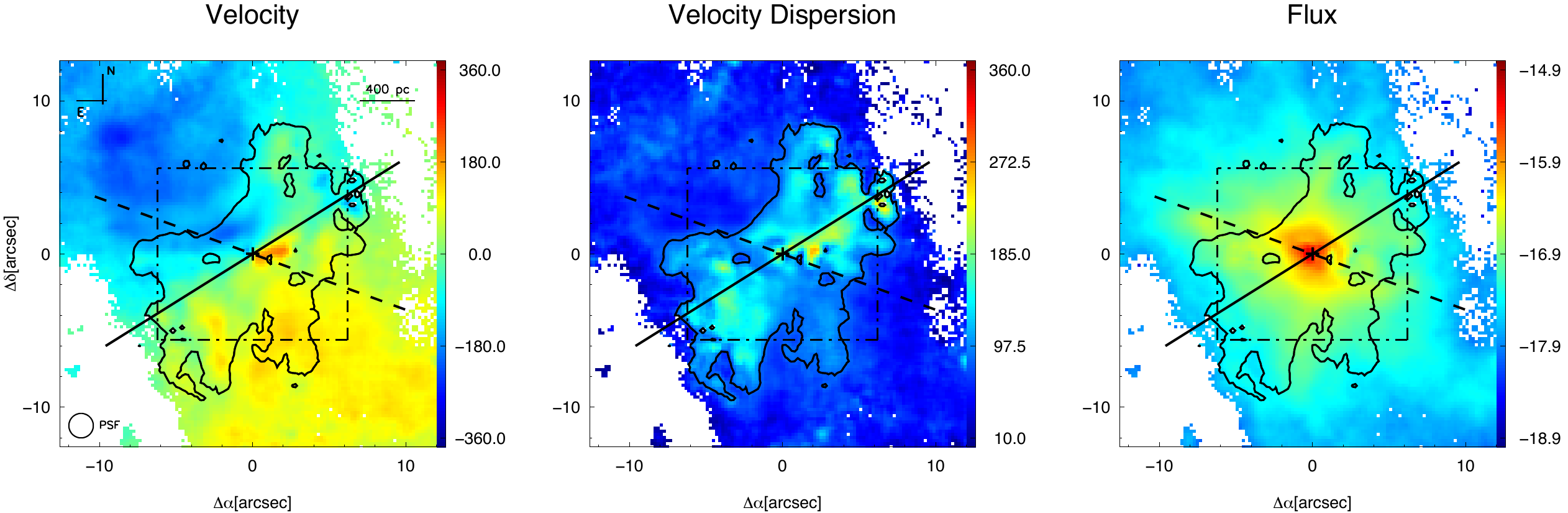}
\caption{The [O\,III]$\lambda$5007 velocity field (\nskms), velocity dispersion (\nskms) and flux intensity (erg\,s$^{-1}$\,cm$^{-2}$) maps for the narrow component. The maps are as in the lower panel of Fig.\,\ref{Fig_combo_OIII}, and are included here as reference. The black solid line indicate the major axis of the stellar rotation (Table\,\ref{T_kinematics}). The dashed lines indicates the orientation of the radio-jet (Table\,\ref{T_properties}).}
\vspace{1cm}
\label{M_OIII_primary_zoom} 		 
\centering
\includegraphics[trim = 1.1cm 6.cm 1.2cm 7.2cm,  clip=true, width=1.\textwidth]{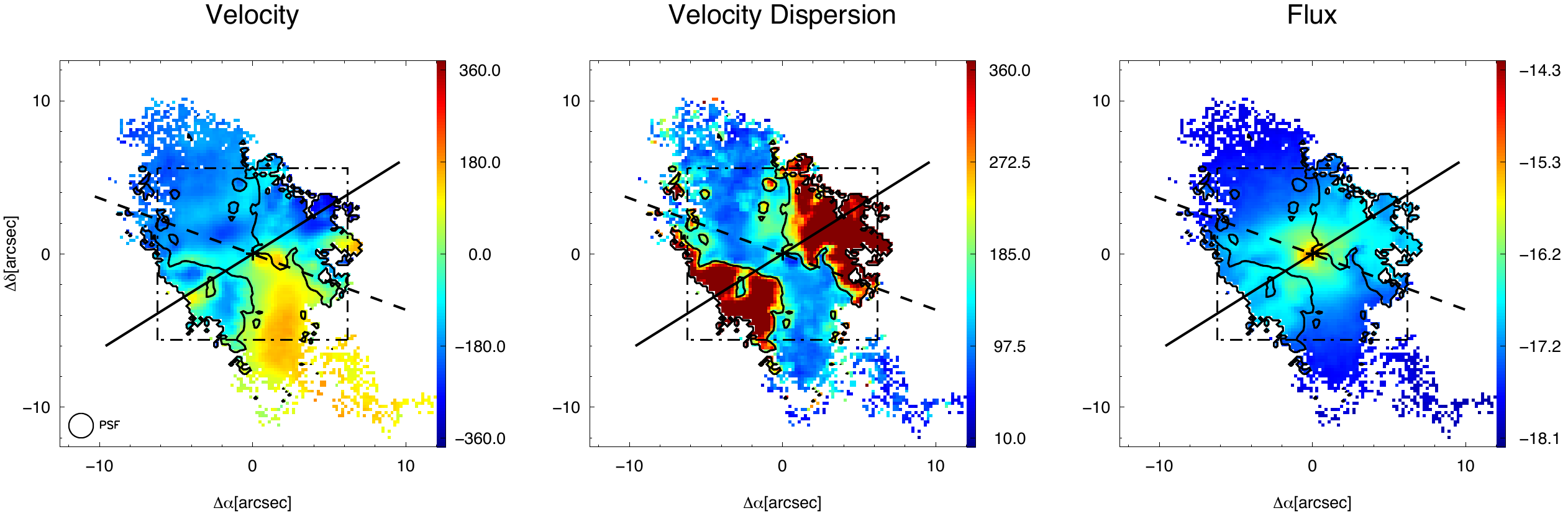}
\caption{As in Fig.\,\ref{M_OIII_primary_zoom} but for [OI]$\lambda$6300.}
\label{M_OI_primary_zoom} 
\vspace{1cm}		 
\centering
\includegraphics[trim = 1.1cm 6.cm 1.2cm 7.2cm,  clip=true, width=1.\textwidth]{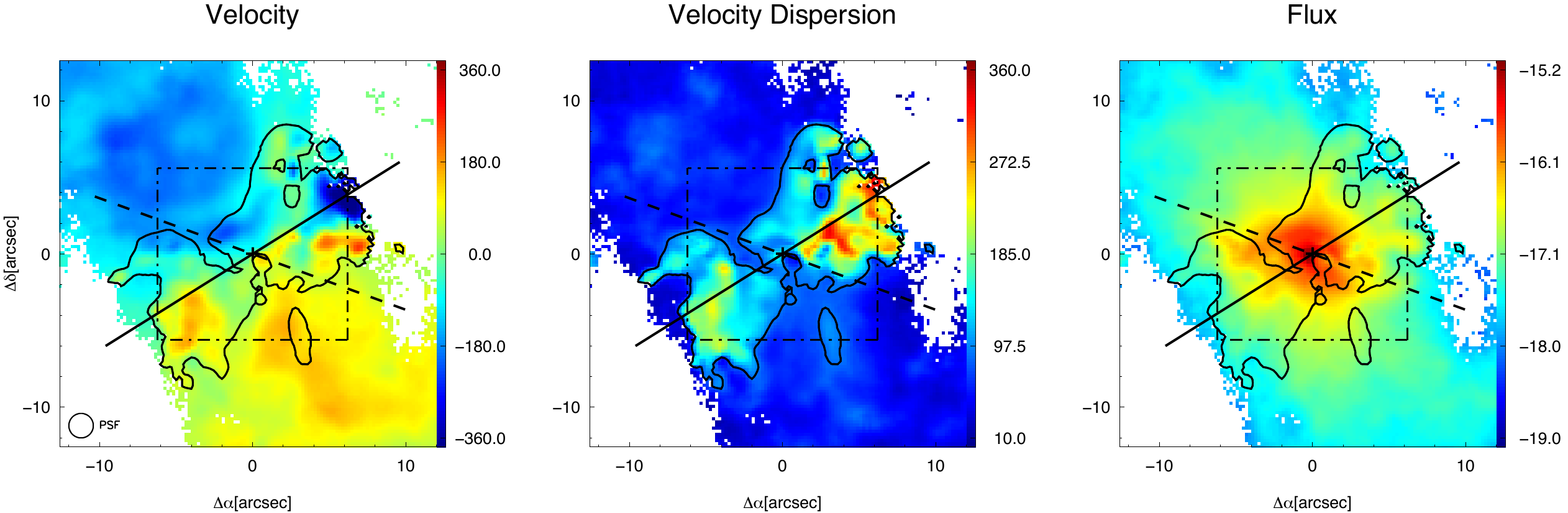}
\caption{As in Fig.\,\ref{M_OIII_primary_zoom} but for H$\alpha$.}
\label{M_Ha_primary_zoom} 		 
\vspace{1cm}
\end{figure*}
\begin{figure*}
\vspace{1cm}
\centering
\includegraphics[trim = 1.1cm 6.cm 1.2cm 7.2cm,  clip=true, width=1.\textwidth]{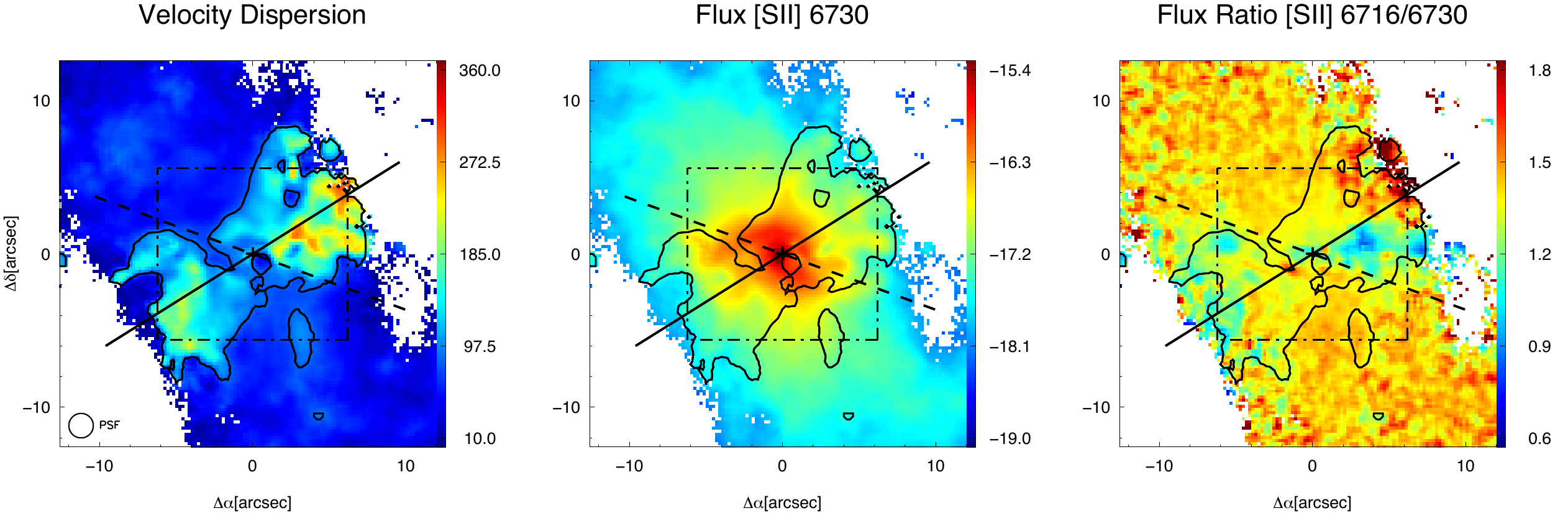}
\caption{Maps of velocity dispersion and flux for the [S\,II]$\lambda$6730 line, and the [S\,II] flux ratio for the primary component. Lines, cross, dashed square and contours are as in Fig.\,\ref{M_OIII_primary_zoom}.}
\label{M_SII_primary_zoom} 
\vspace{1cm}		 
\centering
\includegraphics[trim = 1.1cm 5.8cm 1.cm 7.2cm,  clip=true, width=1.\textwidth]{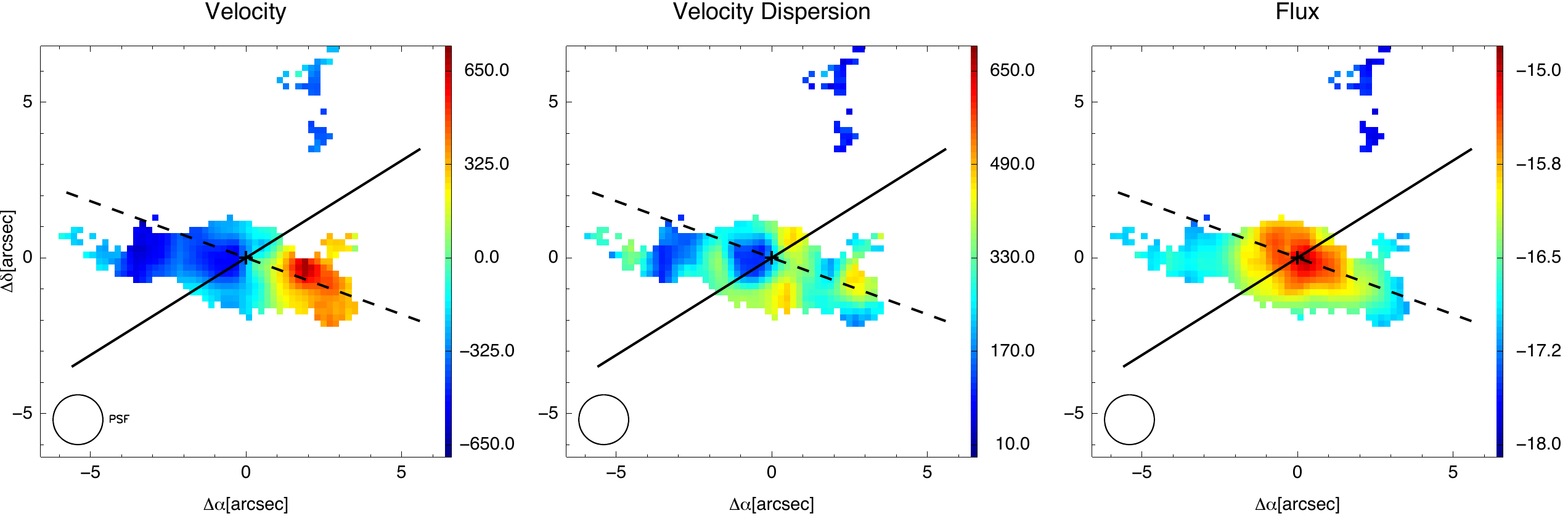}
\caption{As in Fig.\,\ref{M_OIII_primary_zoom} but for the second component and for smaller field of view (similar to the MEGARA footprint).}
\label{M_OIII_second_zoom} 
\vspace{1cm}		 
\centering
\includegraphics[trim = 1.1cm 5.8cm 1.cm 7.2cm,  clip=true, width=1.\textwidth]{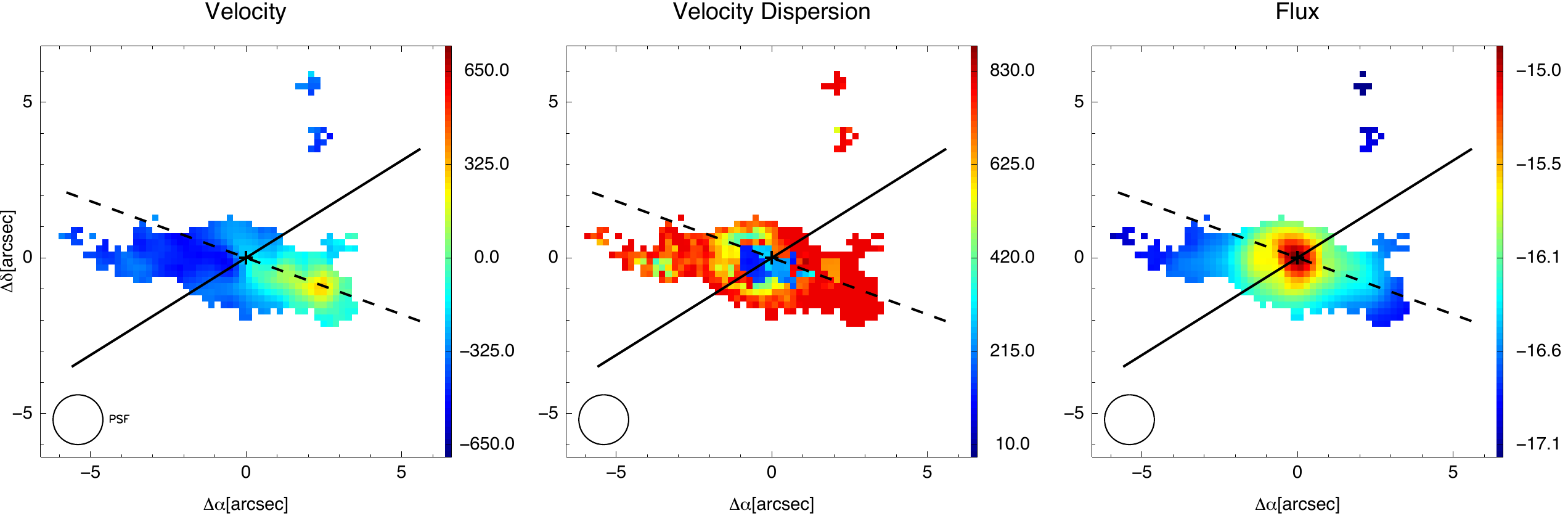}
\caption{As in Fig.\,\ref{M_OIII_second_zoom} but for [OI]$\lambda$6300. }
\label{M_OI_second_zoom} 	
\vspace{1cm}	 
\end{figure*}

\begin{figure*}
\vspace{-0.3cm}	 
\centering
\includegraphics[trim = 1.1cm 5.8cm 1.cm 7.2cm,  clip=true, width=1.\textwidth]{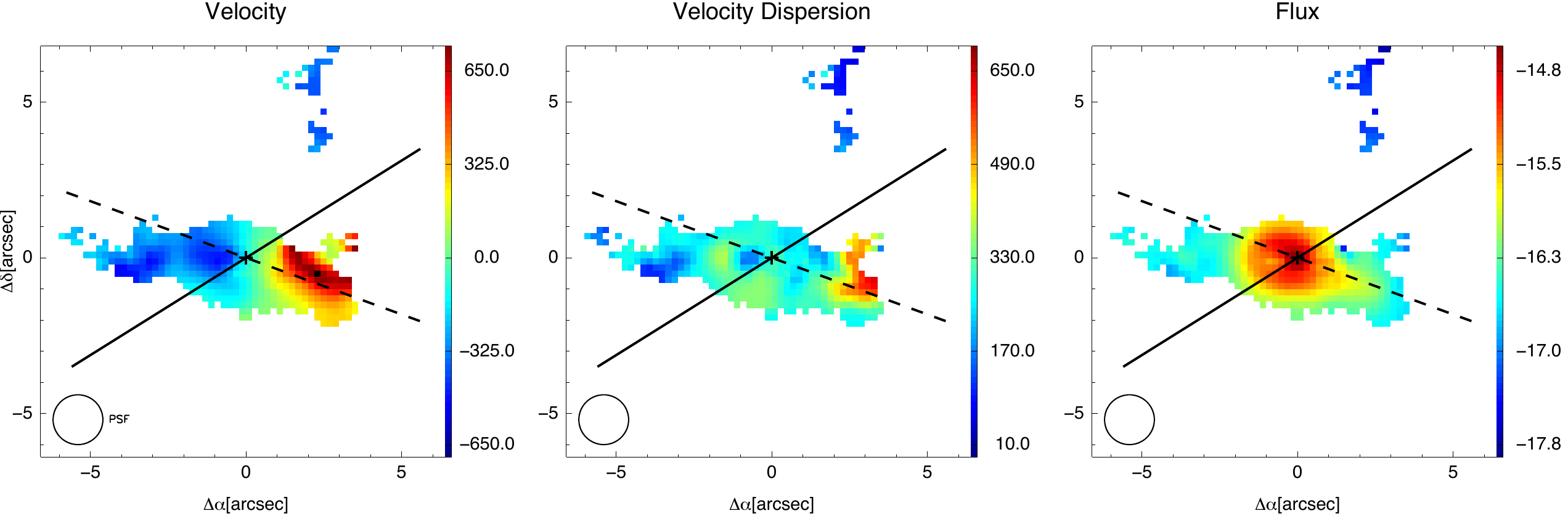}
\caption{As in Fig.\,\ref{M_OIII_second_zoom} but for H$\alpha$. }
\label{M_Ha_second_zoom} 	
\vspace{0.05cm}	 
\centering
\includegraphics[trim = 1.1cm 5.8cm 1.cm 7.2cm,  clip=true, width=1.\textwidth]{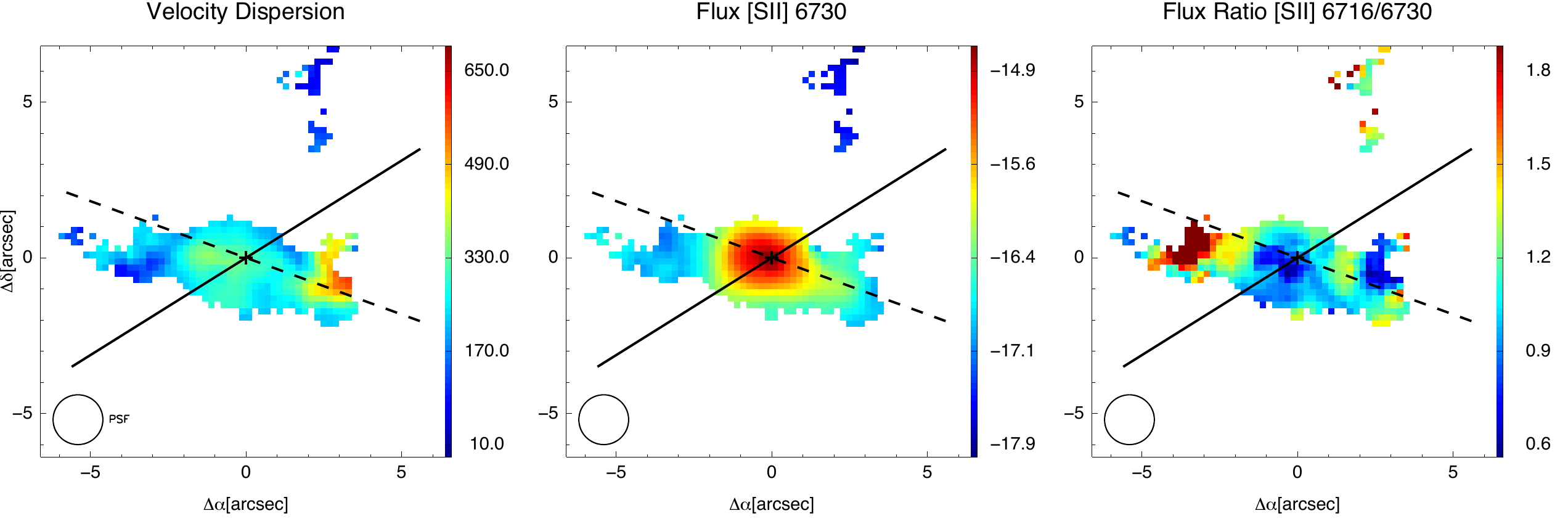}
\caption{As in Fig.\,\ref{M_SII_primary_zoom} but for the second component and for smaller field of view (similar to the MEGARA footprint, see e.g Fig.\,\ref{M_OIII_second_zoom}).}
\label{M_SII_second_zoom} 	
\vspace{0.05cm}	 
\centering
\includegraphics[width=1.\textwidth]{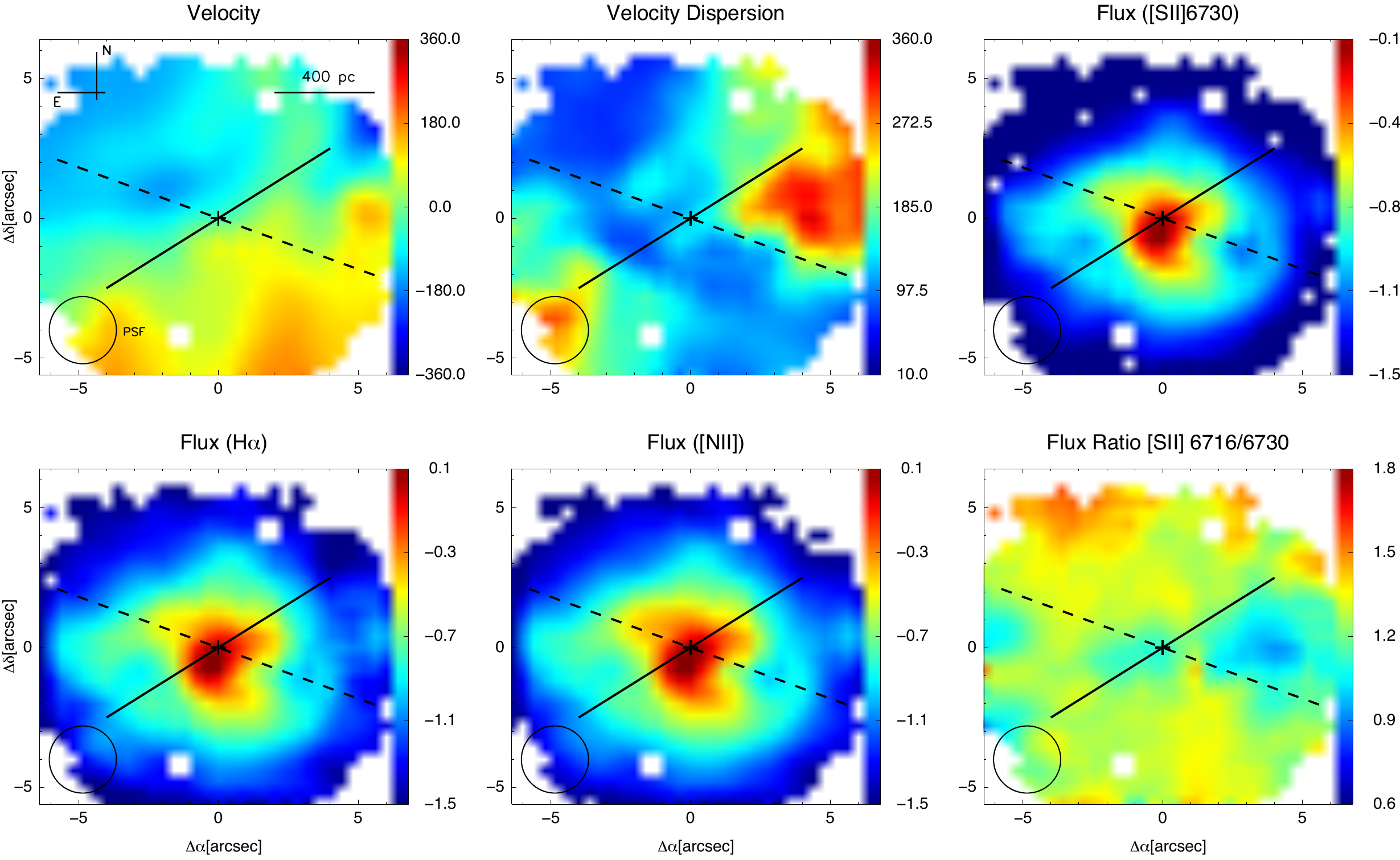}
\caption{Maps of velocity, velocity dispersion and flux maps for the narrow component from MEGARA cube. Flux maps correspond to lines tied together, e.g. [S\,II] doublet and H$\alpha$-[N\,II] complex, see Sect.\,\ref{S_line_mod}. Lines and cross are as in Fig.\,\ref{M_OIII_primary_zoom}.}
\label{M_SII_primary_megara} 
	
\end{figure*}

\begin{figure*}
\vspace{0.25cm}	 
\centering
\includegraphics[width=1.\textwidth]{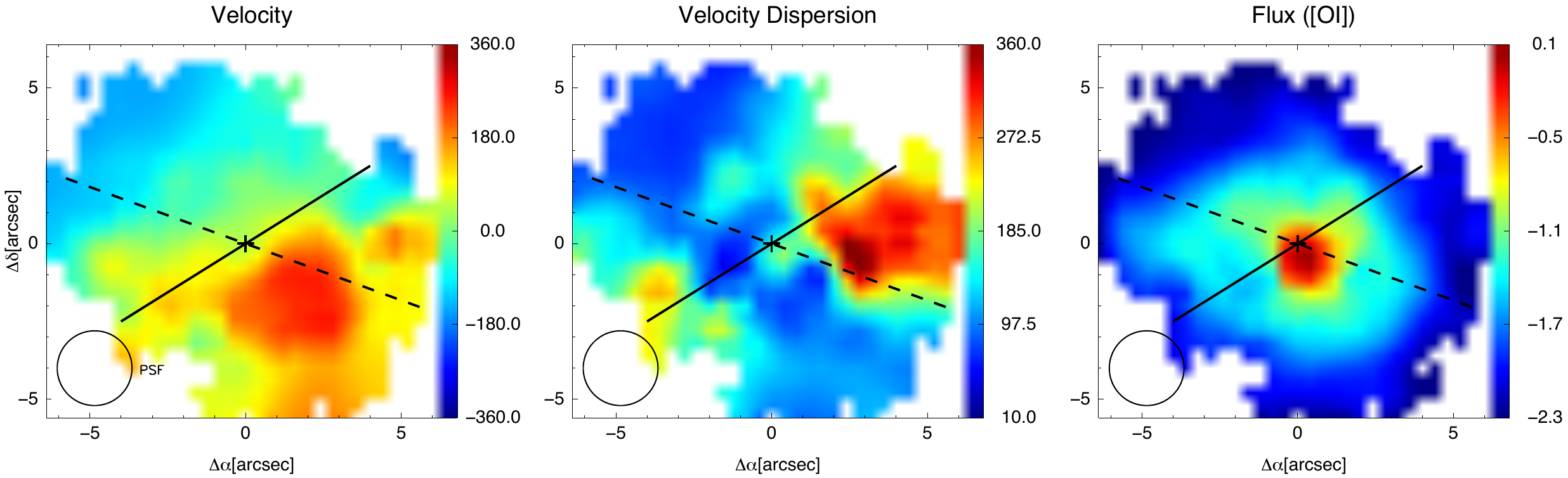}
\caption{The same as Fig.\ref{M_SII_primary_megara} but for [O\,I]. This line has been modelled separately from [S\,II], see Sect.\,\ref{S_line_mod}.  }
\label{M_OI_primary_megara} 
\vspace{0.25cm}	 
\centering
\includegraphics[width=1.\textwidth]{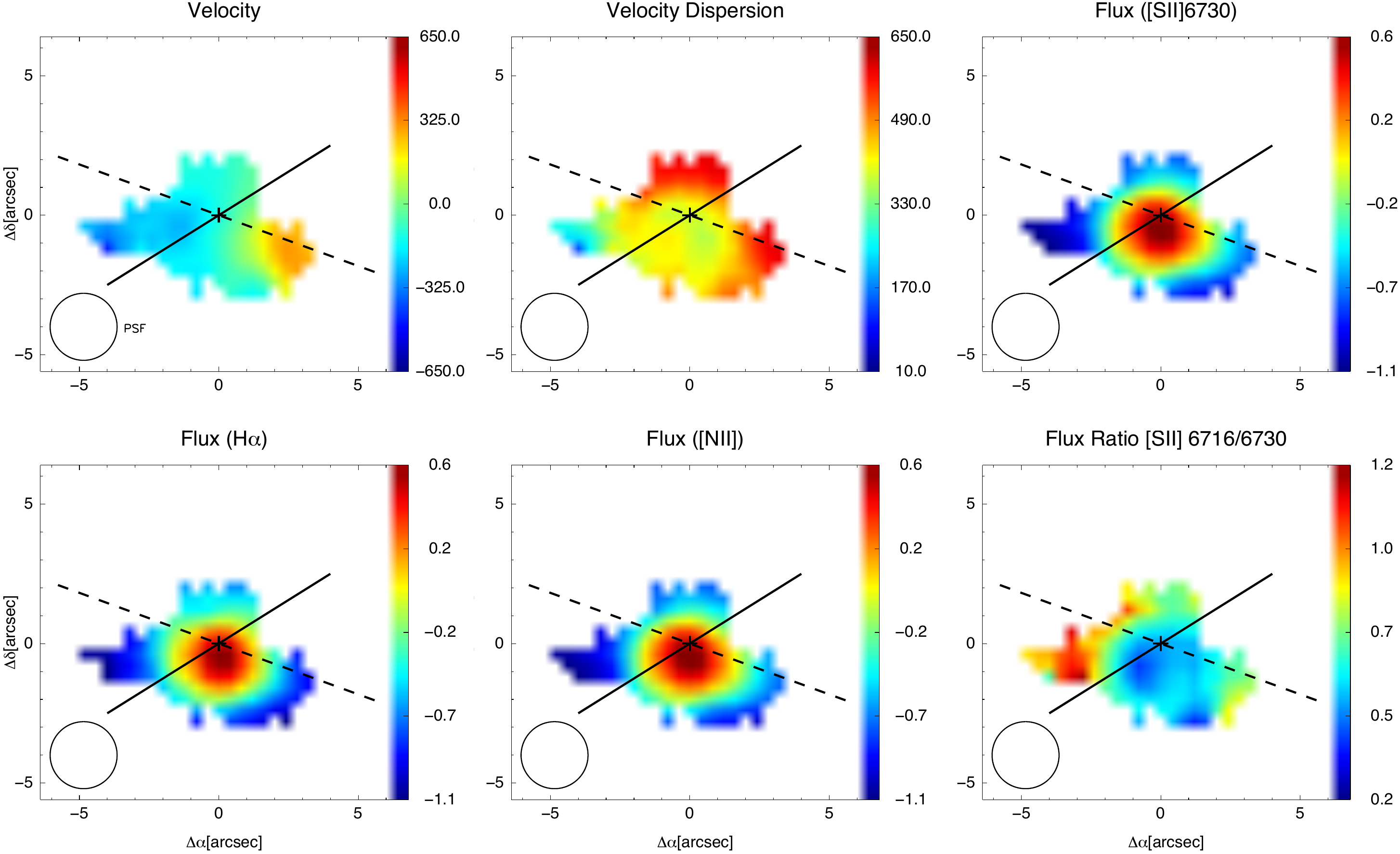}
\caption{As in Fig.\,\ref{M_SII_primary_megara}  but for the second component.}
\label{M_SII_second_megara} 	
\vspace{0.25cm}	 
\centering
\includegraphics[width=1.\textwidth]{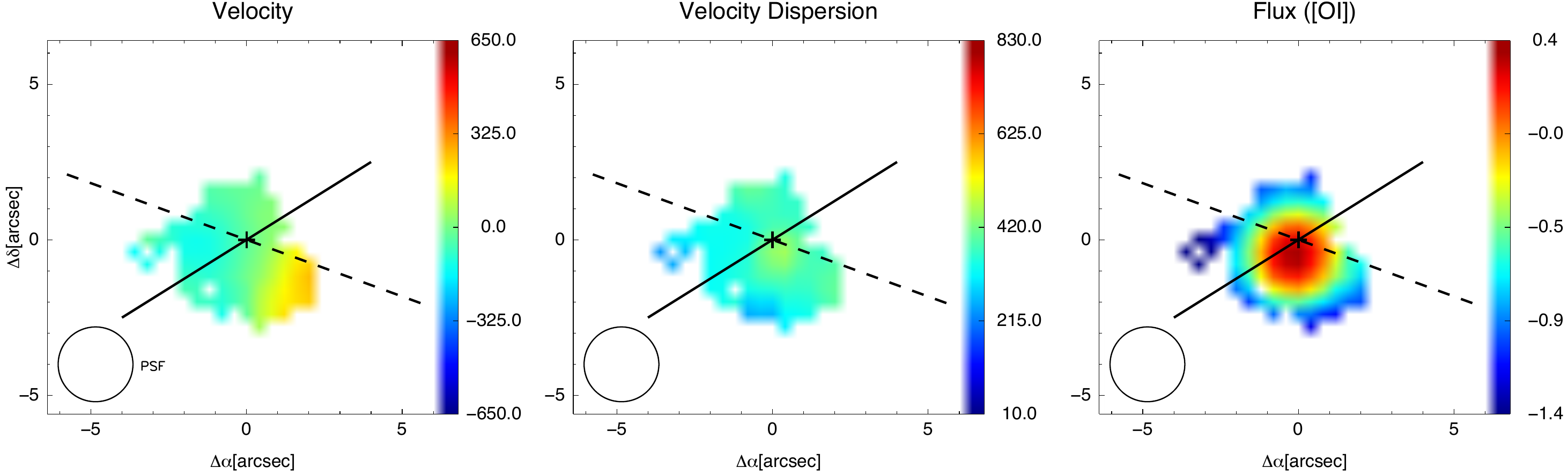}
\caption{As in Fig.\,\ref{M_SII_second_megara}  but for the [O\,I] line. }
\label{M_OI_second_megara} 	
\vspace{0.25cm}	 
\end{figure*}


\begin{figure*}
\vspace{0.5cm}
\centering
\includegraphics[width=1\textwidth]{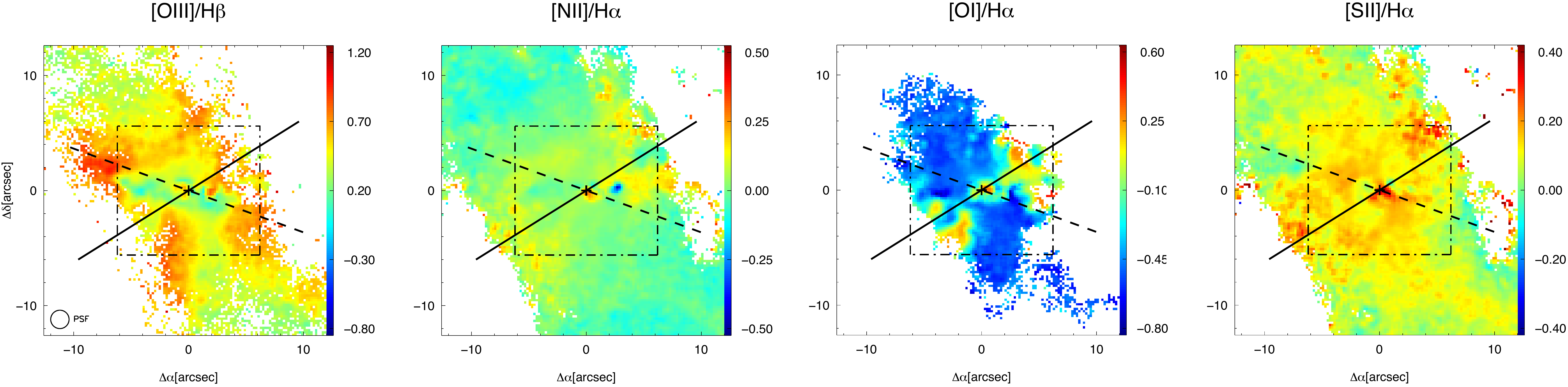}
\caption{Maps of the standard BPT line ratios (labelled on the top) for the narrow component (see also Fig.\,\ref{Fig_BPT_primary}). Lines are as in Fig.\,\ref{M_OIII_primary_zoom}.}
\label{M_BPT_primary_zoom} 	
\vspace{0.5cm}	 
\centering
\includegraphics[width=1.\textwidth]{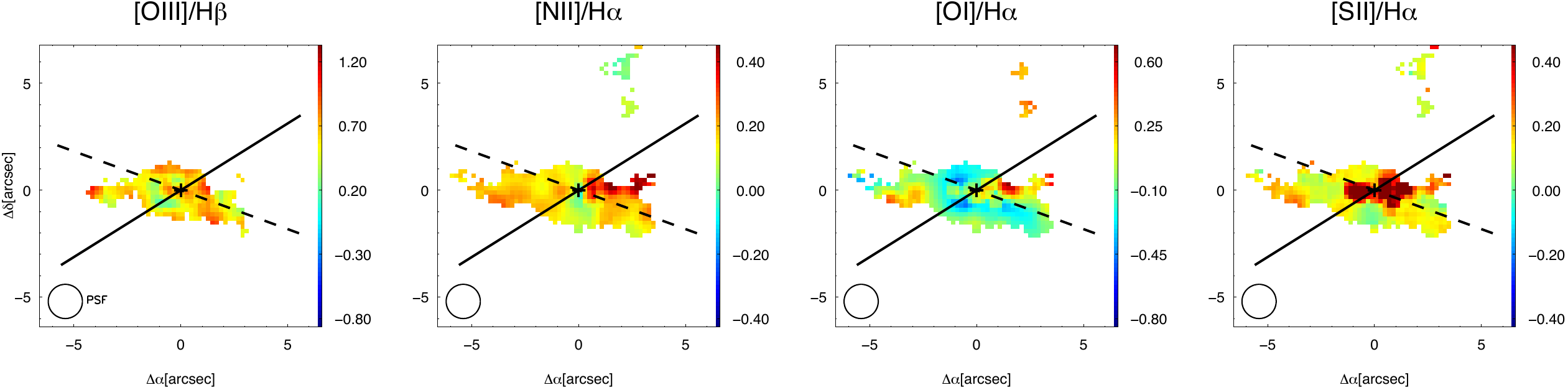}
\caption{As in Fig.\,\ref{M_BPT_primary_zoom}  but for the second component and for smaller field of view (similar to the MEGARA footprint, see e.g Fig.\,\ref{M_OIII_second_zoom}).}
\label{M_BPT_second_zoom} 	
\vspace{0.5cm}
\centering
\includegraphics[trim = 0.1cm 0.1cm 0.1cm 0.1cm,  clip=true,width=1.\textwidth]{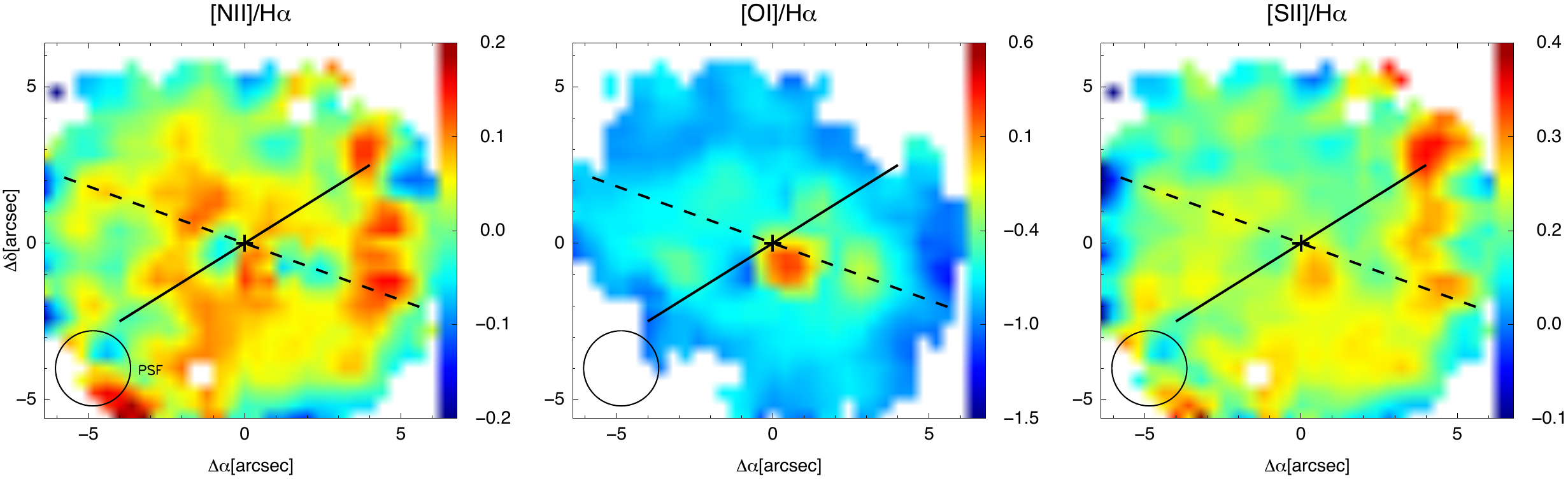}
\caption{Maps of the standard BPT line ratios (labelled on the top) for the narrow component. Lines are as in Fig.\,\ref{M_OIII_primary_zoom}.}
\label{M_BPT_primary_megara} 
\vspace{0.5cm}
\centering
\includegraphics[trim = 0.1cm 0.1cm 0.1cm 0.1cm,  clip=true,width=1.\textwidth]{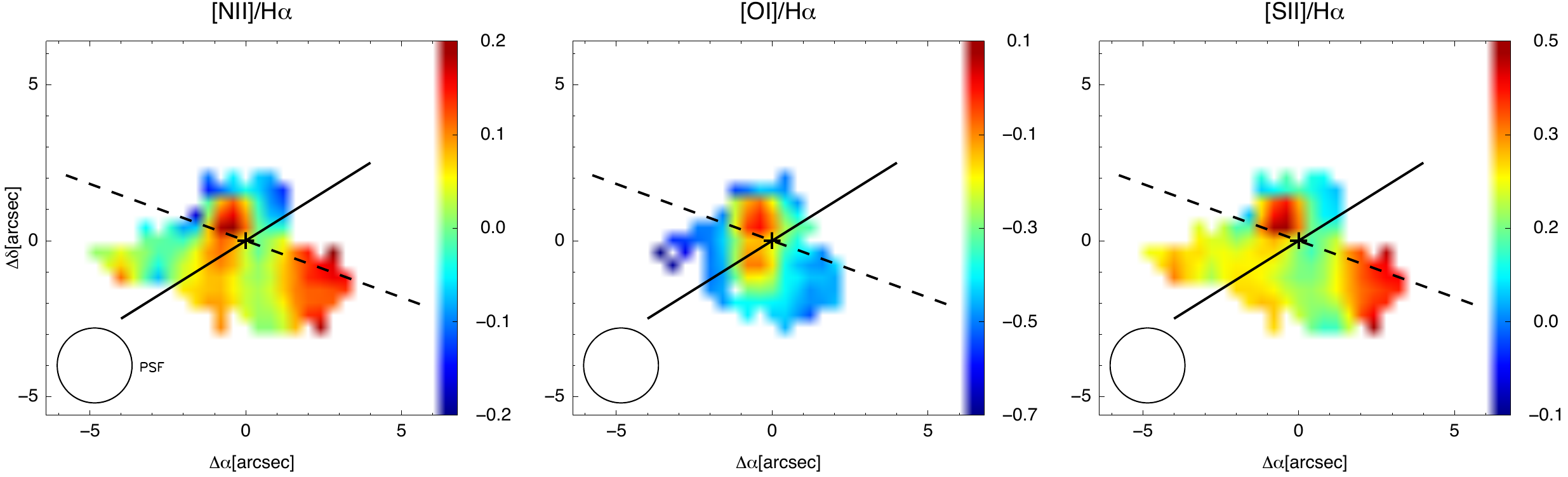}
\caption{As in Fig.\,\ref{M_BPT_primary_megara} but for the second component. The long-dashed line mark the PA of the 20\,cm emission (Table\,\ref{T_properties}) as in Fig.\,\ref{M_OIII_second_zoom}.}
\label{M_BPT_second_megara} 
\vspace{0.5cm}
\end{figure*}

\newpage

\section{Position-Velocity and Position-Dispersion diagrams}
\label{Appendix_C}
This appendix is devoted to the present Position-Velocity (P-V, top) and Position-Dispersion (P-$\sigma$, bottom)  for the primary component used to model emission lines in NGC\,1052 (MUSE data).\\
Both diagrams suggest that the kinematics of the stars is completely decoupled from that of the ionised gas. The stellar P-V and P-$\sigma$ curve show clear signature of a rotating disc, whereas it is not the case for the gas component. Specifically, along both major and minor photometric axis (left and center panels) the ionised gas shows a very perturbed P-V curve as well as an asymmetric and not centrally peaked P-$\sigma$ curve.\\
The curves extracted from the slit aligned according to the radio jet do not show any intriguing feature.
 
\begin{figure*}
\centering
\includegraphics[width=1.\textwidth]{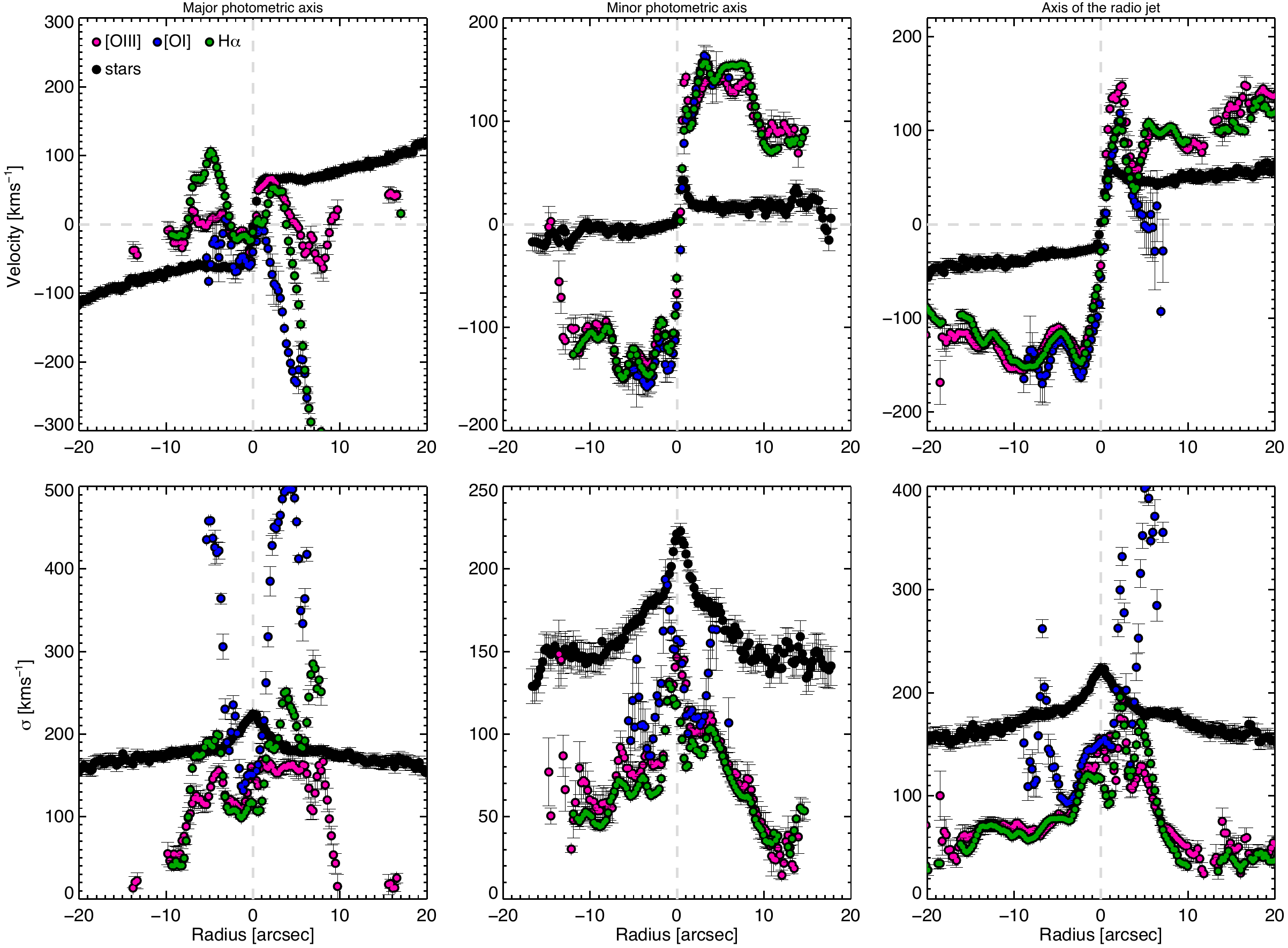}
\caption{Position-Velocity (P-V, top) and Position-Velocity Dispersion (P-$\sigma$, bottom) curves of the stellar (black) and gas component of NGC\,1052 from MUSE data. Specifically, the gas component is probed via [O\,III] (magenta), [O\,I] (blue) and H\,$\alpha$  (green) emission lines. Similarly to Fig.\,\ref{Fig_PVPS} the curves were obtained considering a pseudo-slit of 1$\arcsec$-width aligned according to the major (left) and minor (center) axis of the rotation, as well the axis of the radio jet (right). Position angles are listed in Tables  \ref{T_properties} and \ref{T_kinematics}. Velocities are centred to the kinematic center, and the radius is calculated as the distance from the photometric center. Grey dashed lines show zero-points for position and velocity, as reference.}
\label{P_kin} 
\end{figure*}

\end{appendix}  
\end{document}